%% file: Manuscript.tex
\documentclass[11pt,a4paper]{article}
\pdfoutput=1
\usepackage{jheppub}
\usepackage[T1]{fontenc}

\usepackage{amsmath}
\usepackage{mathtools}
\usepackage{bm}
\usepackage{graphicx}
\usepackage{esvect}
\usepackage{fancyhdr}
\usepackage{graphicx}
\usepackage{graphics}
\usepackage{amssymb}
\usepackage{enumitem}
\usepackage{braket}
\usepackage{xcolor}
\usepackage[colorlinks = true]{hyperref}
\usepackage{subfigure}
\usepackage{tikz}
\usetikzlibrary{decorations.pathreplacing} 

\definecolor{egg}{rgb}{.98,.97,.92}
\definecolor{astroorange}{rgb}{1,.93,.79}
\definecolor{darkorange}{rgb}{1,.89,.6}
\definecolor{dullblue}{rgb}{.29,.47,.77}
\definecolor{grayblue}{rgb}{.98,.98,.98}
\definecolor{fadedblue}{rgb}{.78,.86,.92}
\definecolor{tiffanyblue}{rgb}{.96,1,1}
\definecolor{grayish}{rgb}{.93,.93,.97}
\definecolor{charcoal}{rgb}{.247,.259,.27}
\definecolor{evergreen}{rgb}{.7725,.858,.7647}
\definecolor{dullred}{rgb}{.929,.498,.598}
\definecolor{lavender}{rgb}{.8,.741,.85}

\newcommand{\tr}[1]{\mathrm{Tr}\left\{#1\right\}}
\newcommand{\ptr}[2]{\mathrm{Tr}_{#1}\left\{#2\right\}}

\newcommand{\av}[1]{\underset{\tiny{#1}}{\mathbb{E}}}
\newcommand{\norm}[1]{\left|\left| #1 \right|\right|}
\newcommand{\abs}[1]{\left| #1 \right|}

\newcommand{\xpos}{0} 
\newcommand{\ypos}{0} 
\newcommand{\xrel}{1.4} 
\newcommand{\yrel}{.5} 
\newcommand{\xback}{.4*\xrel} 
\newcommand{\yback}{1.1*\yrel} 
\newcommand{\xleg}{.75*\xrel} 
\newcommand{\yleg}{2*\yrel} 
\newcommand{\yleglong}{2.5*\yrel} 
\newcommand{\Aleggap}{.2*\yleglong} 
\newcommand{\xcircle}{1} 
\newcommand{\ycircle}{2*\yrel} 
\newcommand{\xrecenter}{.15} 
\newcommand{\xglobalshift}{0} 
\newcommand{\yglobalshift}{0} 
\newcommand{\height}{2em} 
\newcommand{\width}{2em} 
\newcommand{\name}{} 
\newcommand{\nodenum}{0} 
\newcommand{\heightsingle}{2em} 
\newcommand{\heightdouble}{4.6em} 
\newcommand{\widthsingle}{2em} 
\newcommand{\widthdouble}{4em} 
\newcommand{\heightstate}{.75em} 
\newcommand{\widthstate}{.75em} 
\newcommand{\circlescale}{.65} 
\newcommand{\rowspace}{2*\yrel} 
\newcommand{\linethick}{.4mm} 

\newtheorem{theorem}{Theorem}
\newtheorem{lemma}{Lemma}
\newtheorem{prop}{Proposition}
\newtheorem{corollary}{Corollary}
\newtheorem{obs}{Observation}

\title{Barren plateaus from learning scramblers with local cost functions}

\author[a]{Roy J. Garcia,}
\author[a,b,c]{Chen Zhao,}
\author[a]{Kaifeng Bu}
\author[a]{and Arthur Jaffe}
\affiliation[a]{Department of Physics, Harvard University, \\Cambridge, Massachusetts 02138, USA}
\affiliation[b]{Academy of Mathematics and Systems Science, Chinese Academy of Sciences, \\Beijing 100190, China}
\affiliation[c]{University of Chinese Academy of Sciences, \\Beijing 100049, China}

\emailAdd{roygarcia@g.harvard.edu}
\emailAdd{chen\_zhao@g.harvard.edu}
\emailAdd{kfbu@fas.harvard.edu}
\emailAdd{jaffe@g.harvard.edu}

\date{\today}

\abstract{
The existence of barren plateaus has recently revealed new training challenges in quantum machine learning (QML). Uncovering the mechanisms behind barren plateaus is essential in understanding the scope of problems that QML can efficiently tackle. Barren plateaus have recently been shown to exist when learning global properties of random unitaries, which is relevant when learning black hole dynamics. Establishing whether local cost functions can circumvent these barren plateaus is pertinent if we hope to apply QML to quantum many-body systems. We prove a no-go theorem showing that local cost functions encounter barren plateaus in learning random unitary properties.}


\arxivnumber{2205.06679}

\begin{document}
\maketitle
\flushbottom

\section{Introduction}
Quantum machine learning (QML) \cite{Biamonte2017} presents a promising avenue for solving optimization problems via quantum computers and is driven by the hope of realizing a quantum advantage \cite{PhysRevX.4.031002,Bravyi2018,Huang2021} over its classical counterpart. There has recently been an interest in proving training guarantees in QML, which resolve whether these models can be scaled up. One widely-used method to train learning models is (stochastic) gradient descent for cost functions. However, learning models based on quantum neural networks suffer from barren plateaus, meaning that gradients of a cost function are exponentially small. Hence, they prevent efficient training with this method at the large scale  \cite{McClean2018}, as they demand an exponential amount of resources. One way to ensure efficiency in training is to identify circumstances under which barren plateaus can be avoided.

Barren plateaus can arise from using global cost functions~\cite{Cerezo2021}, from using deep random circuits~\cite{McClean2018}, from entanglement~\cite{PRXQuantum.2.040316,PhysRevResearch.3.033090}, from noise~\cite{Wang2021}, and from  expressibility~\cite{PRXQuantum.3.010313}. Recently, Holmes et al.~\cite{Holmes2021} showed that barren plateaus can arise when learning random unitaries. These results also extend to learning scramblers~\cite{PhysRevX.9.031048,PhysRevA.94.040302,xu2022scrambling,Landsman2019,PhysRevResearch.3.033155,Garcia2022Resource}, namely unitaries which spread local information, thereby uncovering a novel physical mechanism for engendering barren plateaus. Moreover, an amalgam of results connecting scrambling and QML has recently emerged~\cite{Shen2020,Wu2021}. For instance, scrambling bounds the generalization capability of quantum neural networks \cite{Choudhury_2021} and also bounds certain cost function gradients~\cite{garcia_JHEP}. Importantly, scrambler-induced barren plateaus inhibit learning the dynamics of chaotic quantum systems such as the mixed-field Ising model \cite{PhysRevLett.106.050405,Xu2019}, the kicked Dicke model \cite{PhysRev.93.99,PhysRevA.99.043602,Lewis2019}, the non-integrable Bose-Hubbard model \cite{Bohrdt2017}, the SYK model~\cite{PhysRevLett.70.3339,Kitaev2015}, and black holes (the fastest scramblers known in nature)~\cite{Sekino2008,Shenker2014,Maldacena2016}. This notably poses a challenge to the Hayden-Preskill decoding protocol, which is used to retrieve quantum information thrown into a black hole, as it relies on learning black hole dynamics~\cite{Hayden2007,yoshida2017efficient,Bao2021}. 

The scrambler-induced barren plateaus identified in~\cite{Holmes2021} were found using a global cost function of the form $\bra{\psi}U^\dagger V H V^\dagger U\ket{\psi}$, where $VHV^\dagger$ is a global observable\footnote{It is interesting to explore whether such a global cost function can be made to appear local if the input state has the form $\rho_{\mathrm{in}}\otimes I^{n-1}/2^{n-1}$.}. In this case, a parameterized unitary $U$ is trained to learn a random unitary $V$ globally. However, it has been proven generally that training with arbitrary global cost functions produces barren plateaus for a variety of QML architectures, such as the alternating layered ansatz \cite{Cerezo2021}, the dissipative QNN~\cite{PhysRevLett.128.180505}, and the matrix product state (MPS) architecture~\cite{liu2021presence}. Therefore, local cost functions must be used to accurately characterize srambler-induced barren plateaus. `Local' refers to measurements taken on a subsystem. Local cost functions can be used to access important quantities in quantum many-body physics. For example, local projected outputs of Haar random unitaries and quantum chaotic dynamics have been shown to produce emergent quantum state designs~\cite{cotler2021emergent,Ho2022, choi2021emergent}. Furthermore, local properties of non-integrable systems are instrumental in revealing the periodic behavior of quantum many-body scars~\cite{Bernien2017,Turner2018}. 

Local cost functions stand out as powerful tools to avoid barren plateaus among other approaches, such as initialization strategies~\cite{Grant2019}, correlated parameters~\cite{Volkoff2021}, and entanglement devised mitigation~\cite{PhysRevResearch.3.033090}. The central question we ask is: can local cost functions be used to avoid barren plateaus when learning random unitary properties? A variety of quantum neural network (QNN) architectures, such as the shallow alternating layered ansatz~\cite{Cerezo2021}, the quantum convolutional neural network (QCNN)~\cite{Cong2019,PhysRevX.11.041011} and the dissipative QNN \cite{PhysRevLett.128.180505} can avoid barren plateaus when training with local cost functions which satisfy a special condition. Recently, it was shown that barren plateaus do not exist in the MPS architecture when training with local cost functions defined via observables which are independent of the system size~\cite{liu2021presence}. We extend this barren plateau analysis to arbitrary local cost functions and provide a physically relevant application.

In Theorem~\ref{Theorem}, we show that, when training with arbitrary local cost functions, barren plateaus can exist in the MPS architecture. On the other hand, we prove a condition on the local cost function which allows one to avoid these barren plateaus. Our main contribution is to prove Theorem~\ref{Thm:XEB}, a no-go result, which states that training QML models to learn random unitary properties with local cost functions produces barren plateaus. This implies that local probability distributions of random quantum states also cannot be learned efficiently. Although we prove the existence of these barren plateaus using the MPS architecture, we show how our results can extend to circuit-based architectures. Our no-go theorem indicates that, in the absence of prior knowledge, QML is incompatible with learning local properties of generic quantum many-body systems.

\section{Main results}
\subsection{MPS architecture}
We define a unitarily embedded MPS~\cite{Heferkamp2021, liu2021presence, Zhao_2021} of local dimension $d$, system size $n$ and virtual bond dimension $D$ as:
\begin{equation}\label{Eq:MPSstate}
\begin{tikzpicture}

	\renewcommand{\nodenum}{v0}
    \renewcommand{\name}{$\ket{\psi(\bm{\theta})}=$}
	\renewcommand{\xpos}{-1.5*\xrel}
    \renewcommand{\ypos}{0}
    \renewcommand{\height}{\heightsingle}
    \renewcommand{\width}{\widthsingle}
    \node[] (\nodenum) at (\xpos,\ypos) {\name};
    
	 \renewcommand{\nodenum}{r1u1}
    \renewcommand{\name}{$U_1$}
	\renewcommand{\xpos}{0}
    \renewcommand{\ypos}{0}
    \renewcommand{\height}{\heightsingle}
    \renewcommand{\width}{\widthsingle}
    \node[rectangle, fill=egg, rounded corners, minimum width=\width, minimum height=\height, draw] (\nodenum) at (\xpos,\ypos) {\name};
    
      \renewcommand{\nodenum}{r1u2}
    \renewcommand{\name}{$U_2$}
	\renewcommand{\xpos}{\xrel}
    \renewcommand{\ypos}{0}
    \renewcommand{\height}{\heightsingle}
    \renewcommand{\width}{\widthsingle}
    \node[rectangle, fill=egg, rounded corners, minimum width=\width, minimum height=\height, draw] (\nodenum) at (\xpos,\ypos) {\name};
    
    \renewcommand{\nodenum}{r1u3}
    \renewcommand{\name}{$U_n$}
	\renewcommand{\xpos}{3*\xrel}
    \renewcommand{\ypos}{0}
    \renewcommand{\height}{\heightsingle}
    \renewcommand{\width}{\widthsingle}
    \node[rectangle, fill=egg, rounded corners, minimum width=\width, minimum height=\height, draw] (\nodenum) at (\xpos,\ypos) {\name};

	 \renewcommand{\nodenum}{r1s1}
    \renewcommand{\name}{$\ket{0}$}
	\renewcommand{\xpos}{0}
    \renewcommand{\ypos}{+\yleg}
    \renewcommand{\height}{\heightstate}
    \renewcommand{\width}{\widthstate}
    \node[] (\nodenum) at (\xpos,\ypos) {\name};
    
      \renewcommand{\nodenum}{r1s2}
    \renewcommand{\name}{$\ket{0}$}
	\renewcommand{\xpos}{\xrel}
    \renewcommand{\ypos}{+\yleg}
    \renewcommand{\height}{\heightstate}
    \renewcommand{\width}{\widthstate}
    \node[] (\nodenum) at (\xpos,\ypos) {\name};
    
    \renewcommand{\nodenum}{r1s3}
    \renewcommand{\name}{$\ket{0}$}
	\renewcommand{\xpos}{3*\xrel}
    \renewcommand{\ypos}{+\yleg}
    \renewcommand{\height}{\heightstate}
    \renewcommand{\width}{\widthstate}
    \node[] (\nodenum) at (\xpos,\ypos) {\name};

	 \renewcommand{\nodenum}{r1u1bot}
	\renewcommand{\xpos}{0}
    \renewcommand{\ypos}{-\yleg}
    \coordinate  (\nodenum) at (\xpos,\ypos) {};
    
      \renewcommand{\nodenum}{r1u2bot}
	\renewcommand{\xpos}{\xrel}
    \renewcommand{\ypos}{-\yleg}
    \coordinate (\nodenum) at (\xpos,\ypos) {};
    
    \renewcommand{\nodenum}{r1u3bot}
	\renewcommand{\xpos}{3*\xrel}
    \renewcommand{\ypos}{-\rowspace}
    \coordinate   (\nodenum)  at (\xpos,\ypos) {};

   \renewcommand{\nodenum}{Ellipses}
    \renewcommand{\name}{$\bm{\cdots}$}
	\renewcommand{\xpos}{2.03*\xrel}
    \renewcommand{\ypos}{.5*\yback}
    \node[] (\nodenum) at (\xpos,\ypos) {\name};
    %


    \draw [line width=\linethick,color=dullred]  
    (r1u1)--++(-\xleg,0)
    (r1u1)--(r1u2)--++(\xleg,0)
    (r1u3)--++(-\xleg,0)
    (r1u3)--++(\xleg,0);
    
	\draw [line width=\linethick,color=dullblue]  
	(r1s1)--(r1u1)--(r1u1bot)
	(r1s2)--(r1u2)--(r1u2bot)
    (r1s3)--(r1u3)--(r1u3bot);

\end{tikzpicture}.
\end{equation}
Each $U_i$ is a parameterized $Dd\times Dd$-dimensional unitary. Blue lines indicate physical indices, while the red lines (with implied periodic boundary conditions) indicate virtual indices. Each $U_i$ has the form $U_i=U_i^{(\textrm{poly}(Dd))}\cdots U_i^{(2)}U_i^{(1)}$, where $U_i^{(k)}=e^{-iG_i^{(k)}\theta^{(k)}_i}$ and $G_i^{(k)}$ is a Hermitian operator. We assume that each parameter $\theta^{(k)}_i\in \bm{\theta}$ is randomly initialized such that each $U_i$ forms a unitary 2-design. The values of $\bra{\psi(\bm{\theta})}\psi(\bm{\theta})\rangle$ are exponentially concentrated around unity.

We show how barren plateaus can arise and be avoided when using local cost functions with the MPS architecture in Eq.~\eqref{Eq:MPSstate}. Our  local cost function is 
\begin{equation}\label{Eq:Cost}
	C=\bra{\psi(\bm{\theta})}I_d^{\otimes m-1}\otimes O \otimes I_d^{\otimes n-m}\ket{\psi(\bm{\theta})}\;,
\end{equation}
where $O$ denotes  a local (single-qudit) Hermitian operator on site $m$ and $I_d$ is the identity on a single qudit. The parameters of $\ket{\psi(\bm{\theta})}$ are trained by optimizing $C$ via gradient descent~\cite{JMLR:v12:duchi11a, reddi2019convergence}.  By utilizing the Weingarten calculus \cite{Collins2006}, it can be shown that the average gradient of $C$ vanishes.

\begin{lemma}\label{Lemma:Grad}
The average of $\partial_i^{(k)}C$ over $\bm{\theta}$ vanishes:
\begin{equation}
	\langle \partial_i^{(k)} C\rangle_{\bm{\theta}}=0.
\end{equation}
\end{lemma}

We can show how $\partial_i^{(k)} C$ concentrates about its average via Chebyshev's inequality. Taking $\delta>0$ and randomly initializing $\bm{\theta}$, $\partial_i^{(k)} C$ satisfies the following concentration inequality: 

\begin{equation}
	\mathrm{Prob}_{\bm{\theta}}\left[\abs{\partial_i^{(k)}C}\geq\delta \right]\leq \frac{\mathrm{Var}_{\bm{\theta}}[\partial_i^{(k)} C]}{\delta^2}.
\end{equation}
If the variance vanishes exponentially\footnote{Since we use $O$ to denote an operator, we use $\mathcal{O}$ to denote the mathematical notion of order.} in $n$ for all $\partial_i^{(k)}$, i.e. $\mathrm{Var}_{\bm{\theta}}[\partial_i^{(k)}C]=\mathcal{O}(\mathrm{exp}(-n))$, then by Chebyshev's inequality, the probability of $\abs{\partial_i^{(k)} C}$ being greater than $\delta$ is exponentially small in $n$. This is referred to as a barren plateau. These exponentially vanishing gradients make training via gradient descent exponentially costly. We say that a barren plateau is avoided if there exists at least one partial derivative $\partial_i^{(k)}$ such that the variance decays at worst polynomially, i.e.  $\mathrm{Var}_{\bm{\theta}}[\partial_i^{(k)} C]=\Omega\left(\frac{1}{\mathrm{poly}(n)}\right)$. This scaling allows us to train the QML model in polynomial time. This leads to our result bounding the variance scaling of the MPS architecture.

\begin{theorem}\label{Theorem}
If $\ptr{d}{O}^2$  and $\norm{O}_{\infty}^2$ grow slower than exponential in $n$, then for large $n$ and fixed $m$, the variance of $\partial_i^{(k)}C$ with respect to $\bm{\theta}$ satisfies
\begin{equation}\label{Eq:VarIneq}
	\mathrm{Var}_{\bm{\theta}}[\partial_i^{(k)} C]\leq \epsilon(O) \mathcal{O}\left(\frac{P(D,d)}{Q(D,d)}\right),
\end{equation}
where $
	\epsilon(O)\equiv \norm{O-\tr{O}\frac{I_d}{d}}_{\mathrm{HS}}^2
$. The functions $P(D,d)$ and $Q(D,d)$ are polynomials of $D$ and $d$. Moreover, there exists a partial derivative $\partial_i^{(k)}$ such that Ineq.~\eqref{Eq:VarIneq} becomes an equality.
\end{theorem}

The polynomials $P(D,d)$ and $Q(D,d)$ are independent of $n$. Hence, the variance scaling with respect to $n$ is completely determined by $\epsilon(O)$. In cases where $O$ is exponentially close to the identity (rescaled by $\tr{O}/d$), $\epsilon(O)$ decays exponentially. Hence,  $O$ must be carefully chosen when defining $C$ in Eq.~\eqref{Eq:Cost}. This produces the following two corollaries.

\begin{corollary}\label{Cor:BP}
	If $\epsilon (O) =\mathcal{O}(\mathrm{exp}(-n))$, then the variance upper bound in Ineq.~\eqref{Eq:VarIneq} decays exponentially in $n$, inducing a barren plateau. 
\end{corollary}

\begin{corollary}\label{Cor:NoBP}
If $\epsilon(O)=\Omega\left(\frac{1}{\mathrm{poly}(n)}\right)$, then by Theorem~\ref{Theorem} there exists a partial derivative $\partial_i^{(k)}$ such that $\mathrm{Var}_{\bm{\theta}}[\partial_i^{(k)}C]=\Omega\left(\frac{1}{\mathrm{poly}(n)}\right)$. Hence, there is no barren plateau. 
\end{corollary}

The bounds in Corollaries~\ref{Cor:BP} and \ref{Cor:NoBP} demonstrate how the choice of $O$ in the local cost function definition can be used to remove barren plateaus. Furthermore, these results highlight previous work showing that barren plateaus do not to exist in the MPS architecture when $\epsilon(O)=\Omega(1)$~\cite{liu2021presence}; this condition is widely used in the literature.
With these results, we now turn our attention to the problem of learning random unitary properties.

\subsection{Barren plateaus from random unitaries}
In this section, we prove our main result, Theorem~\ref{Thm:XEB}. This no-go theorem states that barren plateaus are encountered when learning random unitary properties with local cost functions. Before proving this, we first consider the cost function $C$ from Eq.~\eqref{Eq:Cost} to examine the trainability of the MPS architecture when learning properties of an $n$-qudit, Haar random unitary $V$. We let $O$ in Eq.~\eqref{Eq:Cost} depend on $V$. To assess trainability, we find the typical behavior of $\partial_i^{(k)} C$ via Chebyshev's inequality: $\mathrm{Prob}_{\bm{\theta},V}\left[\abs{\partial_i^{(k)}C}\geq\delta \right]\leq \frac{\mathrm{Var}_{\bm{\theta},V}[\partial_i^{(k)}C]}{\delta^2}$. This bounds the probability that $\abs{\partial_i^{(k)} C}$ is larger than $\delta$ when $\bm{\theta}$ is randomly initialized and $V$ is sampled from the Haar measure on the unitary group. To find the variance scaling, we modify Theorem~\ref{Theorem} to obtain the following lemma.

\begin{lemma}\label{Lemma:Scrambling}
Let $O$ depend on an $n$-qudit, Haar random unitary $V$. If $\int_{\mathrm{Haar}}dV\ptr{d}{O}^2$  and $\int_{\mathrm{Haar}}dV \norm{O}_{\infty}^2$ grow slower than exponentially in $n$, then for large $n$ and fixed $m$, the variance of $\partial_i^{(k)}C$ with respect to $\bm{\theta}$ and $V$ satisfies
\begin{equation}\label{Eq:ScramblingVar}
	\mathrm{Var}_{\bm{\theta},V}[\partial_i^{(k)}C]\leq \left[\int_{\mathrm{Haar}} dV \epsilon(O)\right]\mathcal{O}\left(\frac{P(D,d)}{Q(D,d)}\right),
\end{equation}
where $P(D,d)$ and $Q(D,d)$ are polynomials of $D$ and $d$.
\end{lemma}

This lemma indicates that a barren plateau occurs if $\int_{\mathrm{Haar}} dV \epsilon(O)=\mathcal{O}(\mathrm{exp}(-n))$. We now formulate the problem of learning unitary properties by adopting explicit cost functions, namely the cross-entropy and the linear cross-entropy benchmark. When probing random unitaries and scramblers, we are often interested in the output probability distribution on a subsystem of qubits, such that we learn the unitary's non-local structure. For simplicity, we presently consider the case of learning single-qubit probability distributions and discuss the case of learning non-local distributions later on. 

Let $S=\{\ket{\psi_i},\{p_i(V,x)\}\}_{i=1}^{N_S}$ be a training set with $N_S$ training pairs. Input state $\ket{\psi_i}$ is an $n$-qubit computational basis state. The probability of measuring the first qubit in state $\ket{x}\in\{\ket{0},\ket{1}\}$ when state $V\ket{\psi_i}$ is prepared is
\begin{equation}\label{Eq:ProbDis}
	p_i(V,x)=\bra{\psi_i} V^\dagger(\ket{x}\bra{x}\otimes I^{\otimes n-1})V\ket{\psi_i},
\end{equation}
where $I$ is the single-qubit identity. The set $S$ contains local information about how $V$ maps computational basis states. The probabilities are useful for computing expectation values of other local observables. For simplicity, and without loss of generality, we fix the input state to $\ket{\psi_0}=\ket{0}^{\otimes n}$ so that the target probability is $p(V,x)\equiv p_0(V,x)$. The set $S$ allows for training with classical output data. The case of incoherent training with quantum state training pairs is explored in \cite{Holmes2021} when learning $V$ globally.

The output probabilities of the MPS are ${q(x)=\bra{\psi(\bm{\theta})}(\ket{x}\bra{x}\otimes I^{\otimes n-1})}\ket{\psi(\bm{\theta})}$, where $\ket{\psi(\bm{\theta})}$ is given by Eq.~\eqref{Eq:MPSstate}. We train the MPS such that $\{q(x)\}_x$ replicates $\{p(V,x)\}_x$. We stress that we do \textit{not} require that $\ket{\psi(\bm{\theta})}$ reproduce the output state $V\ket{\psi_0}$, which can generally be a highly entangled state.

To measure how well the MPS distribution $\{q(x)\}_x$ approximates the target distribution $\{p(V,x)\}_x$, we adopt the cross-entropy as our cost function:
\begin{equation}\label{Eq:Cross}
	E(V)= -\sum_{x=0}^{1}q(x)\mathrm{ln}[p(V,x)].
\end{equation}
The MPS is trained to optimize the cross-entropy and thereby learn the probability distribution $\{p(V,x)\}_x$. By defining the local observable 
\begin{equation}
	{O_{E}=-\sum_{x=0}^{1}\mathrm{ln}[p(V,x)]\ket{x}\bra{x}},
\end{equation}
we can write the cross-entropy as a local cost function, \linebreak ${E(V)=\bra{\psi(\bm{\theta})} (O_E\otimes I^{\otimes n-1})}\ket{\psi(\bm{\theta})}$, with a form similar to Eq.~\eqref{Eq:Cost}.

Due to the logarithm in its definition, the cross-entropy is tedious to work with analytically. To prove training guarantees, we concurrently consider a closely related cost function, the linear cross-entropy benchmark (XEB) \cite{Arute_supremacy}:
\begin{equation}\label{Eq:XEB}
	\chi(V)= 2\sum_{x=0}^{1}p(V,x)q(x)-1.
\end{equation}
By defining the local observable 
\begin{equation}
	{O_\chi=\sum_{x=0}^1 (2p(V,x)-1)\ket{x}\bra{x}},
\end{equation}
the linear XEB can be written as a local cost function, $\chi=\bra{\psi(\bm{\theta})}(O_\chi\otimes I^{\otimes n-1})\ket{\psi(\bm{\theta})}$. With this cost function, we state the following no-go theorem.

\begin{theorem}\label{Thm:XEB}
Learning local properties of a Haar random unitary using the linear XEB cost function, $\chi$, produces a barren plateau.
\end{theorem}

This theorem can be proved by first showing that ${\int_{\mathrm{Haar}} dV \epsilon(O_\chi)=\frac{2}{2^n+1}}$. This then implies that $\mathrm{Var}_{\bm{\theta},V}[\partial_i^{(k)} \chi]$ vanishes at least exponentially in $n$ by Lemma~\ref{Lemma:Scrambling}, thereby inducing a barren plateau. Theorem~\ref{Thm:XEB} provides a guarantee that a benchmark of the cross-entropy produces exponentially vanishing gradients. This result complements the following observation on trainability using the cross-entropy cost function.

\begin{obs}\label{Observation}
It can be verified numerically that learning local properties of a Haar random unitary with the cross-entropy cost function, $E$, produces a barren plateau.
\end{obs}

Observation~\ref{Observation} comes from numerically showing that ${\int_{\mathrm{Haar}} dV \epsilon(O_E) =\mathcal{O}(\mathrm{exp}(-n))}$. This implies that $\mathrm{Var}_{\bm{\theta},V}[\partial_i^{(k)} E]$ vanishes at least exponentially in $n$ by Lemma~\ref{Lemma:Scrambling}, producing a barren plateau. Both Theorem~\ref{Thm:XEB} and Observation~\ref{Observation} demonstrate that we encounter a barren plateau when attempting to learn local properties of a random unitary with local cost functions. Furthermore, since $V\ket{\psi_0}$ is a Haar random state, our results also imply that we cannot efficiently learn local probability distributions of a large, generic quantum state.

\subsection{Circuit architectures}
\begin{figure}[h!]
\centering
\begin{tikzpicture}

	\renewcommand{\xpos}{-2*\xrel}
	\renewcommand{\ypos}{-3*\rowspace}
	\draw [decorate,decoration={brace,amplitude=5pt},xshift=\xrel,yshift=-\rowspace, line width =.75*\linethick]	 (\xpos,\ypos)--(\xpos,\ypos+4*\rowspace)  node [black,midway,xshift=9pt] {};

\draw [color=black, line width=.75*\linethick]  
    (-1.75*\xrel,\rowspace)--++(3.5*\xrel,0)
    (-1.75*\xrel,0)--++(3.5*\xrel,0)
    (-1.75*\xrel,-\rowspace)--++(3.5*\xrel,0)
    (-1.75*\xrel,-2*\rowspace)--++(3.5*\xrel,0)
    (-1.75*\xrel,-3*\rowspace)--++(3.5*\xrel,0);
    
\draw [color=black, line width=1*\linethick,dashed]  
    (-1.5*\xrel,\yrel+\rowspace)--++(2*\xrel,0)--++(0,-3*\rowspace)--++(\xrel,0)--++(0,-2*\rowspace)--++(-3*\xrel,0)--++(0,5*\rowspace);

	\renewcommand{\nodenum}{state}
    \renewcommand{\name}{$\ket{\psi_0}$}
	\renewcommand{\xpos}{-2.5*\xrel}
    \renewcommand{\ypos}{-\rowspace}
    \renewcommand{\height}{\heightsingle}
    \renewcommand{\width}{\widthsingle}
    \node[] (\nodenum) at (\xpos,\ypos) {\name};
	
	\renewcommand{\nodenum}{U}
    \renewcommand{\name}{$U'$}
	\renewcommand{\xpos}{0*\xrel}
    \renewcommand{\ypos}{-4*\rowspace}
    \renewcommand{\height}{\heightsingle}
    \renewcommand{\width}{\widthsingle}
    \node[] (\nodenum) at (\xpos,\ypos) {\name};
    
	\renewcommand{\nodenum}{A}
    \renewcommand{\name}{$A$}
	\renewcommand{\xpos}{2.5*\xrel}
    \renewcommand{\ypos}{0}
    \renewcommand{\height}{\heightsingle}
    \renewcommand{\width}{\widthsingle}
    \node[] (\nodenum) at (\xpos,\ypos) {\name};

      \renewcommand{\nodenum}{u2}
    \renewcommand{\name}{$U_{S_5}$}
	\renewcommand{\xpos}{-\xrel}
    \renewcommand{\ypos}{-\yrel}
    \renewcommand{\height}{\heightdouble}
    \renewcommand{\width}{\widthsingle}
    \node[rectangle, fill=egg, rounded corners, minimum width=\width, minimum height=\height, draw] (\nodenum) at (\xpos,\ypos) {\name};
     
    \renewcommand{\nodenum}{u3}
    \renewcommand{\name}{$U_{S_6}$}
	\renewcommand{\xpos}{-\xrel}
    \renewcommand{\ypos}{-\yrel-2*\rowspace}
    \renewcommand{\height}{\heightdouble}
    \renewcommand{\width}{\widthsingle}
    \node[rectangle, fill=egg, rounded corners, minimum width=\width, minimum height=\height, draw] (\nodenum) at (\xpos,\ypos) {\name};
    
      \renewcommand{\nodenum}{u2}
    \renewcommand{\name}{$U_{S_3}$}
	\renewcommand{\xpos}{0*\xrel}
    \renewcommand{\ypos}{+\yrel}
    \renewcommand{\height}{\heightdouble}
    \renewcommand{\width}{\widthsingle}
    \node[rectangle, fill=egg, rounded corners, minimum width=\width, minimum height=\height, draw] (\nodenum) at (\xpos,\ypos) {\name};
    
      \renewcommand{\nodenum}{u2}
    \renewcommand{\name}{$U_{S_4}$}
	\renewcommand{\xpos}{0*\xrel}
    \renewcommand{\ypos}{+\yrel-2*\rowspace}
    \renewcommand{\height}{\heightdouble}
    \renewcommand{\width}{\widthsingle}
    \node[rectangle, fill=egg, rounded corners, minimum width=\width, minimum height=\height, draw] (\nodenum) at (\xpos,\ypos) {\name};
        
      \renewcommand{\nodenum}{u2}
    \renewcommand{\name}{$U_{S_O}$}
	\renewcommand{\xpos}{\xrel}
    \renewcommand{\ypos}{-\yrel}
    \renewcommand{\height}{\heightdouble}
    \renewcommand{\width}{\widthsingle}
    \node[rectangle, fill=egg, rounded corners, minimum width=\width, minimum height=\height, draw] (\nodenum) at (\xpos,\ypos) {\name};
     
    \renewcommand{\nodenum}{u3}
    \renewcommand{\name}{$U_{S_2}$}
	\renewcommand{\xpos}{\xrel}
    \renewcommand{\ypos}{-\yrel-2*\rowspace}
    \renewcommand{\height}{\heightdouble}
    \renewcommand{\width}{\widthsingle}
    \node[rectangle, fill=egg, rounded corners, minimum width=\width, minimum height=\height, draw] (\nodenum) at (\xpos,\ypos) {\name};
    
      \renewcommand{\nodenum}{Measure}
    \renewcommand{\name}{
    \includegraphics[scale=.05]{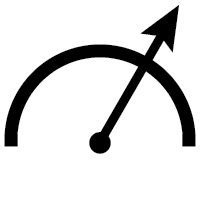}}
	\renewcommand{\xpos}{2*\xrel}
    \renewcommand{\ypos}{0}
    \renewcommand{\height}{\heightsingle}
    \renewcommand{\width}{\widthsingle}
    \node[rectangle, fill=white, rounded corners, minimum width=\width, minimum height=\height, draw] (\nodenum) at (\xpos,\ypos) {\name};

\end{tikzpicture}
\caption{A quantum circuit with a brick-layer architecture. The  parameterized unitary $U$ acts on an input state $\ket{\psi_0}$;  we perform a measurement on system $A$. The unitary $U$ can be written as $U=U_{S_O}U'$, where $U_{S_O}$ is the unitary which acts non-trivially on system $S_O$ such that $A\subseteq S_O$. We define $U'$ as the unitary composed of the remaining unitaries in the architecture.}
    \label{Fig:Circuit}
\end{figure}
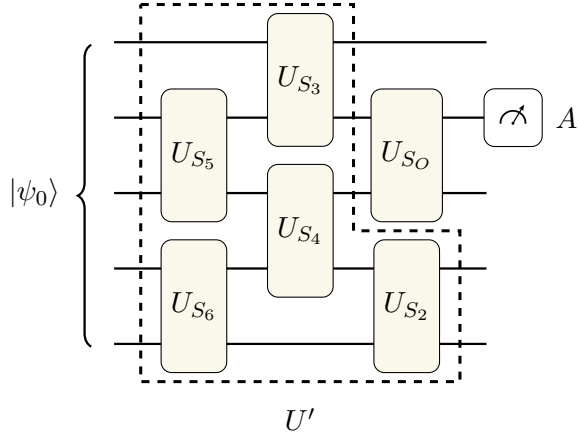

Although we have adopted the MPS architecture to study barren plateaus, we show how this phenomenon can also arise for variational quantum circuit architectures. Define the cost function ${C_c=\bra{\psi_0}U^\dagger (O\otimes I^{A'})U\ket{\psi_0}}$, where $U(\bm{\theta})$ is a parameterized unitary, $\ket{\psi_0}$ is an $n$-qubit state, $O$ is a local Hermitian operator on subsystem $A$ of dimension $d_A$, and $I^{A'}$ is the identity on $A'$, the complement of $A$. We define $U$ to have the general form $U=\prod_{i=1}^{L}U_{S_i}$, where ${U_{S_i}=(e^{-i \theta_i^{\mathrm{polyn}(n)} V_i^{\mathrm{poly}(n)}}\cdots e^{-i \theta_i^1 V_i^1})\otimes I^{S'_i}}$ acts non-trivially on system $S_i$ and acts the identity on its complement $S_i'$. $V_i$ is a Hermitian operator. Each parameter $\theta_i\in \bm{\theta}$ is random such that $U_{S_i}$ forms a 2-design on system $S_i$. We assume that $U$ can be written as $U=U_{S_O}U'$ where $U_{S_O}$ acts non-trivially on the support of $O$ and $U'$ contains the remaining unitaries in $U$. See figure~\ref{Fig:Circuit} for an example. 

We let $\partial_k C_c$ denote the derivative of $C_c$ with respect to $\theta_k$. The following proposition establishes the variance scaling of this derivative.

\begin{prop}\label{Prop:Circuit}
The average of $\partial_k C_c$ over $\bm{\theta}$ satisfies $\langle \partial_k C_c \rangle_{\bm{\theta}}=0$. The variance of $\partial_k C_c$ satisfies
\begin{equation}
\begin{split}
\mathrm{Var}_{\bm{\theta}}[\partial_k C_c]
 	=\epsilon(O)F,
\end{split}
\end{equation}
where $\epsilon(O)=\norm{O-\frac{I^A}{d_A}\tr{O}}_{\mathrm{HS}}^2$. $F$ is an average over the circuit architecture. If ${\epsilon(O)=\mathcal{O}(\mathrm{exp}(-n))}$ and $F$ grows at most polynomially in $n$, $F=\mathcal{O}(\mathrm{poly}(n))$, then the variance vanishes exponentially in $n$. This induces a barren plateau.
\end{prop}

Proposition~ \ref{Prop:Circuit} demonstrates that barren plateaus can arise due to the scaling of $\epsilon(O)$ with respect to $n$. The value of $F$ depends on the particular architecture used. The assumption that $U$ can be written as $U=U_{S_O}U'$ is fairly general, as it is compatible with the alternating layered ansatz and the QCNN. When training with the single-qubit cross-entropy and the linear XEB cost functions, the average value of $\epsilon(O)$ decays exponentially in $n$. By Proposition~ \ref{Prop:Circuit}, this can produce a barren plateau, given that $F$ has at most polynomial scaling. Furthermore, Proposition~ \ref{Prop:Circuit} holds even when $O$ acts on a large, non-local subsystem $A$. In this case, the cross-entropy is used to learn the output probability distribution on $A$. Hence, training inefficiencies can still arise even when learning the non-local structure of a random unitary. This is especially relevant when probing scramblers.

\section{Numerical results}

\begin{figure}[t!]
    \centering
    \subfigure{\includegraphics[scale=.45]{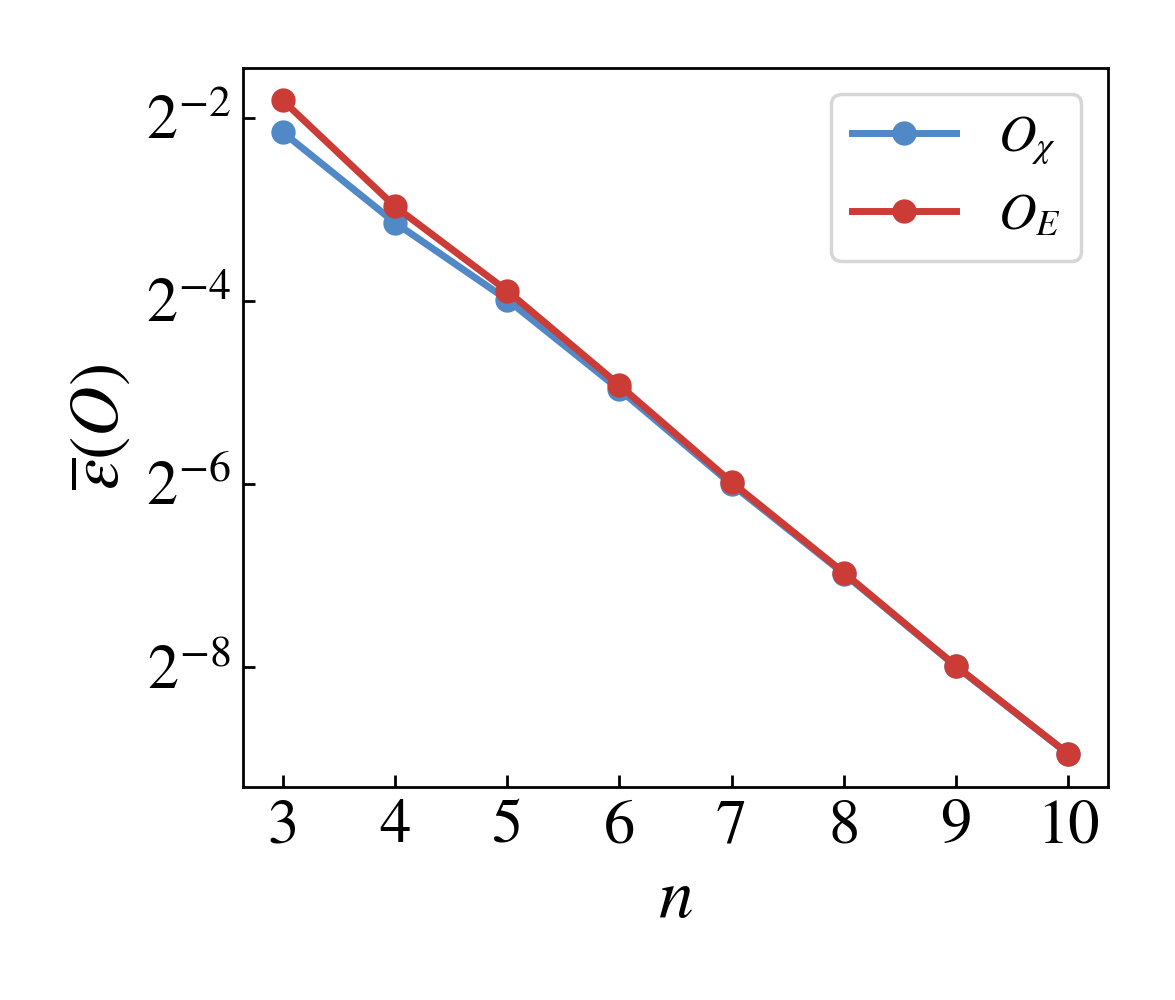}}
        \subfigure{\includegraphics[scale=.45]{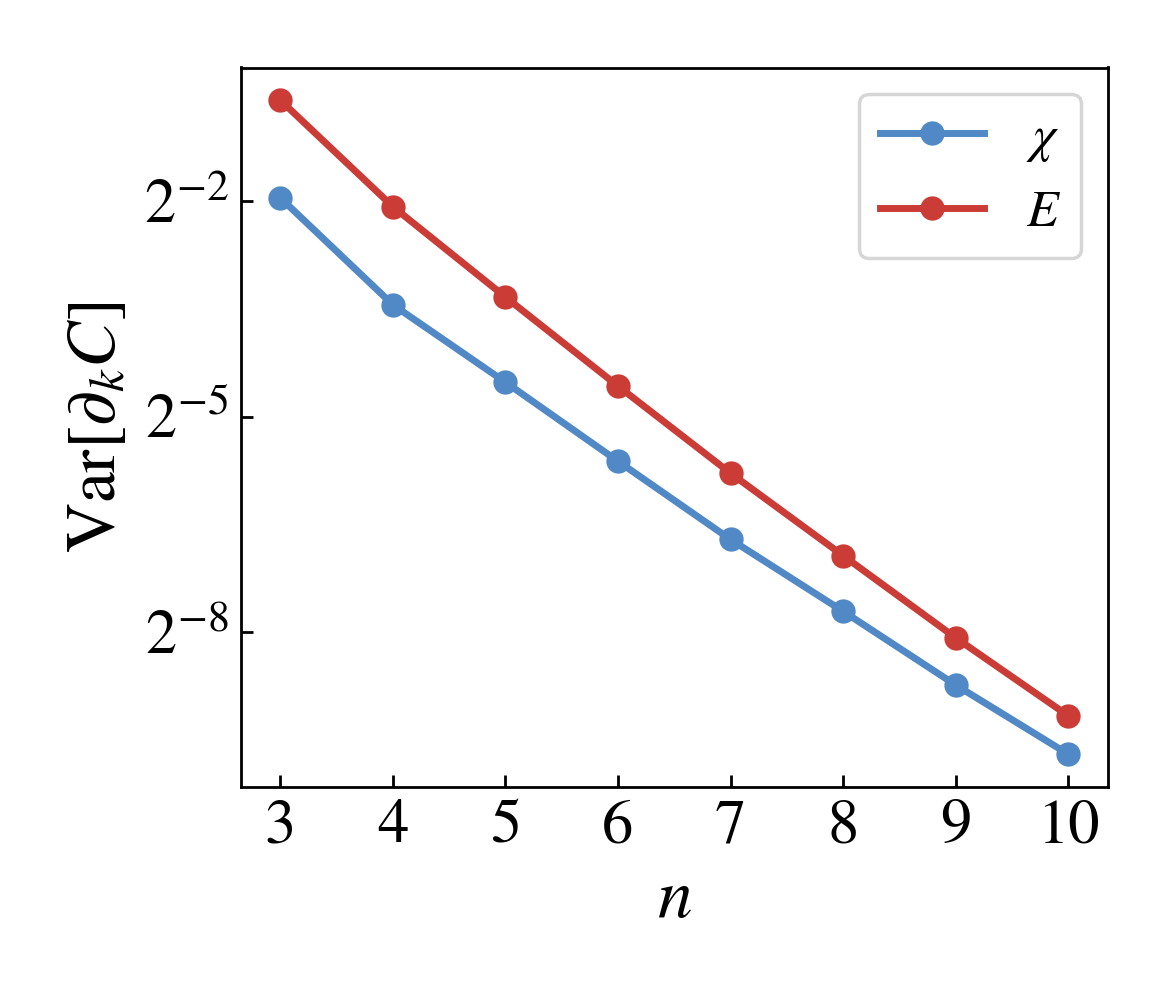}}
    \caption{(Left) Plot of $\overline{\varepsilon}(O)$ defined in Eq.~\eqref{eq:Oe} against qubit number $n$ for ${O\in \{O_E,O_\chi\}}$. 
    For both values of $O$, $\overline{\varepsilon}(O)$ decays exponentially when $n$ is sufficiently large. (Right) Sample plot of the variance $\mathrm{Var}_{\bm{\theta},V}[\partial_i^{(k)}C]$ where $C\in\{\chi,E\}$ and $D=d=2$. We use 5000 Haar random unitaries to empirically compute $\overline{\varepsilon}(O)$ and the variance. Both variance plots approach exponential decay when $n$ is sufficiently large.} 
    \label{Fig:Epsilon}
\end{figure}
For simplicity, we define 
\begin{equation}\label{eq:Oe}
    {\overline{\varepsilon}(O)=\int_{\mathrm{Haar}}dV \epsilon(O)}.
\end{equation}
We substantiate Observation~\ref{Observation} by figure~\ref{Fig:Epsilon}, where we provide numerical values for (left) $\overline{\varepsilon}(O)$ for $O\in\{O_\chi,O_E\}$ and (right) $\mathrm{Var}[\partial_i^{(k)} C]$ for $C\in \{\chi, E\}$. Here, $O_E$ and $O_{\chi}$ are local observables, which define the cross-entropy and linear XEB, respectively. We  compute $\overline{\varepsilon}(O)$ by averaging $\epsilon(O)$ over 5000 unitaries randomly sampled from the Haar measure on the unitary group. In figure~\ref{Fig:Epsilon} (right), the variance in Eq.~\eqref{Eq:VarNumerics} is numerically computed, which corresponds to the case where the derivative and the local observable are located on the same site in the MPS architecture; we refer to appendix~\ref{Appendix:ProofThm1} for further details on this equation. 

Figure~\ref{Fig:Epsilon} (left) indicates that $\overline{\varepsilon}(O_{\chi})$ and $\overline{\varepsilon}(O_E)$ decay exponential for sufficiently large~$n$. Hence, by Lemma~\ref{Lemma:Scrambling}, the variances $\mathrm{Var}_{\bm{\theta},V}[\partial_i^{(k)} \chi]$ and $\mathrm{Var}_{\bm{\theta},V}[\partial_i^{(k)} E]$ must decay at least exponentially in this limit. Figure~\ref{Fig:Epsilon} (right) confirms that $\mathrm{Var}[\partial_i^{(k)} \chi]$ and $\mathrm{Var}[\partial_i^{(k)}E]$ indeed approach exponential decay when $n$ becomes large, validating Theorem~\ref{Thm:XEB} and Observation~\ref{Observation}, respectively. Moreover, for large $n$, figure~\ref{Fig:Epsilon} shows that $\overline{\varepsilon}(O_\chi)$ and $\overline{\varepsilon}(O_E)$ are in agreement and $\mathrm{Var}[\partial_i^{(k)}\chi]$ approaches $\mathrm{Var}[\partial_i^{(k)} E]$. The discrepancy between the two variance plots arises from an extra exponentially decaying term in the expression for $\mathrm{Var}[\partial_i^{(k)} E]$.  Figure~\ref{Fig:Epsilon} demonstrates that, in the context of identifying barren plateaus, $\chi$ exhibits similar behavior as $E$.

\section{Conclusion}
We prove a condition under which barren plateaus may arise and be removed when training the MPS architecture. Using this condition, we prove a no-go theorem by showing that barren plateaus exist when learning random unitary properties with local cost functions. We conjecture that these barren plateaus should also exist for quantum circuit architectures. Our results reveal a barrier impeding an application of QML to learn efficiently the dynamics of generic quantum many-body systems. 

This suggests the following question: can local cost functions be used to avoid barren plateaus when learning the dynamics of integrable systems? Integrable systems are typically not scramblers, implying that they may potentially be efficiently learned with QML.

\begin{acknowledgments}
The authors are grateful for insightful comments from discussions with Zo\"{e} Holmes and Xun Gao, as well as for helpful correspondence with Li-Wei Yu. This work was supported in part by  ARO Grant W911NF-19-1-0302,  ARO MURI Grant W911NF-20-1-0082, and  NSF Eager Grant 2037687. CZ was also supported by the NKRDP Grant No. 2018YFA0704705.

\end{acknowledgments}

\bibliography{Bibliography}

\clearpage

\appendix
In this appendix, we give the proofs of the claims in the main body of our paper. We prove Lemma~\ref{Lemma:Grad} in appendix~\ref{Appendix:Lemma1}. We give a succinct proof of Theorem~\ref{Theorem} in appendix~\ref{Appendix:ProofThm1} and an extended proof in appendix~\ref{Appendix:Extended}. We prove Theorem~\ref{Thm:XEB} in appendix~\ref{Appendix:Theorem2}. In appendix~\ref{Appendix:Observation}, we give a condition to support Observation~\ref{Observation}. In appendix~\ref{Appendix:Prop}, we prove Proposition~\ref{Prop:Circuit}. We state some identities in appendix~\ref{Sec:Identities} and give their proofs in appendix~\ref{Appendix:ProofIdentities}.

\raggedbottom
\input ProofLemma1.tex
\input ProofTheorem1.tex

\input ProofTheorem2.tex

\input ProofObservation1.tex

\input ProofProposition1.tex

\input Identities.tex

\input ExtendedProofTheorem1.tex

\input ProofIdentities.tex

\end{document}

%% file: ProofLemma1.tex
\section{Proof of Lemma~\ref{Lemma:Grad}}\label{Appendix:Lemma1}
The cost function is defined as
\begin{equation}
	C=\bra{\psi(\bm{\theta)}}I_d^{\otimes m-1}\otimes O \otimes I_d^{\otimes n-m}\ket{\psi(\bm{\theta})},
\end{equation} 
where $\ket{\psi(\bm{\theta})}$ is a parameterized, $n$-qudit MPS state and $O$ is a Hermitian operator acting on site $m$. The cost function can be expressed diagrammatically as
\begin{equation}
\begin{tikzpicture}

	\renewcommand{\nodenum}{v0}
    \renewcommand{\name}{$C=$}
	\renewcommand{\xpos}{-1.5*\xrel}
    \renewcommand{\ypos}{0}
    \renewcommand{\height}{\heightsingle}
    \renewcommand{\width}{\widthsingle}
    \node[] (\nodenum) at (\xpos,\ypos) {\name};
    
     \renewcommand{\nodenum}{r2u1}
    \renewcommand{\name}{$\overline{U}_1$}
	\renewcommand{\xpos}{\xback}
    \renewcommand{\ypos}{\yback}
    \renewcommand{\height}{\heightsingle}
    \renewcommand{\width}{\widthsingle}
    \node[rectangle, fill=egg, rounded corners, minimum width=\width, minimum height=\height, draw] (\nodenum) at (\xpos,\ypos) {\name};
    
      \renewcommand{\nodenum}{r2u2}
    \renewcommand{\name}{$\overline{U}_2$}
	\renewcommand{\xpos}{\xrel+\xback}
    \renewcommand{\ypos}{\yback}
    \renewcommand{\height}{\heightsingle}
    \renewcommand{\width}{\widthsingle}
    \node[rectangle, fill=egg, rounded corners, minimum width=\width, minimum height=\height, draw] (\nodenum) at (\xpos,\ypos) {\name};
    
     \renewcommand{\nodenum}{r2u3}
    \renewcommand{\name}{$\overline{U}_m$}
	\renewcommand{\xpos}{3*\xrel+\xback}
    \renewcommand{\ypos}{\yback}
    \renewcommand{\height}{\heightsingle}
    \renewcommand{\width}{\widthsingle}
    \node[rectangle, fill=egg, rounded corners, minimum width=\width, minimum height=\height, draw] (\nodenum) at (\xpos,\ypos) {\name};
	
	 \renewcommand{\nodenum}{r2u4}
    \renewcommand{\name}{$\overline{U}_n$}
	\renewcommand{\xpos}{5*\xrel+\xback}
    \renewcommand{\ypos}{\yback}
    \renewcommand{\height}{\heightsingle}
    \renewcommand{\width}{\widthsingle}
    \node[rectangle, fill=egg, rounded corners, minimum width=\width, minimum height=\height, draw] (\nodenum) at (\xpos,\ypos) {\name};
    
	 \renewcommand{\nodenum}{r1u1}
    \renewcommand{\name}{$U_1$}
	\renewcommand{\xpos}{0}
    \renewcommand{\ypos}{0}
    \renewcommand{\height}{\heightsingle}
    \renewcommand{\width}{\widthsingle}
    \node[rectangle, fill=egg, rounded corners, minimum width=\width, minimum height=\height, draw] (\nodenum) at (\xpos,\ypos) {\name};
    
      \renewcommand{\nodenum}{r1u2}
    \renewcommand{\name}{$U_2$}
	\renewcommand{\xpos}{\xrel}
    \renewcommand{\ypos}{0}
    \renewcommand{\height}{\heightsingle}
    \renewcommand{\width}{\widthsingle}
    \node[rectangle, fill=egg, rounded corners, minimum width=\width, minimum height=\height, draw] (\nodenum) at (\xpos,\ypos) {\name};
    
    \renewcommand{\nodenum}{r1u3}
    \renewcommand{\name}{$U_m$}
	\renewcommand{\xpos}{3*\xrel}
    \renewcommand{\ypos}{0}
    \renewcommand{\height}{\heightsingle}
    \renewcommand{\width}{\widthsingle}
    \node[rectangle, fill=egg, rounded corners, minimum width=\width, minimum height=\height, draw] (\nodenum) at (\xpos,\ypos) {\name};
    
        \renewcommand{\nodenum}{r1u4}
    \renewcommand{\name}{$U_n$}
	\renewcommand{\xpos}{5*\xrel}
    \renewcommand{\ypos}{0}
    \renewcommand{\height}{\heightsingle}
    \renewcommand{\width}{\widthsingle}
    \node[rectangle, fill=egg, rounded corners, minimum width=\width, minimum height=\height, draw] (\nodenum) at (\xpos,\ypos) {\name};
    
    \renewcommand{\nodenum}{r2s1}
    \renewcommand{\name}{}
	\renewcommand{\xpos}{\xback}
    \renewcommand{\ypos}{\yback+\yleg}
    \renewcommand{\height}{\heightstate}
    \renewcommand{\width}{\widthstate}
    \node[rectangle, fill=dullblue, rounded corners=.2em, minimum width=\width, minimum height=\height, draw] (\nodenum) at (\xpos,\ypos) {\name};
    
          \renewcommand{\nodenum}{r2s2}
    \renewcommand{\name}{}
	\renewcommand{\xpos}{\xrel+\xback}
    \renewcommand{\ypos}{\yback+\yleg}
    \renewcommand{\height}{\heightstate}
    \renewcommand{\width}{\widthstate}
    \node[rectangle, fill=dullblue, rounded corners=.2em, minimum width=\width, minimum height=\height, draw] (\nodenum) at (\xpos,\ypos) {\name};
    
         \renewcommand{\nodenum}{r2s3}
    \renewcommand{\name}{}
	\renewcommand{\xpos}{3*\xrel+\xback}
    \renewcommand{\ypos}{\yback+\yleg}
    \renewcommand{\height}{\heightstate}
    \renewcommand{\width}{\widthstate}
    \node[rectangle, fill=dullblue, rounded corners=.2em, minimum width=\width, minimum height=\height, draw] (\nodenum) at (\xpos,\ypos) {\name};
    
         \renewcommand{\nodenum}{r2s4}
    \renewcommand{\name}{}
	\renewcommand{\xpos}{5*\xrel+\xback}
    \renewcommand{\ypos}{\yback+\yleg}
    \renewcommand{\height}{\heightstate}
    \renewcommand{\width}{\widthstate}
    \node[rectangle, fill=dullblue, rounded corners=.2em, minimum width=\width, minimum height=\height, draw] (\nodenum) at (\xpos,\ypos) {\name};
    	
	 \renewcommand{\nodenum}{r1s1}
    \renewcommand{\name}{}
	\renewcommand{\xpos}{0}
    \renewcommand{\ypos}{+\yleg}
    \renewcommand{\height}{\heightstate}
    \renewcommand{\width}{\widthstate}
    \node[rectangle, fill=dullblue, rounded corners=.2em, minimum width=\width, minimum height=\height, draw] (\nodenum) at (\xpos,\ypos) {\name};
    
      \renewcommand{\nodenum}{r1s2}
    \renewcommand{\name}{}
	\renewcommand{\xpos}{\xrel}
    \renewcommand{\ypos}{+\yleg}
    \renewcommand{\height}{\heightstate}
    \renewcommand{\width}{\widthstate}
    \node[rectangle, fill=dullblue, rounded corners=.2em, minimum width=\width, minimum height=\height, draw] (\nodenum) at (\xpos,\ypos) {\name};
    
    \renewcommand{\nodenum}{r1s3}
    \renewcommand{\name}{}
	\renewcommand{\xpos}{3*\xrel}
    \renewcommand{\ypos}{+\yleg}
    \renewcommand{\height}{\heightstate}
    \renewcommand{\width}{\widthstate}
    \node[rectangle, fill=dullblue, rounded corners=.2em, minimum width=\width, minimum height=\height, draw] (\nodenum) at (\xpos,\ypos) {\name};
    
        \renewcommand{\nodenum}{r1s4}
    \renewcommand{\name}{}
	\renewcommand{\xpos}{5*\xrel}
    \renewcommand{\ypos}{+\yleg}
    \renewcommand{\height}{\heightstate}
    \renewcommand{\width}{\widthstate}
    \node[rectangle, fill=dullblue, rounded corners=.2em, minimum width=\width, minimum height=\height, draw] (\nodenum) at (\xpos,\ypos) {\name};

    \renewcommand{\nodenum}{O}
    \renewcommand{\name}{$O$}
	\renewcommand{\xpos}{3*\xrel}
    \renewcommand{\ypos}{-\rowspace}
    \renewcommand{\height}{\heightsingle}
    \renewcommand{\width}{\widthsingle}
    \node[rectangle, fill=egg, rounded corners, minimum width=\width, minimum height=\height, draw] (\nodenum) at (\xpos,\ypos) {\name};
    

     \renewcommand{\nodenum}{r2u1bot}
	\renewcommand{\xpos}{\xback}
    \renewcommand{\ypos}{\yback-\yleg}
    \coordinate (\nodenum) at (\xpos,\ypos) {};
    
      \renewcommand{\nodenum}{r2u2bot}
	\renewcommand{\xpos}{\xrel+\xback}
    \renewcommand{\ypos}{\yback-\yleg}
    \coordinate  (\nodenum) at (\xpos,\ypos) {};
    
     \renewcommand{\nodenum}{r2u3bot}
	\renewcommand{\xpos}{3*\xrel+\xback}
    \renewcommand{\ypos}{\yback-\rowspace-\yleg}
    \coordinate  (\nodenum) at (\xpos,\ypos) {};
	
    	\renewcommand{\nodenum}{r2u4bot}
	\renewcommand{\xpos}{5*\xrel+\xback}
    \renewcommand{\ypos}{\yback-\yleg}
    \coordinate  (\nodenum) at (\xpos,\ypos) {};
	
	 \renewcommand{\nodenum}{r1u1bot}
	\renewcommand{\xpos}{0}
    \renewcommand{\ypos}{-\yleg}
    \coordinate  (\nodenum) at (\xpos,\ypos) {};
    
      \renewcommand{\nodenum}{r1u2bot}
	\renewcommand{\xpos}{\xrel}
    \renewcommand{\ypos}{-\yleg}
    \coordinate (\nodenum) at (\xpos,\ypos) {};
    
    \renewcommand{\nodenum}{r1u3bot}
	\renewcommand{\xpos}{3*\xrel}
    \renewcommand{\ypos}{-\rowspace-\yleg}
    \coordinate   (\nodenum)  at (\xpos,\ypos) {};
    
    \renewcommand{\nodenum}{r1u4bot}
	\renewcommand{\xpos}{5*\xrel}
    \renewcommand{\ypos}{-\yleg}
    \coordinate (\nodenum)   at (\xpos,\ypos) {};

   \renewcommand{\nodenum}{Ellipses}
    \renewcommand{\name}{$\bm{\cdots}$}
	\renewcommand{\xpos}{2.25*\xrel}
    \renewcommand{\ypos}{.5*\yback}
    \node[] (\nodenum) at (\xpos,\ypos) {\name};
   \renewcommand{\nodenum}{Ellipses}
    \renewcommand{\name}{$\bm{\cdots}$}
	\renewcommand{\xpos}{4.25*\xrel}
    \renewcommand{\ypos}{.5*\yback}
    \node[] (\nodenum) at (\xpos,\ypos) {\name};
    %
    

    \draw [line width=\linethick,color=dullred]  
    (r1u1)--++(-\xleg,0)
    (r1u1)--(r1u2)--++(\xleg,0);

     \draw [line width=\linethick,color=dullred]  
    (r2u1)--++(-\xleg,0)
    (r2u1)--(r2u2)--++(\xleg,0);
          
	\draw [line width=\linethick,color=dullred]  
    (r1u3)--++(-\xleg,0)
    (r1u3)--++(\xleg,0);
    
    \draw [line width=\linethick,color=dullred]  
    (r2u3)--++(-\xleg,0)
    (r2u3)--++(\xleg,0);
          
	\draw [line width=\linethick,color=dullred]  
    (r1u4)--++(-\xleg,0)
    (r1u4)--++(\xleg,0);
              
	\draw [line width=\linethick,color=dullred]  
    (r2u4)--++(-\xleg,0)
    (r2u4)--++(\xleg,0);
    
     \draw [line width=\linethick,color=dullblue]  (r2s1)--(r2u1)--(r2u1bot)--(r1u1bot)--(r1u1)--(r1s1) ;
     \draw [line width=\linethick,color=dullblue]  (r2s2)--(r2u2)--(r2u2bot)--(r1u2bot)--(r1u2)--(r1s2) ;
     \draw [line width=\linethick,color=dullblue]  (r2s3)--(r2u3)--(r2u3bot)--(r1u3bot)--(O)--(r1u3)--(r1s3) ;
      \draw [line width=\linethick,color=dullblue]  (r2s4)--(r2u4)--(r2u4bot)--(r1u4bot)--(r1u4)--(r1s4) ;
    
\end{tikzpicture},
\end{equation}
where blue boxes correspond to a fixed state $\ket{0}$,
\begin{equation}
\begin{tikzpicture}

\renewcommand{\nodenum}{s1}
    \renewcommand{\name}{}
	\renewcommand{\xpos}{0}
    \renewcommand{\ypos}{0}
    \renewcommand{\height}{\heightstate}
    \renewcommand{\width}{\widthstate}
    \node[rectangle, fill=dullblue, rounded corners=.2em, minimum width=\width, minimum height=\height, draw] (\nodenum) at (\xpos,\ypos) {\name};
    
	\renewcommand{\nodenum}{Equal}
    \renewcommand{\name}{$=$}
	\renewcommand{\xpos}{\xrel}
    \renewcommand{\ypos}{-.5*\yleg}
    \renewcommand{\height}{\heightsingle}
    \renewcommand{\width}{\widthsingle}
    \node[] (\nodenum) at (\xpos,\ypos) {\name};

	\renewcommand{\nodenum}{psi}
    \renewcommand{\name}{$\ket{0}$}
	\renewcommand{\xpos}{2*\xrel}
    \renewcommand{\ypos}{0}
    \renewcommand{\height}{\heightsingle}
    \renewcommand{\width}{\widthsingle}
    \node[] (\nodenum) at (\xpos,\ypos) {\name};
    
    \draw [line width=\linethick,color=dullblue]  
    (s1)--++(0,-\yleg)
    (psi)--++(0,-\yleg) ;
    
\end{tikzpicture}.
\end{equation}
Other single-qudit pure states can be used, but we fix each to the $\ket{0}$ state for simplicity. Each $U_i$ has the form $U_i=U_i^{(\textrm{poly}(Dd))}\cdots U_i^{(2)}U_i^{(1)}$, where $U_i^{(k)}=e^{-iG_i^{(k)}\theta_i^{(k)}}$ and $G_i^{(k)}$ is a Hermitian operator. We assume that each $\theta_i^{(k)}$ is randomly initialized such that each $U_i$ forms a unitary 2-design. We use the notation $\overline{U}_i=U_i^*$ to denote the conjugate. Each $U_i$ is a unitary of dimension $Dd\times Dd$. Blue lines correspond to physical indices and red lines correspond to virtual indices. We assume periodic boundary conditions for the red lines. 

We define $U_{i+}=U_i^{(k)}\cdots U_i^{(1)}$ and $U_{i-}=U_i^{(\textrm{poly}(Dd))}\cdots U_i^{(k+1)}$. The value of $k$ will determine which unitary, $U_+$ or $U_-$, forms  a unitary t-design. Either only one forms a t-design or both form a t-design. We define the following notation for the partial derivative: $\partial^{(k)}_i=\frac{\partial}{\partial \theta_i^{(k)}}$. For simplicity of notation, we will set $U_i\rightarrow U$, $G_{i}^{(k)}\rightarrow G$, $\theta_i^{(k)}\rightarrow \theta_k$, $\partial_i^{(k)}\rightarrow \partial_k$, and $G^0=I$.

The following identity will be useful:
\begin{equation}
\begin{split}
	\int dU \partial_k (U\otimes \overline{U})
	&=\int dU( \partial_k U\otimes \overline{U}+U\otimes \partial_k \overline{U})\\
	&=\int dU_- dU_+( U_-(-iG)U_+\otimes \overline{U}_- \overline{U}_++U_-U_+\otimes \overline{U}_-(i\overline{G})\overline{U}_+)\\
	&=\int dU_- dU_+i( -U_-GU_+\otimes \overline{U}_-\overline{G}^0\overline{U}_++U_-G^0U_+\otimes \overline{U}_-\overline{G} \overline{U}_+)\\
	&=\sum_{\alpha=0}^{1}i( -1)^{\alpha} \int dU_- dU_+ (U_-G^{\alpha}U_+\otimes \overline{U}_-\overline{G}^{1-\alpha}\overline{U}_+).
\end{split}
\end{equation}
 In diagrammatic form, this is

\begin{equation}
\begin{tikzpicture}

	
     \renewcommand{\nodenum}{r1u1Red}
	\renewcommand{\xpos}{-\xrecenter}
    \renewcommand{\ypos}{\yleg}
    \coordinate (\nodenum) at (\xpos,\ypos) {};

	\renewcommand{\nodenum}{r1u1Blue}
	\renewcommand{\xpos}{\xrecenter}
    \renewcommand{\ypos}{\yleg}
    \coordinate (\nodenum) at (\xpos,\ypos) {};

	\renewcommand{\nodenum}{r2u1Red}
	\renewcommand{\xpos}{-\xrecenter}
    \renewcommand{\ypos}{-2*\rowspace-\yleg}
    \coordinate (\nodenum) at (\xpos,\ypos) {};
   
	\renewcommand{\nodenum}{r2u1Blue}
	\renewcommand{\xpos}{\xrecenter}
    \renewcommand{\ypos}{-2*\rowspace-\yleg}
    \coordinate (\nodenum) at (\xpos,\ypos) {}; 
    
     \renewcommand{\nodenum}{r1u2Blue}
	\renewcommand{\xpos}{\xrel-\xrecenter}
    \renewcommand{\ypos}{\yleg}
    \coordinate (\nodenum) at (\xpos,\ypos) {};

	\renewcommand{\nodenum}{r1u2Red}
	\renewcommand{\xpos}{\xrel+\xrecenter}
    \renewcommand{\ypos}{\yleg}
    \coordinate (\nodenum) at (\xpos,\ypos) {};

	\renewcommand{\nodenum}{r2u2Blue}
	\renewcommand{\xpos}{\xrel-\xrecenter}
    \renewcommand{\ypos}{-2*\rowspace-\yleg}
    \coordinate (\nodenum) at (\xpos,\ypos) {};
   
	\renewcommand{\nodenum}{r2u2Red}
	\renewcommand{\xpos}{\xrel+\xrecenter}
    \renewcommand{\ypos}{-2*\rowspace-\yleg}
    \coordinate (\nodenum) at (\xpos,\ypos) {}; 
    
	\draw [line width=\linethick,color=dullred]  
	(r1u1Red)--++(0,-\yleg)
	(r1u2Red)--++(0,-\yleg)
	(r2u1Red)--++(0,\yleg)
	(r2u2Red)--++(0,\yleg);
	
	\draw [line width=\linethick,color=dullblue]  
	(r1u1Blue)--++(0,-\yleg)
	(r1u2Blue)--++(0,-\yleg)
	(r2u1Blue)--++(0,\yleg)
	(r2u2Blue)--++(0,\yleg);

	
	\renewcommand{\nodenum}{v0}
    \renewcommand{\name}{$\int dU \partial_k (U\otimes \overline{U})=\sum_{\alpha=0}^{1}i( -1)^{\alpha} \int dU_- dU_+$}
	\renewcommand{\xpos}{-3*\xrel}
    \renewcommand{\ypos}{-1*\rowspace}
    \renewcommand{\height}{\heightsingle}
    \renewcommand{\width}{\widthsingle}
    \node[] (\nodenum) at (\xpos,\ypos) {\name};
   
     \renewcommand{\nodenum}{r1u1}
    \renewcommand{\name}{$U_-$}
	\renewcommand{\xpos}{0}
    \renewcommand{\ypos}{0}
    \renewcommand{\height}{\heightsingle}
    \renewcommand{\width}{\widthsingle}
    \node[rectangle, fill=egg, rounded corners, minimum width=\width, minimum height=\height, draw] (\nodenum) at (\xpos,\ypos) {\name};
    
    \renewcommand{\nodenum}{r1u2}
    \renewcommand{\name}{$\overline{U}_-$}
	\renewcommand{\xpos}{\xrel}
    \renewcommand{\ypos}{0}
    \renewcommand{\height}{\heightsingle}
    \renewcommand{\width}{\widthsingle}
    \node[rectangle, fill=egg, rounded corners, minimum width=\width, minimum height=\height, draw] (\nodenum) at (\xpos,\ypos) {\name};

     \renewcommand{\nodenum}{G1}
    \renewcommand{\name}{$G^\alpha$}
	\renewcommand{\xpos}{0}
    \renewcommand{\ypos}{-\rowspace}
    \renewcommand{\height}{\heightsingle}
    \renewcommand{\width}{\widthsingle}
    \node[rectangle, fill=egg, rounded corners, minimum width=\width, minimum height=\height, draw] (\nodenum) at (\xpos,\ypos) {\name};
    
     \renewcommand{\nodenum}{G2}
    \renewcommand{\name}{$\overline{G}^{1-\alpha}$}
	\renewcommand{\xpos}{\xrel}
    \renewcommand{\ypos}{-\rowspace}
    \renewcommand{\height}{\heightsingle}
    \renewcommand{\width}{\widthsingle}
    \node[rectangle, fill=egg, rounded corners, minimum width=\width, minimum height=\height, draw] (\nodenum) at (\xpos,\ypos) {\name};
    
     \renewcommand{\nodenum}{r2u1}
    \renewcommand{\name}{$U_+$}
	\renewcommand{\xpos}{0}
    \renewcommand{\ypos}{-2*\rowspace}
    \renewcommand{\height}{\heightsingle}
    \renewcommand{\width}{\widthsingle}
    \node[rectangle, fill=egg, rounded corners, minimum width=\width, minimum height=\height, draw] (\nodenum) at (\xpos,\ypos) {\name};
    
     \renewcommand{\nodenum}{r2u2}
    \renewcommand{\name}{$\overline{U}_+$}
	\renewcommand{\xpos}{\xrel}
    \renewcommand{\ypos}{-2*\rowspace}
    \renewcommand{\height}{\heightsingle}
    \renewcommand{\width}{\widthsingle}
    \node[rectangle, fill=egg, rounded corners, minimum width=\width, minimum height=\height, draw] (\nodenum) at (\xpos,\ypos) {\name};
    
     \draw [line width=\linethick,color=black]  (r1u1)--(G1)--(r2u1) ;
     
     \draw [line width=\linethick,color=black]  (r1u2)--(G2)--(r2u2) ;

\end{tikzpicture}.
\end{equation}
Black lines correspond to physical \textit{and} virtual indices of dimension $Dd$.

We now turn our attention to computing the average of $\partial_k C$. Without loss of generality, we assume that the derivative is taken on the first qudit site. Define $\langle \cdot \rangle\equiv\int dU_1U_2\cdots U_n(\cdot)=\int dU_{all}$. In the main text, we use the notation $\langle \cdot \rangle_{\bm{\theta}}$ for the average, but we omit the $\bm{\theta}$ subscript here. We will show that $\partial_k C=0$. The diagram for the average derivative is

\begin{equation}
\begin{tikzpicture}

	\renewcommand{\nodenum}{v0}
    \renewcommand{\name}{$\langle \partial_k C\rangle =\sum_{\alpha=0}^{1}i( -1)^{\alpha} \int dU_{\mathrm{all}}$}
	\renewcommand{\xpos}{-2.7*\xrel}
    \renewcommand{\ypos}{0}
    \renewcommand{\height}{\heightsingle}
    \renewcommand{\width}{\widthsingle}
    \node[] (\nodenum) at (\xpos,\ypos) {\name};
    
     \renewcommand{\nodenum}{r2u-}
    \renewcommand{\name}{$\overline{U}_-$}
	\renewcommand{\xpos}{\xback}
    \renewcommand{\ypos}{\yback}
    \renewcommand{\height}{\heightsingle}
    \renewcommand{\width}{\widthsingle}
    \node[rectangle, fill=egg, rounded corners, minimum width=\width, minimum height=\height, draw] (\nodenum) at (\xpos,\ypos) {\name};
        
     \renewcommand{\nodenum}{G2}
    \renewcommand{\name}{$\overline{G}^{1-\alpha}$}
	\renewcommand{\xpos}{\xback}
    \renewcommand{\ypos}{\yback-\rowspace}
    \renewcommand{\height}{\heightsingle}
    \renewcommand{\width}{\widthsingle}
    \node[rectangle, fill=egg, rounded corners, minimum width=\width, minimum height=\height, draw] (\nodenum) at (\xpos,\ypos) {\name};
    
         \renewcommand{\nodenum}{r2u+}
    \renewcommand{\name}{$\overline{U}_+$}
	\renewcommand{\xpos}{\xback}
    \renewcommand{\ypos}{\yback-2*\rowspace}
    \renewcommand{\height}{\heightsingle}
    \renewcommand{\width}{\widthsingle}
    \node[rectangle, fill=egg, rounded corners, minimum width=\width, minimum height=\height, draw] (\nodenum) at (\xpos,\ypos) {\name};
    
      \renewcommand{\nodenum}{r2u2}
    \renewcommand{\name}{$\overline{U}_2$}
	\renewcommand{\xpos}{\xrel+\xback}
    \renewcommand{\ypos}{\yback}
    \renewcommand{\height}{\heightsingle}
    \renewcommand{\width}{\widthsingle}
    \node[rectangle, fill=egg, rounded corners, minimum width=\width, minimum height=\height, draw] (\nodenum) at (\xpos,\ypos) {\name};
    
     \renewcommand{\nodenum}{r2u3}
    \renewcommand{\name}{$\overline{U}_m$}
	\renewcommand{\xpos}{3*\xrel+\xback}
    \renewcommand{\ypos}{\yback}
    \renewcommand{\height}{\heightsingle}
    \renewcommand{\width}{\widthsingle}
    \node[rectangle, fill=egg, rounded corners, minimum width=\width, minimum height=\height, draw] (\nodenum) at (\xpos,\ypos) {\name};
	
	 \renewcommand{\nodenum}{r2u4}
    \renewcommand{\name}{$\overline{U}_n$}
	\renewcommand{\xpos}{5*\xrel+\xback}
    \renewcommand{\ypos}{\yback}
    \renewcommand{\height}{\heightsingle}
    \renewcommand{\width}{\widthsingle}
    \node[rectangle, fill=egg, rounded corners, minimum width=\width, minimum height=\height, draw] (\nodenum) at (\xpos,\ypos) {\name};
    
	 \renewcommand{\nodenum}{r1u-}
    \renewcommand{\name}{$U_-$}
	\renewcommand{\xpos}{0}
    \renewcommand{\ypos}{0}
    \renewcommand{\height}{\heightsingle}
    \renewcommand{\width}{\widthsingle}
    \node[rectangle, fill=egg, rounded corners, minimum width=\width, minimum height=\height, draw] (\nodenum) at (\xpos,\ypos) {\name};
    
	 \renewcommand{\nodenum}{G1}
    \renewcommand{\name}{$G^\alpha$}
	\renewcommand{\xpos}{0}
    \renewcommand{\ypos}{-\rowspace}
    \renewcommand{\height}{\heightsingle}
    \renewcommand{\width}{\widthsingle}
    \node[rectangle, fill=egg, rounded corners, minimum width=\width, minimum height=\height, draw] (\nodenum) at (\xpos,\ypos) {\name};
    
	 \renewcommand{\nodenum}{r1u+}
    \renewcommand{\name}{$U_+$}
	\renewcommand{\xpos}{0}
    \renewcommand{\ypos}{-2*\rowspace}
    \renewcommand{\height}{\heightsingle}
    \renewcommand{\width}{\widthsingle}
    \node[rectangle, fill=egg, rounded corners, minimum width=\width, minimum height=\height, draw] (\nodenum) at (\xpos,\ypos) {\name};
    
      \renewcommand{\nodenum}{r1u2}
    \renewcommand{\name}{$U_2$}
	\renewcommand{\xpos}{\xrel}
    \renewcommand{\ypos}{0}
    \renewcommand{\height}{\heightsingle}
    \renewcommand{\width}{\widthsingle}
    \node[rectangle, fill=egg, rounded corners, minimum width=\width, minimum height=\height, draw] (\nodenum) at (\xpos,\ypos) {\name};
    
    \renewcommand{\nodenum}{r1u3}
    \renewcommand{\name}{$U_m$}
	\renewcommand{\xpos}{3*\xrel}
    \renewcommand{\ypos}{0}
    \renewcommand{\height}{\heightsingle}
    \renewcommand{\width}{\widthsingle}
    \node[rectangle, fill=egg, rounded corners, minimum width=\width, minimum height=\height, draw] (\nodenum) at (\xpos,\ypos) {\name};

        \renewcommand{\nodenum}{r1u4}
    \renewcommand{\name}{$U_n$}
	\renewcommand{\xpos}{5*\xrel}
    \renewcommand{\ypos}{0}
    \renewcommand{\height}{\heightsingle}
    \renewcommand{\width}{\widthsingle}
    \node[rectangle, fill=egg, rounded corners, minimum width=\width, minimum height=\height, draw] (\nodenum) at (\xpos,\ypos) {\name};
    
    \renewcommand{\nodenum}{r2s1}
    \renewcommand{\name}{}
	\renewcommand{\xpos}{\xback}
    \renewcommand{\ypos}{\yback+\yleg}
    \renewcommand{\height}{\heightstate}
    \renewcommand{\width}{\widthstate}
    \node[rectangle, fill=dullblue, rounded corners=.2em, minimum width=\width, minimum height=\height, draw] (\nodenum) at (\xpos,\ypos) {\name};
    
          \renewcommand{\nodenum}{r2s2}
    \renewcommand{\name}{}
	\renewcommand{\xpos}{\xrel+\xback}
    \renewcommand{\ypos}{\yback+\yleg}
    \renewcommand{\height}{\heightstate}
    \renewcommand{\width}{\widthstate}
    \node[rectangle, fill=dullblue, rounded corners=.2em, minimum width=\width, minimum height=\height, draw] (\nodenum) at (\xpos,\ypos) {\name};
    
         \renewcommand{\nodenum}{r2s3}
    \renewcommand{\name}{}
	\renewcommand{\xpos}{3*\xrel+\xback}
    \renewcommand{\ypos}{\yback+\yleg}
    \renewcommand{\height}{\heightstate}
    \renewcommand{\width}{\widthstate}
    \node[rectangle, fill=dullblue, rounded corners=.2em, minimum width=\width, minimum height=\height, draw] (\nodenum) at (\xpos,\ypos) {\name};
    
         \renewcommand{\nodenum}{r2s4}
    \renewcommand{\name}{}
	\renewcommand{\xpos}{5*\xrel+\xback}
    \renewcommand{\ypos}{\yback+\yleg}
    \renewcommand{\height}{\heightstate}
    \renewcommand{\width}{\widthstate}
    \node[rectangle, fill=dullblue, rounded corners=.2em, minimum width=\width, minimum height=\height, draw] (\nodenum) at (\xpos,\ypos) {\name};
    	
	 \renewcommand{\nodenum}{r1s1}
    \renewcommand{\name}{}
	\renewcommand{\xpos}{0}
    \renewcommand{\ypos}{+\yleg}
    \renewcommand{\height}{\heightstate}
    \renewcommand{\width}{\widthstate}
    \node[rectangle, fill=dullblue, rounded corners=.2em, minimum width=\width, minimum height=\height, draw] (\nodenum) at (\xpos,\ypos) {\name};
    
      \renewcommand{\nodenum}{r1s2}
    \renewcommand{\name}{}
	\renewcommand{\xpos}{\xrel}
    \renewcommand{\ypos}{+\yleg}
    \renewcommand{\height}{\heightstate}
    \renewcommand{\width}{\widthstate}
    \node[rectangle, fill=dullblue, rounded corners=.2em, minimum width=\width, minimum height=\height, draw] (\nodenum) at (\xpos,\ypos) {\name};
    
    \renewcommand{\nodenum}{r1s3}
    \renewcommand{\name}{}
	\renewcommand{\xpos}{3*\xrel}
    \renewcommand{\ypos}{+\yleg}
    \renewcommand{\height}{\heightstate}
    \renewcommand{\width}{\widthstate}
    \node[rectangle, fill=dullblue, rounded corners=.2em, minimum width=\width, minimum height=\height, draw] (\nodenum) at (\xpos,\ypos) {\name};
    
        \renewcommand{\nodenum}{r1s4}
    \renewcommand{\name}{}
	\renewcommand{\xpos}{5*\xrel}
    \renewcommand{\ypos}{+\yleg}
    \renewcommand{\height}{\heightstate}
    \renewcommand{\width}{\widthstate}
    \node[rectangle, fill=dullblue, rounded corners=.2em, minimum width=\width, minimum height=\height, draw] (\nodenum) at (\xpos,\ypos) {\name};

    \renewcommand{\nodenum}{O}
    \renewcommand{\name}{$O$}
	\renewcommand{\xpos}{3*\xrel}
    \renewcommand{\ypos}{-\rowspace}
    \renewcommand{\height}{\heightsingle}
    \renewcommand{\width}{\widthsingle}
    \node[rectangle, fill=egg, rounded corners, minimum width=\width, minimum height=\height, draw] (\nodenum) at (\xpos,\ypos) {\name};
    

     \renewcommand{\nodenum}{r2u+bot}
	\renewcommand{\xpos}{\xback}
    \renewcommand{\ypos}{\yback-2*\rowspace-\yleg}
    \coordinate (\nodenum) at (\xpos,\ypos) {};
    
      \renewcommand{\nodenum}{r2u2bot}
	\renewcommand{\xpos}{\xrel+\xback}
    \renewcommand{\ypos}{\yback-\yleg}
    \coordinate  (\nodenum) at (\xpos,\ypos) {};
    
     \renewcommand{\nodenum}{r2u3bot}
	\renewcommand{\xpos}{3*\xrel+\xback}
    \renewcommand{\ypos}{\yback-\rowspace-\yleg}
    \coordinate  (\nodenum) at (\xpos,\ypos) {};
	
    	\renewcommand{\nodenum}{r2u4bot}
	\renewcommand{\xpos}{5*\xrel+\xback}
    \renewcommand{\ypos}{\yback-\yleg}
    \coordinate  (\nodenum) at (\xpos,\ypos) {};
	
	 \renewcommand{\nodenum}{r1u+bot}
	\renewcommand{\xpos}{0}
    \renewcommand{\ypos}{-2*\rowspace-\yleg}
    \coordinate  (\nodenum) at (\xpos,\ypos) {};
    
      \renewcommand{\nodenum}{r1u2bot}
	\renewcommand{\xpos}{\xrel}
    \renewcommand{\ypos}{-\yleg}
    \coordinate (\nodenum) at (\xpos,\ypos) {};
    
    \renewcommand{\nodenum}{r1u3bot}
	\renewcommand{\xpos}{3*\xrel}
    \renewcommand{\ypos}{-\rowspace-\yleg}
    \coordinate   (\nodenum)  at (\xpos,\ypos) {};
    
    \renewcommand{\nodenum}{r1u4bot}
	\renewcommand{\xpos}{5*\xrel}
    \renewcommand{\ypos}{-\yleg}
    \coordinate (\nodenum)   at (\xpos,\ypos) {};

   \renewcommand{\nodenum}{Ellipses}
    \renewcommand{\name}{$\bm{\cdots}$}
	\renewcommand{\xpos}{2.25*\xrel}
    \renewcommand{\ypos}{.5*\yback}
    \node[] (\nodenum) at (\xpos,\ypos) {\name};
   \renewcommand{\nodenum}{Ellipses}
    \renewcommand{\name}{$\bm{\cdots}$}
	\renewcommand{\xpos}{4.25*\xrel}
    \renewcommand{\ypos}{.5*\yback}
    \node[] (\nodenum) at (\xpos,\ypos) {\name};
    %
    
    
    \draw [line width=\linethick,color=dullred]  
    (r1u-)--++(-\xleg,0)
    (r1u+)--(r1u2)--++(\xleg,0);

     \draw [line width=\linethick,color=dullred]  
    (r2u-)--++(-\xleg,0)
    (r2u+)--++(.5*\xrel,\rowspace)
    (r2u2)--++(\xleg,0);
          
	\draw [line width=\linethick,color=dullred]  
    (r1u3)--++(-\xleg,0)
    (r1u3)--++(\xleg,0);
    
    \draw [line width=\linethick,color=dullred]  
    (r2u3)--++(-\xleg,0)
    (r2u3)--++(\xleg,0);
          
	\draw [line width=\linethick,color=dullred]  
    (r1u4)--++(-\xleg,0)
    (r1u4)--++(\xleg,0);
              
	\draw [line width=\linethick,color=dullred]  
    (r2u4)--++(-\xleg,0)
    (r2u4)--++(\xleg,0);
    
     \draw [line width=\linethick,color=dullblue]  (r2s1)--(r2u-);
     \draw [line width=\linethick,color=black] (r2u-)--(G2)--(r2u+);
     \draw [line width=\linethick,color=dullblue] (r2u+)--(r2u+bot)--(r1u+bot)--(r1u+);
     \draw [line width=\linethick,color=black] (r1u+) --(G1)--(r1u-);
    \draw [line width=\linethick,color=dullblue]  (r1u-) --(r1s1) ;
    
     \draw [line width=\linethick,color=dullblue]  (r2s2)--(r2u2)--(r2u2bot)--(r1u2bot)--(r1u2)--(r1s2) ;
     \draw [line width=\linethick,color=dullblue]  (r2s3)--(r2u3)--(r2u3bot)--(r1u3bot)--(O)--(r1u3)--(r1s3) ;
      \draw [line width=\linethick,color=dullblue]  (r2s4)--(r2u4)--(r2u4bot)--(r1u4bot)--(r1u4)--(r1s4) ;
    
\end{tikzpicture}.
\end{equation}
When integrating over the unitaries, there are three possibilities: only $U_-$ forms a 2-design, only $U_+$ forms a 2-design, or both $U_-$ and $U_+$ form 2-designs. We consider the first case where only $U_-$ forms a 2-design. We evaluate the diagram on the first site by integrating over $U_-$ and using the Weingarten calculus:

\begin{equation}
\begin{tikzpicture}

	
     \renewcommand{\nodenum}{r1u1}
	\renewcommand{\xpos}{0}
    \renewcommand{\ypos}{0}
    \coordinate (\nodenum) at (\xpos,\ypos) {};
    
     \renewcommand{\nodenum}{r1u1top}
	\renewcommand{\xpos}{0}
    \renewcommand{\ypos}{\yleg}
    \coordinate (\nodenum) at (\xpos,\ypos) {};

	\renewcommand{\nodenum}{r2u1Red}
	\renewcommand{\xpos}{-\xrecenter}
    \renewcommand{\ypos}{-2*\rowspace-\yleglong}
    \coordinate (\nodenum) at (\xpos,\ypos) {};
   
	\renewcommand{\nodenum}{r2u1Blue}
	\renewcommand{\xpos}{\xrecenter}
    \renewcommand{\ypos}{-2*\rowspace-\yleg}
    \coordinate (\nodenum) at (\xpos,\ypos) {}; 
    
     \renewcommand{\nodenum}{r1u2}
	\renewcommand{\xpos}{\xrel}
    \renewcommand{\ypos}{0}
    \coordinate (\nodenum) at (\xpos,\ypos) {};

     \renewcommand{\nodenum}{r1u2top}
	\renewcommand{\xpos}{\xrel}
    \renewcommand{\ypos}{\yleg}
    \coordinate (\nodenum) at (\xpos,\ypos) {};
    
	\renewcommand{\nodenum}{r2u2Blue}
	\renewcommand{\xpos}{\xrel-\xrecenter}
    \renewcommand{\ypos}{-2*\rowspace-\yleg}
    \coordinate (\nodenum) at (\xpos,\ypos) {};
   
	\renewcommand{\nodenum}{r2u2Red}
	\renewcommand{\xpos}{\xrel+\xrecenter}
    \renewcommand{\ypos}{-2*\rowspace-\yleglong}
    \coordinate (\nodenum) at (\xpos,\ypos) {}; 
    

	\draw [line width=\linethick,color=dullred]  
	(r2u1Red)--++(0,\yleg)
	(r2u2Red)--++(0,\yleg);
	
	\draw [line width=\linethick,color=dullblue]  
	(r2u1Blue)--++(0,\yleg)
	(r2u2Blue)--++(0,\yleg);
		
	\draw [line width=\linethick,color=dullblue] 
	(r2u1Blue)--(r2u2Blue);
	
	
	\renewcommand{\nodenum}{v0}
    \renewcommand{\name}{$\sum_{\alpha=0}^{1}i( -1)^{\alpha} \int dU_+$}
	\renewcommand{\xpos}{-1.75*\xrel}
    \renewcommand{\ypos}{-1*\rowspace}
    \renewcommand{\height}{\heightsingle}
    \renewcommand{\width}{\widthsingle}
    \node[] (\nodenum) at (\xpos,\ypos) {\name};

     \renewcommand{\nodenum}{G1}
    \renewcommand{\name}{$G^\alpha$}
	\renewcommand{\xpos}{0}
    \renewcommand{\ypos}{-\rowspace}
    \renewcommand{\height}{\heightsingle}
    \renewcommand{\width}{\widthsingle}
    \node[rectangle, fill=egg, rounded corners, minimum width=\width, minimum height=\height, draw] (\nodenum) at (\xpos,\ypos) {\name};
    
     \renewcommand{\nodenum}{G2}
    \renewcommand{\name}{$\overline{G}^{1-\alpha}$}
	\renewcommand{\xpos}{\xrel}
    \renewcommand{\ypos}{-\rowspace}
    \renewcommand{\height}{\heightsingle}
    \renewcommand{\width}{\widthsingle}
    \node[rectangle, fill=egg, rounded corners, minimum width=\width, minimum height=\height, draw] (\nodenum) at (\xpos,\ypos) {\name};
    
     \renewcommand{\nodenum}{r2u1}
    \renewcommand{\name}{$U_+$}
	\renewcommand{\xpos}{0}
    \renewcommand{\ypos}{-2*\rowspace}
    \renewcommand{\height}{\heightsingle}
    \renewcommand{\width}{\widthsingle}
    \node[rectangle, fill=egg, rounded corners, minimum width=\width, minimum height=\height, draw] (\nodenum) at (\xpos,\ypos) {\name};
    
     \renewcommand{\nodenum}{r2u2}
    \renewcommand{\name}{$\overline{U}_+$}
	\renewcommand{\xpos}{\xrel}
    \renewcommand{\ypos}{-2*\rowspace}
    \renewcommand{\height}{\heightsingle}
    \renewcommand{\width}{\widthsingle}
    \node[rectangle, fill=egg, rounded corners, minimum width=\width, minimum height=\height, draw] (\nodenum) at (\xpos,\ypos) {\name};

     \draw [line width=\linethick,color=black]  (r2u2)--(G2)--(r1u2)--(r1u1)--(G1)--(r2u1) ;
      \draw [line width=\linethick,color=black]  (r1u1top)--++(0,-.5*\yleg)--++(\xrel,0)--(r1u2top);

    
	
     \renewcommand{\nodenum}{r1u3}
	\renewcommand{\xpos}{4.5*\xrel}
    \renewcommand{\ypos}{0}
    \coordinate (\nodenum) at (\xpos,\ypos) {};
    
     \renewcommand{\nodenum}{r1u3top}
	\renewcommand{\xpos}{4.5*\xrel}
    \renewcommand{\ypos}{\yleg}
    \coordinate (\nodenum) at (\xpos,\ypos) {};

	\renewcommand{\nodenum}{r2u3Red}
	\renewcommand{\xpos}{4.5*\xrel-\xrecenter}
    \renewcommand{\ypos}{-2*\rowspace-\yleglong}
    \coordinate (\nodenum) at (\xpos,\ypos) {};
   
	\renewcommand{\nodenum}{r2u3Blue}
	\renewcommand{\xpos}{4.5*\xrel+\xrecenter}
    \renewcommand{\ypos}{-2*\rowspace-\yleg}
    \coordinate (\nodenum) at (\xpos,\ypos) {}; 
    
     \renewcommand{\nodenum}{r1u4}
	\renewcommand{\xpos}{5.5*\xrel}
    \renewcommand{\ypos}{0}
    \coordinate (\nodenum) at (\xpos,\ypos) {};

     \renewcommand{\nodenum}{r1u4top}
	\renewcommand{\xpos}{5.5*\xrel}
    \renewcommand{\ypos}{\yleg}
    \coordinate (\nodenum) at (\xpos,\ypos) {};
    
	\renewcommand{\nodenum}{r2u4Blue}
	\renewcommand{\xpos}{5.5*\xrel-\xrecenter}
    \renewcommand{\ypos}{-2*\rowspace-\yleg}
    \coordinate (\nodenum) at (\xpos,\ypos) {};
   
	\renewcommand{\nodenum}{r2u4Red}
	\renewcommand{\xpos}{5.5*\xrel+\xrecenter}
    \renewcommand{\ypos}{-2*\rowspace-\yleglong}
    \coordinate (\nodenum) at (\xpos,\ypos) {}; 
    

	\draw [line width=\linethick,color=dullred]  
	(r2u3Red)--++(0,\yleg)
	(r2u4Red)--++(0,\yleg);
	
	\draw [line width=\linethick,color=dullblue]  
	(r2u3Blue)--++(0,\yleg)
	(r2u4Blue)--++(0,\yleg);

	\draw [line width=\linethick,color=dullblue] 
	(r2u3Blue)--(r2u4Blue);
	
	
	\renewcommand{\nodenum}{Equal1}
    \renewcommand{\name}{$=\sum_{\alpha=0}^1i(-1)^\alpha  \int dU_+$}
	\renewcommand{\xpos}{2.8*\xrel}
    \renewcommand{\ypos}{-1*\rowspace}
    \renewcommand{\height}{\heightsingle}
    \renewcommand{\width}{\widthsingle}
    \node[] (\nodenum) at (\xpos,\ypos) {\name};

     \renewcommand{\nodenum}{G3}
    \renewcommand{\name}{$G$}
	\renewcommand{\xpos}{4.5*\xrel}
    \renewcommand{\ypos}{-\rowspace}
    \renewcommand{\height}{\heightsingle}
    \renewcommand{\width}{\widthsingle}
    \node[rectangle, fill=egg, rounded corners, minimum width=\width, minimum height=\height, draw] (\nodenum) at (\xpos,\ypos) {\name};
    
     \renewcommand{\nodenum}{r2u3}
    \renewcommand{\name}{$U_+$}
	\renewcommand{\xpos}{4.5*\xrel}
    \renewcommand{\ypos}{-2*\rowspace}
    \renewcommand{\height}{\heightsingle}
    \renewcommand{\width}{\widthsingle}
    \node[rectangle, fill=egg, rounded corners, minimum width=\width, minimum height=\height, draw] (\nodenum) at (\xpos,\ypos) {\name};
    
     \renewcommand{\nodenum}{r2u4}
    \renewcommand{\name}{$\overline{U}_+$}
	\renewcommand{\xpos}{5.5*\xrel}
    \renewcommand{\ypos}{-2*\rowspace}
    \renewcommand{\height}{\heightsingle}
    \renewcommand{\width}{\widthsingle}
    \node[rectangle, fill=egg, rounded corners, minimum width=\width, minimum height=\height, draw] (\nodenum) at (\xpos,\ypos) {\name};

     \draw [line width=\linethick,color=black]  (r2u4)--(r1u4)--(r1u3)--(G3)--(r2u3) ;
      \draw [line width=\linethick,color=black]  (r1u3top)--++(0,-.5*\yleg)--++(\xrel,0)--(r1u4top);

\end{tikzpicture}.
\end{equation}

This produces $\langle \partial_k C\rangle=0$ because the diagram on the right-hand side is independent of $\alpha$ and $\sum_{\alpha=0}^1i(-1)^\alpha=0$. Now take the case where only $U_+$ forms a 2-design. The diagram on the first site evaluates to

\begin{equation}
\begin{tikzpicture}

	
     \renewcommand{\nodenum}{r1u1Red}
	\renewcommand{\xpos}{-\xrecenter}
    \renewcommand{\ypos}{\yleg}
    \coordinate (\nodenum) at (\xpos,\ypos) {};

	\renewcommand{\nodenum}{r1u1Blue}
	\renewcommand{\xpos}{\xrecenter}
    \renewcommand{\ypos}{\yleg}
    \coordinate (\nodenum) at (\xpos,\ypos) {};

	\renewcommand{\nodenum}{r2u1}
	\renewcommand{\xpos}{0}
    \renewcommand{\ypos}{-2*\rowspace}
    \coordinate (\nodenum) at (\xpos,\ypos) {};
    
	\renewcommand{\nodenum}{r2u1bot}
	\renewcommand{\xpos}{0}
    \renewcommand{\ypos}{-2*\rowspace-\yleg}
    \coordinate (\nodenum) at (\xpos,\ypos) {};
    
     \renewcommand{\nodenum}{r1u2Blue}
	\renewcommand{\xpos}{\xrel-\xrecenter}
    \renewcommand{\ypos}{\yleg}
    \coordinate (\nodenum) at (\xpos,\ypos) {};

	\renewcommand{\nodenum}{r1u2Red}
	\renewcommand{\xpos}{\xrel+\xrecenter}
    \renewcommand{\ypos}{\yleg}
    \coordinate (\nodenum) at (\xpos,\ypos) {};

	\renewcommand{\nodenum}{r2u2}
	\renewcommand{\xpos}{\xrel}
    \renewcommand{\ypos}{-2*\rowspace}
    \coordinate (\nodenum) at (\xpos,\ypos) {};
   
	\renewcommand{\nodenum}{r2u2bot}
	\renewcommand{\xpos}{\xrel}
    \renewcommand{\ypos}{-2*\rowspace-\yleg}
    \coordinate (\nodenum) at (\xpos,\ypos) {};
    
	\draw [line width=\linethick,color=dullred]  
	(r1u1Red)--++(0,-\yleg)
	(r1u2Red)--++(0,-\yleg);
	
	\draw [line width=\linethick,color=dullblue]  
	(r1u1Blue)--++(0,-\yleg)
	(r1u2Blue)--++(0,-\yleg);

	
	\renewcommand{\nodenum}{v0}
    \renewcommand{\name}{$\sum_{\alpha=0}^{1}i( -1)^{\alpha} \int dU_- $}
	\renewcommand{\xpos}{-1.5*\xrel}
    \renewcommand{\ypos}{-1*\rowspace}
    \renewcommand{\height}{\heightsingle}
    \renewcommand{\width}{\widthsingle}
    \node[] (\nodenum) at (\xpos,\ypos) {\name};
   
   \renewcommand{\nodenum}{s1}
    \renewcommand{\name}{}
	\renewcommand{\xpos}{\xrecenter}
    \renewcommand{\ypos}{\yleg}
    \renewcommand{\height}{\heightstate}
    \renewcommand{\width}{\widthstate}
    \node[rectangle, fill=dullblue, rounded corners=.2em, minimum width=\width, minimum height=\height, draw] (\nodenum) at (\xpos,\ypos) {\name};
       
   \renewcommand{\nodenum}{s2}
    \renewcommand{\name}{}
	\renewcommand{\xpos}{\xrel-\xrecenter}
    \renewcommand{\ypos}{\yleg}
    \renewcommand{\height}{\heightstate}
    \renewcommand{\width}{\widthstate}
    \node[rectangle, fill=dullblue, rounded corners=.2em, minimum width=\width, minimum height=\height, draw] (\nodenum) at (\xpos,\ypos) {\name};
    
     \renewcommand{\nodenum}{r1u1}
    \renewcommand{\name}{$U_-$}
	\renewcommand{\xpos}{0}
    \renewcommand{\ypos}{0}
    \renewcommand{\height}{\heightsingle}
    \renewcommand{\width}{\widthsingle}
    \node[rectangle, fill=egg, rounded corners, minimum width=\width, minimum height=\height, draw] (\nodenum) at (\xpos,\ypos) {\name};
    
    \renewcommand{\nodenum}{r1u2}
    \renewcommand{\name}{$\overline{U}_-$}
	\renewcommand{\xpos}{\xrel}
    \renewcommand{\ypos}{0}
    \renewcommand{\height}{\heightsingle}
    \renewcommand{\width}{\widthsingle}
    \node[rectangle, fill=egg, rounded corners, minimum width=\width, minimum height=\height, draw] (\nodenum) at (\xpos,\ypos) {\name};

     \renewcommand{\nodenum}{G1}
    \renewcommand{\name}{$G^\alpha$}
	\renewcommand{\xpos}{0}
    \renewcommand{\ypos}{-\rowspace}
    \renewcommand{\height}{\heightsingle}
    \renewcommand{\width}{\widthsingle}
    \node[rectangle, fill=egg, rounded corners, minimum width=\width, minimum height=\height, draw] (\nodenum) at (\xpos,\ypos) {\name};
    
     \renewcommand{\nodenum}{G2}
    \renewcommand{\name}{$\overline{G}^{1-\alpha}$}
	\renewcommand{\xpos}{\xrel}
    \renewcommand{\ypos}{-\rowspace}
    \renewcommand{\height}{\heightsingle}
    \renewcommand{\width}{\widthsingle}
    \node[rectangle, fill=egg, rounded corners, minimum width=\width, minimum height=\height, draw] (\nodenum) at (\xpos,\ypos) {\name};

     \draw [line width=\linethick,color=black]  (r1u1)--(G1)--(G1) --(r2u1) --(r2u2)--(G2)--(r1u2) ;
         \draw [line width=\linethick,color=black]  (r2u1bot)--++(0,.5*\yleg)--++(\xrel,0)--(r2u2bot);
    

	
     \renewcommand{\nodenum}{r1u3Red}
	\renewcommand{\xpos}{4.5*\xrel-\xrecenter}
    \renewcommand{\ypos}{\yleg}
    \coordinate (\nodenum) at (\xpos,\ypos) {};

	\renewcommand{\nodenum}{r1u3Blue}
	\renewcommand{\xpos}{4.5*\xrel+\xrecenter}
    \renewcommand{\ypos}{\yleg}
    \coordinate (\nodenum) at (\xpos,\ypos) {};

	\renewcommand{\nodenum}{r2u3}
	\renewcommand{\xpos}{4.5*\xrel}
    \renewcommand{\ypos}{-2*\rowspace}
    \coordinate (\nodenum) at (\xpos,\ypos) {};
    
	\renewcommand{\nodenum}{r2u3bot}
	\renewcommand{\xpos}{4.5*\xrel}
    \renewcommand{\ypos}{-2*\rowspace-\yleg}
    \coordinate (\nodenum) at (\xpos,\ypos) {};
    
     \renewcommand{\nodenum}{r1u4Blue}
	\renewcommand{\xpos}{5.5*\xrel-\xrecenter}
    \renewcommand{\ypos}{\yleg}
    \coordinate (\nodenum) at (\xpos,\ypos) {};

	\renewcommand{\nodenum}{r1u4Red}
	\renewcommand{\xpos}{5.5*\xrel+\xrecenter}
    \renewcommand{\ypos}{\yleg}
    \coordinate (\nodenum) at (\xpos,\ypos) {};

	\renewcommand{\nodenum}{r2u4}
	\renewcommand{\xpos}{5.5*\xrel}
    \renewcommand{\ypos}{-2*\rowspace}
    \coordinate (\nodenum) at (\xpos,\ypos) {};
   
	\renewcommand{\nodenum}{r2u4bot}
	\renewcommand{\xpos}{4.5*\xrel+\xrel}
    \renewcommand{\ypos}{-2*\rowspace-\yleg}
    \coordinate (\nodenum) at (\xpos,\ypos) {};
    
	\draw [line width=\linethick,color=dullred]  
	(r1u3Red)--++(0,-\yleg)
	(r1u4Red)--++(0,-\yleg);
	
	\draw [line width=\linethick,color=dullblue]  
	(r1u3Blue)--++(0,-\yleg)
	(r1u4Blue)--++(0,-\yleg);

	
	\renewcommand{\nodenum}{Equality1}
    \renewcommand{\name}{$=\sum_{\alpha=0}^{1}i( -1)^{\alpha} \int dU_- $}
	\renewcommand{\xpos}{2.8*\xrel}
    \renewcommand{\ypos}{-1*\rowspace}
    \renewcommand{\height}{\heightsingle}
    \renewcommand{\width}{\widthsingle}
    \node[] (\nodenum) at (\xpos,\ypos) {\name};
   
   \renewcommand{\nodenum}{s3}
    \renewcommand{\name}{}
	\renewcommand{\xpos}{4.5*\xrel+\xrecenter}
    \renewcommand{\ypos}{\yleg}
    \renewcommand{\height}{\heightstate}
    \renewcommand{\width}{\widthstate}
    \node[rectangle, fill=dullblue, rounded corners=.2em, minimum width=\width, minimum height=\height, draw] (\nodenum) at (\xpos,\ypos) {\name};
       
   \renewcommand{\nodenum}{s4}
    \renewcommand{\name}{}
	\renewcommand{\xpos}{5.5*\xrel-\xrecenter}
    \renewcommand{\ypos}{\yleg}
    \renewcommand{\height}{\heightstate}
    \renewcommand{\width}{\widthstate}
    \node[rectangle, fill=dullblue, rounded corners=.2em, minimum width=\width, minimum height=\height, draw] (\nodenum) at (\xpos,\ypos) {\name};
    
     \renewcommand{\nodenum}{r1u3}
    \renewcommand{\name}{$U_-$}
	\renewcommand{\xpos}{4.5*\xrel}
    \renewcommand{\ypos}{0}
    \renewcommand{\height}{\heightsingle}
    \renewcommand{\width}{\widthsingle}
    \node[rectangle, fill=egg, rounded corners, minimum width=\width, minimum height=\height, draw] (\nodenum) at (\xpos,\ypos) {\name};
    
    \renewcommand{\nodenum}{r1u4}
    \renewcommand{\name}{$\overline{U}_-$}
	\renewcommand{\xpos}{5.5*\xrel}
    \renewcommand{\ypos}{0}
    \renewcommand{\height}{\heightsingle}
    \renewcommand{\width}{\widthsingle}
    \node[rectangle, fill=egg, rounded corners, minimum width=\width, minimum height=\height, draw] (\nodenum) at (\xpos,\ypos) {\name};

     \renewcommand{\nodenum}{G3}
    \renewcommand{\name}{$G$}
	\renewcommand{\xpos}{4.5*\xrel}
    \renewcommand{\ypos}{-\rowspace}
    \renewcommand{\height}{\heightsingle}
    \renewcommand{\width}{\widthsingle}
    \node[rectangle, fill=egg, rounded corners, minimum width=\width, minimum height=\height, draw] (\nodenum) at (\xpos,\ypos) {\name};

     \draw [line width=\linethick,color=black]  (r1u3)--(G3)--(r2u3) --(r2u4)--(r1u4) ;
     
         \draw [line width=\linethick,color=black]  (r2u3bot)--++(0,.5*\yleg)--++(\xrel,0)--(r2u4bot);

\end{tikzpicture}.
\end{equation}
This also produces $\langle \partial_k C\rangle=0$. The case where both $U_-$ and $U_+$ form 2-designs also produces $\langle \partial_k C\rangle=0$. Therefore, when either $U_-$ or $U_+$ form a 2-design,
\begin{equation}
	\langle \partial_k C\rangle=0.
\end{equation}
In fact, the above holds even if $U_+$ or $U_-$ form a 1-design.

%% file: ProofTheorem1.tex
\section{Proof of Theorem~\ref{Theorem}}\label{Appendix:ProofThm1}
The variance of $\partial_kC$ is
\begin{equation}
	\mathrm{Var}[\partial_kC]=\langle (\partial_k C)^2\rangle-\langle \partial_k C\rangle^2.
\end{equation}
Since $\langle \partial_k C\rangle=0$, then
\begin{equation}
	\mathrm{Var}[\partial_kC]=\langle (\partial_k C)^2\rangle.
\end{equation}

We now introduce the Weingarten calculus to evaluate the average. When $U_i$ forms a 2-design, the following identity holds:

\begin{equation}\label{Eq:HaarIntegral}
.
\end{equation}

On the right-hand side, each leg represents four legs on the left-hand side. The green dot represents the sum of permutation operators

\begin{equation}
%
.
\end{equation}
These permutation operators represent the possible ways we can contract the indices for four input legs. The dashed line in Eq.~\eqref{Eq:HaarIntegral} represents the weight given by the Weingarten function. The four possible values this weight can have are given by
\begin{equation}\label{Eq:Dashed}
	%
.
\end{equation}

We also define the following useful diagrams:

\begin{equation}\label{Eq:GreenU}
%
.
\end{equation}
When $U_{\pm}$ forms a 2-design, Eq.~\eqref{Eq:GreenU} reduces to Eq.~\eqref{Eq:HaarIntegral}.
The following identities will be useful:

\begin{equation}\label{Eq:StateConstraction}
%
\end{equation}
where $\Omega=d,D,Dd$, depending on whether the legs are blue, red, or black respectively.

The variance can then be expressed diagrammatically as
\begin{equation}
%
.
\end{equation}
From Eq.~\eqref{Eq:StateConstraction}, it follows that the variance can be simplified to
\begin{equation}
%
,
\end{equation}
where we define $\Delta=m-1$ as the distance between the derivative site and the site on which $O$ acts. We refer to the case where $O$ and the derivative act on different sites as the off-site case. In the on-site case, where the local operator $O$ acts on the derivative site, the variance diagram is
\begin{equation}
%
.
\end{equation}

The variance takes on one of six possible forms, depending on whether the derivative and $O$ act on the same site, and whether $U_-$, $U_+$ or both form 2-designs.

\subsection{Off-site case}
Before evaluating the variance, it will be useful to define the following constants:
\begin{equation}
\begin{split}
	q&=(Dd)^2-1,\\
	\xi&=\frac{1}{q}D(d^2-1),\\
	\eta&=\frac{1}{q}d(D^2-1),\\
	\Gamma_{L}&=\frac{1-\eta^{L}}{1-\eta},
\end{split}
\end{equation}
and
\begin{equation}
\begin{split}
	C_1&=2\int dU_+\left[-\ptr{d}{\ptr{D}{U_+^\dagger GU_+}^2}+D\tr{G^2}\right],\\
	C_2&=2\int dU_-  \left[-\tr{\rho G\rho G}+\tr{ G^2 \rho^2}\right],\\
	C_3&=2\int dU_-  \left[-\tr{\rho G}^2+D\tr{\rho G^2 }\right],\\
	C_4&=2[-\tr{G}^2+Dd\tr{G^2}],\\
	C_5&=2\int dU_+  \tr{\sigma G[G,\sigma]},\\
	C_6&=2\int dU_+ \Big[-\ptr{d}{\left(\ptr{D}{U_+^\dagger GU_+}O\right)^2}+D\ptr{d}{\ptr{D}{U_+^\dagger G^2U_+}O^2}\Big].
\end{split}
\end{equation}
In the above, we define $\rho=U_-^\dagger (I_D\otimes \ket{0}\bra{0})U_-$ and $\sigma=U_+ (I_D\otimes O)U_+^\dagger$. We assume that $G$ is independent of $n$. $C_5$ and $C_6$ depend on $O$ and can therefore depend on $n$. $C_1,C_2,C_3$, and $C_4$ are independent of $n$.

We evaluate the variance in the off-site case, when the derivative is not taken on the site where $O$ acts. Consider the case where $U_-$ forms  a 2-design, but $U_+$ does not. 
The variance is

\begin{equation}
\begin{tikzpicture}

     \renewcommand{\nodenum}{u-edge}
	\renewcommand{\xpos}{-.5*\heightsingle}
    \renewcommand{\ypos}{0}
    \coordinate (\nodenum) at (\xpos,\ypos) {};
    
     \renewcommand{\nodenum}{u+edge}
	\renewcommand{\xpos}{.5*\heightsingle}
    \renewcommand{\ypos}{-2*\rowspace}
    \coordinate (\nodenum) at (\xpos,\ypos) {};
    
   
   \renewcommand{\nodenum}{r1s1}
    \renewcommand{\name}{}
	\renewcommand{\xpos}{0}
    \renewcommand{\ypos}{2*\yleg}
    \renewcommand{\height}{\heightstate}
    \renewcommand{\width}{\widthstate}
    \node[rectangle, fill=lavender, rounded corners=.2em, minimum width=\width, minimum height=\height, draw] (\nodenum) at (\xpos,\ypos) {\name};
   
   \renewcommand{\nodenum}{Var}
    \renewcommand{\name}{$\mathrm{Var}[\partial_k C]=$}
	\renewcommand{\xpos}{-1.5*\xrel}
    \renewcommand{\ypos}{-\rowspace}
    \node[] (\nodenum) at (\xpos,\ypos) {\name};
    
   \renewcommand{\nodenum}{S1}
    \renewcommand{\name}{$S$}
	\renewcommand{\xpos}{0}
    \renewcommand{\ypos}{-2*\rowspace-\yleg}
    \node[] (\nodenum) at (\xpos,\ypos) {\name};
    
   \renewcommand{\nodenum}{S2}
    \renewcommand{\name}{$S$}
	\renewcommand{\xpos}{\xcircle}
    \renewcommand{\ypos}{-\rowspace-\yleg}
    \node[] (\nodenum) at (\xpos,\ypos) {\name};
        
   \renewcommand{\nodenum}{S3}
    \renewcommand{\name}{$S$}
	\renewcommand{\xpos}{2*\xcircle}
    \renewcommand{\ypos}{-\rowspace-\yleg}
    \node[] (\nodenum) at (\xpos,\ypos) {\name};
            
   \renewcommand{\nodenum}{S4}
    \renewcommand{\name}{$S$}
	\renewcommand{\xpos}{3.5*\xcircle}
    \renewcommand{\ypos}{-2*\rowspace-\yleg}
    \renewcommand{\height}{\heightstate}
    \renewcommand{\width}{\widthstate}
    \node[] (\nodenum) at (\xpos,\ypos) {\name};
                
   \renewcommand{\nodenum}{S5}
    \renewcommand{\name}{$S$}
	\renewcommand{\xpos}{5*\xcircle}
    \renewcommand{\ypos}{-\rowspace-\yleg}
    \renewcommand{\height}{\heightstate}
    \renewcommand{\width}{\widthstate}
    \node[] (\nodenum) at (\xpos,\ypos) {\name};
    
    \renewcommand{\nodenum}{r1c0}
    \renewcommand{\name}{}
	\renewcommand{\xpos}{0}
    \renewcommand{\ypos}{\ycircle}
    \node[circle, scale=\circlescale, fill=evergreen, draw] (\nodenum) at (\xpos,\ypos) {\name};
    
    \renewcommand{\nodenum}{r2c0}
    \renewcommand{\name}{}
	\renewcommand{\xpos}{0}
    \renewcommand{\ypos}{0}
    \node[circle, scale=\circlescale, fill=evergreen, draw] (\nodenum) at (\xpos,\ypos) {\name};
    
    \renewcommand{\nodenum}{G}
    \renewcommand{\name}{$G$}
	\renewcommand{\xpos}{0}
    \renewcommand{\ypos}{-\rowspace}
    \renewcommand{\height}{\heightsingle}
    \renewcommand{\width}{\widthsingle}
    \node[rectangle, fill=lavender, rounded corners, minimum width=\width, minimum height=\height, draw] (\nodenum) at (\xpos,\ypos) {\name};
    
    \renewcommand{\nodenum}{u+}
    \renewcommand{\name}{$U_{+}$}
	\renewcommand{\xpos}{0}
    \renewcommand{\ypos}{-2*\rowspace}
    \renewcommand{\height}{\heightsingle}
    \renewcommand{\width}{\widthsingle}
    \node[rectangle, fill=evergreen, rounded corners, minimum width=\width, minimum height=\height, draw] (\nodenum) at (\xpos,\ypos) {\name};
    	    
    	\renewcommand{\nodenum}{O}
    \renewcommand{\name}{$O$}
	\renewcommand{\xpos}{3.5*\xcircle}
    \renewcommand{\ypos}{-2*\rowspace}
    \renewcommand{\height}{\heightsingle}
    \renewcommand{\width}{\widthsingle}
    \node[rectangle, fill=lavender, rounded corners, minimum width=\width, minimum height=\height, draw] (\nodenum) at (\xpos,\ypos) {\name};
    
    \renewcommand{\nodenum}{r1c1}
    \renewcommand{\name}{}
	\renewcommand{\xpos}{\xcircle}
    \renewcommand{\ypos}{0}
    \node[circle, scale=\circlescale, fill=evergreen, draw] (\nodenum) at (\xpos,\ypos) {\name};
    
    \renewcommand{\nodenum}{r2c1}
    \renewcommand{\name}{}
	\renewcommand{\xpos}{\xcircle}
    \renewcommand{\ypos}{-\ycircle}
    \node[circle, scale=\circlescale, fill=evergreen, draw] (\nodenum) at (\xpos,\ypos) {\name};
        
    \renewcommand{\nodenum}{r1c2}
    \renewcommand{\name}{}
	\renewcommand{\xpos}{2*\xcircle}
    \renewcommand{\ypos}{0}
    \node[circle, scale=\circlescale, fill=evergreen, draw] (\nodenum) at (\xpos,\ypos) {\name};
    
    \renewcommand{\nodenum}{r2c2}
    \renewcommand{\name}{}
	\renewcommand{\xpos}{2*\xcircle}
    \renewcommand{\ypos}{-\ycircle}
    \node[circle, scale=\circlescale, fill=evergreen, draw] (\nodenum) at (\xpos,\ypos) {\name};
        
    \renewcommand{\nodenum}{r1c3}
    \renewcommand{\name}{}
	\renewcommand{\xpos}{3.5*\xcircle}
    \renewcommand{\ypos}{0}
    \node[circle, scale=\circlescale, fill=evergreen, draw] (\nodenum) at (\xpos,\ypos) {\name};
    
    \renewcommand{\nodenum}{r2c3}
    \renewcommand{\name}{}
	\renewcommand{\xpos}{3.5*\xcircle}
    \renewcommand{\ypos}{-\ycircle}
    \node[circle, scale=\circlescale, fill=evergreen, draw] (\nodenum) at (\xpos,\ypos) {\name};
    
    \renewcommand{\nodenum}{r1c4}
    \renewcommand{\name}{}
	\renewcommand{\xpos}{5*\xcircle}
    \renewcommand{\ypos}{0}
    \node[circle, scale=\circlescale, fill=evergreen, draw] (\nodenum) at (\xpos,\ypos) {\name};
    
    \renewcommand{\nodenum}{r2c4}
    \renewcommand{\name}{}
	\renewcommand{\xpos}{5*\xcircle}
    \renewcommand{\ypos}{-\ycircle}
    \node[circle, scale=\circlescale, fill=evergreen, draw] (\nodenum) at (\xpos,\ypos) {\name};

   \renewcommand{\nodenum}{Ellipses}
    \renewcommand{\name}{$\bm{\cdots}$}
	\renewcommand{\xpos}{2.8*\xcircle}
    \renewcommand{\ypos}{-1.5*\yrel}
    \node[] (\nodenum) at (\xpos,\ypos) {\name};
   \renewcommand{\nodenum}{Ellipses}
    \renewcommand{\name}{$\bm{\cdots}$}
	\renewcommand{\xpos}{4.3*\xcircle}
    \renewcommand{\ypos}{-1.5*\yrel}
    \node[] (\nodenum) at (\xpos,\ypos) {\name};
    \renewcommand{\nodenum}{L1}
    \renewcommand{\name}{$\Delta-1$}
	\renewcommand{\xpos}{2*\xcircle}
    \renewcommand{\ypos}{1.5*\yrel}
    \renewcommand{\height}{\heightsingle}
    \renewcommand{\width}{\widthsingle}
    \node[] (\nodenum) at (\xpos,\ypos) {\name};
    
    \renewcommand{\nodenum}{L1}
    \renewcommand{\name}{$n-\Delta-1$}
	\renewcommand{\xpos}{4.5*\xcircle}
    \renewcommand{\ypos}{1.5*\yrel}
    \renewcommand{\height}{\heightsingle}
    \renewcommand{\width}{\widthsingle}
    \node[] (\nodenum) at (\xpos,\ypos) {\name};
    
	\renewcommand{\xpos}{\xcircle}
	\renewcommand{\ypos}{.6*\yrel}
	\draw [decorate,decoration={brace,amplitude=5pt},xshift=\xrel,yshift=-\rowspace]	 (\xpos,\ypos)--(\xpos+1.75*\xcircle,\ypos)  node [black,midway,xshift=9pt] {};
    %
	\renewcommand{\xpos}{4*\xcircle}
	\renewcommand{\ypos}{.6*\yrel}
	\draw [decorate,decoration={brace,amplitude=5pt},xshift=\xrel,yshift=-\rowspace]	 (\xpos,\ypos)--(\xpos+\xcircle,\ypos)  node [black,midway,xshift=9pt] {};
	
    \draw [line width=\linethick,color=dullred]  
    (r1c0)--++(-\xcircle,-\ycircle)
    (u+edge)--(r1c1)
    (r2c1)--(r1c2)
    (r2c2)--++(.5*\xcircle,.5*\ycircle)
    (r1c3)--++(-.5*\xcircle,-.5*\ycircle)
    (r2c3)--++(.5*\xcircle,.5*\ycircle)
    (r1c4)--++(-.5*\xcircle,-.5*\ycircle)
    (r2c4)--++(.5*\xcircle,.5*\ycircle);
    
    \draw [line width=\linethick,color=dullblue]  
    (r1c0)--(r1s1);
    
     \draw [line width=\linethick,color=dullblue]  
    (u+)--(S1)
    (r2c1)--(S2)
    (r2c2)--(S3)
    (r2c3)--(O)--(S4)
    (r2c4)--(S5);
    
    \draw [line width=\linethick, color=black]  
    (r2c0)--(G)--(u+);
    
    \draw [line width=\linethick, dashed, color=black]  
    (r1c0)--(r2c0)
    (r1c1)--(r2c1)
    (r1c2)--(r2c2)
    (r1c3)--(r2c3)
    (r1c4)--(r2c4);
\end{tikzpicture}.
\end{equation}
This evaluates to
\begin{equation}
	\mathrm{Var}[\partial_k C]=\frac{C_1\eta^{\Delta-1}}{q^2}\left[\epsilon(O)\left(-\frac{1}{d}+D\xi\Gamma_{n-\Delta-1}+D^2\eta^{n-\Delta-1}\right)+\ptr{d}{O}^2\frac{(D^2-1)\eta^{n-\Delta-1}}{d}\right],
\end{equation}
where we define 
\begin{equation}
	\epsilon(O)\equiv \norm{O-\ptr{d}{O}\frac{I_d}{d}}_{\mathrm{HS}}^2.
\end{equation}
We assume $\ptr{d}{O}^2$ grows slower than exponentially in $n$. When $n$ is large, the variance becomes 
\begin{equation}
	\mathrm{Var}[\partial_k C]=\epsilon(O)\frac{C_1\eta^{\Delta-1}}{q^2}\left(-\frac{1}{d}+\frac{D\xi}{1-\eta} \right),
\end{equation}
where we assume $d\geq 2$. 

In the off-site case where only $U_+$ forms a 2-design, the variance is
\begin{equation}
\begin{tikzpicture}

     \renewcommand{\nodenum}{u-edge}
	\renewcommand{\xpos}{-.5*\heightsingle}
    \renewcommand{\ypos}{0}
    \coordinate (\nodenum) at (\xpos,\ypos) {};
    
   
   \renewcommand{\nodenum}{r1s1}
    \renewcommand{\name}{}
	\renewcommand{\xpos}{0}
    \renewcommand{\ypos}{\yleg}
    \renewcommand{\height}{\heightstate}
    \renewcommand{\width}{\widthstate}
    \node[rectangle, fill=lavender, rounded corners=.2em, minimum width=\width, minimum height=\height, draw] (\nodenum) at (\xpos,\ypos) {\name};
   
   \renewcommand{\nodenum}{Var}
    \renewcommand{\name}{$\mathrm{Var}[\partial_k C]=$}
	\renewcommand{\xpos}{-1.5*\xrel}
    \renewcommand{\ypos}{-\rowspace}
    \node[] (\nodenum) at (\xpos,\ypos) {\name};
    
   \renewcommand{\nodenum}{S1}
    \renewcommand{\name}{$S$}
	\renewcommand{\xpos}{0}
    \renewcommand{\ypos}{-2*\rowspace-\ycircle-\yleg}
    \node[] (\nodenum) at (\xpos,\ypos) {\name};
    
   \renewcommand{\nodenum}{S2}
    \renewcommand{\name}{$S$}
	\renewcommand{\xpos}{\xcircle}
    \renewcommand{\ypos}{-\rowspace-\yleg}
    \node[] (\nodenum) at (\xpos,\ypos) {\name};
        
   \renewcommand{\nodenum}{S3}
    \renewcommand{\name}{$S$}
	\renewcommand{\xpos}{2*\xcircle}
    \renewcommand{\ypos}{-\rowspace-\yleg}
    \node[] (\nodenum) at (\xpos,\ypos) {\name};
            
   \renewcommand{\nodenum}{S4}
    \renewcommand{\name}{$S$}
	\renewcommand{\xpos}{3.5*\xcircle}
    \renewcommand{\ypos}{-2*\rowspace-\yleg}
    \renewcommand{\height}{\heightstate}
    \renewcommand{\width}{\widthstate}
    \node[] (\nodenum) at (\xpos,\ypos) {\name};
                
   \renewcommand{\nodenum}{S5}
    \renewcommand{\name}{$S$}
	\renewcommand{\xpos}{5*\xcircle}
    \renewcommand{\ypos}{-\rowspace-\yleg}
    \renewcommand{\height}{\heightstate}
    \renewcommand{\width}{\widthstate}
    \node[] (\nodenum) at (\xpos,\ypos) {\name};
    
    \renewcommand{\nodenum}{u-}
    \renewcommand{\name}{$U_{-}$}
	\renewcommand{\xpos}{0}
    \renewcommand{\ypos}{0}
    \renewcommand{\height}{\heightsingle}
    \renewcommand{\width}{\widthsingle}
    \node[rectangle, fill=evergreen, rounded corners, minimum width=\width, minimum height=\height, draw] (\nodenum) at (\xpos,\ypos) {\name};
    
    \renewcommand{\nodenum}{G}
    \renewcommand{\name}{$G$}
	\renewcommand{\xpos}{0}
    \renewcommand{\ypos}{-\rowspace}
    \renewcommand{\height}{\heightsingle}
    \renewcommand{\width}{\widthsingle}
    \node[rectangle, fill=lavender, rounded corners, minimum width=\width, minimum height=\height, draw] (\nodenum) at (\xpos,\ypos) {\name};
        
    \renewcommand{\nodenum}{r1c0}
    \renewcommand{\name}{}
	\renewcommand{\xpos}{0}
    \renewcommand{\ypos}{-2*\rowspace}
    \node[circle, scale=\circlescale, fill=evergreen, draw] (\nodenum) at (\xpos,\ypos) {\name};
            
    \renewcommand{\nodenum}{r2c0}
    \renewcommand{\name}{}
	\renewcommand{\xpos}{0}
    \renewcommand{\ypos}{-2*\rowspace-\ycircle}
    \node[circle, scale=\circlescale, fill=evergreen, draw] (\nodenum) at (\xpos,\ypos) {\name};
    	    
    	\renewcommand{\nodenum}{O}
    \renewcommand{\name}{$O$}
	\renewcommand{\xpos}{3.5*\xcircle}
    \renewcommand{\ypos}{-2*\rowspace}
    \renewcommand{\height}{\heightsingle}
    \renewcommand{\width}{\widthsingle}
    \node[rectangle, fill=lavender, rounded corners, minimum width=\width, minimum height=\height, draw] (\nodenum) at (\xpos,\ypos) {\name};
    
    \renewcommand{\nodenum}{r1c1}
    \renewcommand{\name}{}
	\renewcommand{\xpos}{\xcircle}
    \renewcommand{\ypos}{0}
    \node[circle, scale=\circlescale, fill=evergreen, draw] (\nodenum) at (\xpos,\ypos) {\name};
    
    \renewcommand{\nodenum}{r2c1}
    \renewcommand{\name}{}
	\renewcommand{\xpos}{\xcircle}
    \renewcommand{\ypos}{-\ycircle}
    \node[circle, scale=\circlescale, fill=evergreen, draw] (\nodenum) at (\xpos,\ypos) {\name};
        
    \renewcommand{\nodenum}{r1c2}
    \renewcommand{\name}{}
	\renewcommand{\xpos}{2*\xcircle}
    \renewcommand{\ypos}{0}
    \node[circle, scale=\circlescale, fill=evergreen, draw] (\nodenum) at (\xpos,\ypos) {\name};
    
    \renewcommand{\nodenum}{r2c2}
    \renewcommand{\name}{}
	\renewcommand{\xpos}{2*\xcircle}
    \renewcommand{\ypos}{-\ycircle}
    \node[circle, scale=\circlescale, fill=evergreen, draw] (\nodenum) at (\xpos,\ypos) {\name};
        
    \renewcommand{\nodenum}{r1c3}
    \renewcommand{\name}{}
	\renewcommand{\xpos}{3.5*\xcircle}
    \renewcommand{\ypos}{0}
    \node[circle, scale=\circlescale, fill=evergreen, draw] (\nodenum) at (\xpos,\ypos) {\name};
    
    \renewcommand{\nodenum}{r2c3}
    \renewcommand{\name}{}
	\renewcommand{\xpos}{3.5*\xcircle}
    \renewcommand{\ypos}{-\ycircle}
    \node[circle, scale=\circlescale, fill=evergreen, draw] (\nodenum) at (\xpos,\ypos) {\name};
    
    \renewcommand{\nodenum}{r1c4}
    \renewcommand{\name}{}
	\renewcommand{\xpos}{5*\xcircle}
    \renewcommand{\ypos}{0}
    \node[circle, scale=\circlescale, fill=evergreen, draw] (\nodenum) at (\xpos,\ypos) {\name};
    
    \renewcommand{\nodenum}{r2c4}
    \renewcommand{\name}{}
	\renewcommand{\xpos}{5*\xcircle}
    \renewcommand{\ypos}{-\ycircle}
    \node[circle, scale=\circlescale, fill=evergreen, draw] (\nodenum) at (\xpos,\ypos) {\name};

   \renewcommand{\nodenum}{Ellipses}
    \renewcommand{\name}{$\bm{\cdots}$}
	\renewcommand{\xpos}{2.8*\xcircle}
    \renewcommand{\ypos}{-1.5*\yrel}
    \node[] (\nodenum) at (\xpos,\ypos) {\name};
   \renewcommand{\nodenum}{Ellipses}
    \renewcommand{\name}{$\bm{\cdots}$}
	\renewcommand{\xpos}{4.3*\xcircle}
    \renewcommand{\ypos}{-1.5*\yrel}
    \node[] (\nodenum) at (\xpos,\ypos) {\name};
    \renewcommand{\nodenum}{L1}
    \renewcommand{\name}{$\Delta-1$}
	\renewcommand{\xpos}{2*\xcircle}
    \renewcommand{\ypos}{1.5*\yrel}
    \renewcommand{\height}{\heightsingle}
    \renewcommand{\width}{\widthsingle}
    \node[] (\nodenum) at (\xpos,\ypos) {\name};
    
    \renewcommand{\nodenum}{L1}
    \renewcommand{\name}{$n-\Delta-1$}
	\renewcommand{\xpos}{4.5*\xcircle}
    \renewcommand{\ypos}{1.5*\yrel}
    \renewcommand{\height}{\heightsingle}
    \renewcommand{\width}{\widthsingle}
    \node[] (\nodenum) at (\xpos,\ypos) {\name};
    
	\renewcommand{\xpos}{\xcircle}
	\renewcommand{\ypos}{.6*\yrel}
	\draw [decorate,decoration={brace,amplitude=5pt},xshift=\xrel,yshift=-\rowspace]	 (\xpos,\ypos)--(\xpos+1.75*\xcircle,\ypos)  node [black,midway,xshift=9pt] {};
    %
	\renewcommand{\xpos}{4*\xcircle}
	\renewcommand{\ypos}{.6*\yrel}
	\draw [decorate,decoration={brace,amplitude=5pt},xshift=\xrel,yshift=-\rowspace]	 (\xpos,\ypos)--(\xpos+\xcircle,\ypos)  node [black,midway,xshift=9pt] {};
    \draw [line width=\linethick,color=dullred]  
    (u-edge)--++(-.5*\xcircle,-.5*\ycircle)
    (r2c0)--(r1c1)
    (r2c1)--(r1c2)
    (r2c2)--++(.5*\xcircle,.5*\ycircle)
    (r1c3)--++(-.5*\xcircle,-.5*\ycircle)
    (r2c3)--++(.5*\xcircle,.5*\ycircle)
    (r1c4)--++(-.5*\xcircle,-.5*\ycircle)
    (r2c4)--++(.5*\xcircle,.5*\ycircle);
    
    \draw [line width=\linethick,color=dullblue]  
    (u-)--(r1s1);
    
     \draw [line width=\linethick,color=dullblue]  
    (r2c0)--(S1)
    (r2c1)--(S2)
    (r2c2)--(S3)
    (r2c3)--(O)--(S4)
    (r2c4)--(S5);
    
    \draw [line width=\linethick, color=black]  
    (u-)--(G)--(r1c0);
    
    \draw [line width=\linethick, dashed, color=black]  
    (r1c0)--(r2c0)
    (r1c1)--(r2c1)
    (r1c2)--(r2c2)
    (r1c3)--(r2c3)
    (r1c4)--(r2c4);
\end{tikzpicture},
\end{equation}
where
\begin{equation}
\begin{split}
	\mathrm{Var}[\partial_k C]=&\frac{\eta^{\Delta}}{q^2}\Bigg[\epsilon(O)
\left(C_2Dd^2-C_3+\left(-\frac{C_2d}{D}+C_3d\right)\Big(D\xi \Gamma_{n-\Delta-2}+D^2\eta^{n-\Delta-2}\Big)\right)\\
&\hspace{10mm}+\ptr{d}{O}^2\left(-\frac{C_2}{D}+C_3\right)(D^2-1)\eta^{n-\Delta-2}\Bigg].
\end{split}
\end{equation}
When $n$ is large, this converges to
\begin{equation}
	\mathrm{Var}[\partial_k C]=\epsilon(O)\frac{\eta^{\Delta}}{q^2}\left[C_2Dd^2-C_3+d\left(-\frac{C_2}{D}+C_3\right)\left(\frac{D\xi}{1-\eta}\right)\right].
\end{equation}

In the off-site case where both $U_-$ and $U_+$ form 2-designs, the variance is
\begin{equation}
\begin{tikzpicture}	

   
   \renewcommand{\nodenum}{r1s1}
    \renewcommand{\name}{}
	\renewcommand{\xpos}{0}
    \renewcommand{\ypos}{\ycircle+\yleg}
    \renewcommand{\height}{\heightstate}
    \renewcommand{\width}{\widthstate}
    \node[rectangle, fill=lavender, rounded corners=.2em, minimum width=\width, minimum height=\height, draw] (\nodenum) at (\xpos,\ypos) {\name};
   
   \renewcommand{\nodenum}{Var}
    \renewcommand{\name}{$\mathrm{Var}[\partial_k C]=$}
	\renewcommand{\xpos}{-1.5*\xrel}
    \renewcommand{\ypos}{-\rowspace}
    \node[] (\nodenum) at (\xpos,\ypos) {\name};
    
   \renewcommand{\nodenum}{S1}
    \renewcommand{\name}{$S$}
	\renewcommand{\xpos}{0}
    \renewcommand{\ypos}{-2*\rowspace-2*\yleg}
    \node[] (\nodenum) at (\xpos,\ypos) {\name};
    
   \renewcommand{\nodenum}{S2}
    \renewcommand{\name}{$S$}
	\renewcommand{\xpos}{\xcircle}
    \renewcommand{\ypos}{-\rowspace-\yleg}
    \node[] (\nodenum) at (\xpos,\ypos) {\name};
        
   \renewcommand{\nodenum}{S3}
    \renewcommand{\name}{$S$}
	\renewcommand{\xpos}{2*\xcircle}
    \renewcommand{\ypos}{-\rowspace-\yleg}
    \node[] (\nodenum) at (\xpos,\ypos) {\name};
            
   \renewcommand{\nodenum}{S4}
    \renewcommand{\name}{$S$}
	\renewcommand{\xpos}{3.5*\xcircle}
    \renewcommand{\ypos}{-2*\rowspace-\yleg}
    \renewcommand{\height}{\heightstate}
    \renewcommand{\width}{\widthstate}
    \node[] (\nodenum) at (\xpos,\ypos) {\name};
                
   \renewcommand{\nodenum}{S5}
    \renewcommand{\name}{$S$}
	\renewcommand{\xpos}{5*\xcircle}
    \renewcommand{\ypos}{-\rowspace-\yleg}
    \renewcommand{\height}{\heightstate}
    \renewcommand{\width}{\widthstate}
    \node[] (\nodenum) at (\xpos,\ypos) {\name};
    
    \renewcommand{\nodenum}{r1c0}
    \renewcommand{\name}{}
	\renewcommand{\xpos}{0}
    \renewcommand{\ypos}{\ycircle}
    \node[circle, scale=\circlescale, fill=evergreen, draw] (\nodenum) at (\xpos,\ypos) {\name};
    
        \renewcommand{\nodenum}{r2c0}
    \renewcommand{\name}{}
	\renewcommand{\xpos}{0}
    \renewcommand{\ypos}{0}
    \node[circle, scale=\circlescale, fill=evergreen, draw] (\nodenum) at (\xpos,\ypos) {\name};
        
        \renewcommand{\nodenum}{r3c0}
    \renewcommand{\name}{}
	\renewcommand{\xpos}{0}
    \renewcommand{\ypos}{-2*\rowspace}
    \node[circle, scale=\circlescale, fill=evergreen, draw] (\nodenum) at (\xpos,\ypos) {\name};
    
        \renewcommand{\nodenum}{r4c0}
    \renewcommand{\name}{}
	\renewcommand{\xpos}{0}
    \renewcommand{\ypos}{-2*\rowspace-\ycircle}
    \node[circle, scale=\circlescale, fill=evergreen, draw] (\nodenum) at (\xpos,\ypos) {\name};
    
    \renewcommand{\nodenum}{G}
    \renewcommand{\name}{$G$}
	\renewcommand{\xpos}{0}
    \renewcommand{\ypos}{-\rowspace}
    \renewcommand{\height}{\heightsingle}
    \renewcommand{\width}{\widthsingle}
    \node[rectangle, fill=lavender, rounded corners, minimum width=\width, minimum height=\height, draw] (\nodenum) at (\xpos,\ypos) {\name};

    	\renewcommand{\nodenum}{O}
    \renewcommand{\name}{$O$}
	\renewcommand{\xpos}{3.5*\xcircle}
    \renewcommand{\ypos}{-2*\rowspace}
    \renewcommand{\height}{\heightsingle}
    \renewcommand{\width}{\widthsingle}
    \node[rectangle, fill=lavender, rounded corners, minimum width=\width, minimum height=\height, draw] (\nodenum) at (\xpos,\ypos) {\name};
    
    \renewcommand{\nodenum}{r1c1}
    \renewcommand{\name}{}
	\renewcommand{\xpos}{\xcircle}
    \renewcommand{\ypos}{0}
    \node[circle, scale=\circlescale, fill=evergreen, draw] (\nodenum) at (\xpos,\ypos) {\name};
    
    \renewcommand{\nodenum}{r2c1}
    \renewcommand{\name}{}
	\renewcommand{\xpos}{\xcircle}
    \renewcommand{\ypos}{-\ycircle}
    \node[circle, scale=\circlescale, fill=evergreen, draw] (\nodenum) at (\xpos,\ypos) {\name};
        
    \renewcommand{\nodenum}{r1c2}
    \renewcommand{\name}{}
	\renewcommand{\xpos}{2*\xcircle}
    \renewcommand{\ypos}{0}
    \node[circle, scale=\circlescale, fill=evergreen, draw] (\nodenum) at (\xpos,\ypos) {\name};
    
    \renewcommand{\nodenum}{r2c2}
    \renewcommand{\name}{}
	\renewcommand{\xpos}{2*\xcircle}
    \renewcommand{\ypos}{-\ycircle}
    \node[circle, scale=\circlescale, fill=evergreen, draw] (\nodenum) at (\xpos,\ypos) {\name};
        
    \renewcommand{\nodenum}{r1c3}
    \renewcommand{\name}{}
	\renewcommand{\xpos}{3.5*\xcircle}
    \renewcommand{\ypos}{0}
    \node[circle, scale=\circlescale, fill=evergreen, draw] (\nodenum) at (\xpos,\ypos) {\name};
    
    \renewcommand{\nodenum}{r2c3}
    \renewcommand{\name}{}
	\renewcommand{\xpos}{3.5*\xcircle}
    \renewcommand{\ypos}{-\ycircle}
    \node[circle, scale=\circlescale, fill=evergreen, draw] (\nodenum) at (\xpos,\ypos) {\name};
    
    \renewcommand{\nodenum}{r1c4}
    \renewcommand{\name}{}
	\renewcommand{\xpos}{5*\xcircle}
    \renewcommand{\ypos}{0}
    \node[circle, scale=\circlescale, fill=evergreen, draw] (\nodenum) at (\xpos,\ypos) {\name};
    
    \renewcommand{\nodenum}{r2c4}
    \renewcommand{\name}{}
	\renewcommand{\xpos}{5*\xcircle}
    \renewcommand{\ypos}{-\ycircle}
    \node[circle, scale=\circlescale, fill=evergreen, draw] (\nodenum) at (\xpos,\ypos) {\name};

   \renewcommand{\nodenum}{Ellipses}
    \renewcommand{\name}{$\bm{\cdots}$}
	\renewcommand{\xpos}{2.8*\xcircle}
    \renewcommand{\ypos}{-1.5*\yrel}
    \node[] (\nodenum) at (\xpos,\ypos) {\name};
   \renewcommand{\nodenum}{Ellipses}
    \renewcommand{\name}{$\bm{\cdots}$}
	\renewcommand{\xpos}{4.3*\xcircle}
    \renewcommand{\ypos}{-1.5*\yrel}
    \node[] (\nodenum) at (\xpos,\ypos) {\name};
    \renewcommand{\nodenum}{L1}
    \renewcommand{\name}{$\Delta-1$}
	\renewcommand{\xpos}{2*\xcircle}
    \renewcommand{\ypos}{1.5*\yrel}
    \renewcommand{\height}{\heightsingle}
    \renewcommand{\width}{\widthsingle}
    \node[] (\nodenum) at (\xpos,\ypos) {\name};
    
    \renewcommand{\nodenum}{L1}
    \renewcommand{\name}{$n-\Delta-1$}
	\renewcommand{\xpos}{4.5*\xcircle}
    \renewcommand{\ypos}{1.5*\yrel}
    \renewcommand{\height}{\heightsingle}
    \renewcommand{\width}{\widthsingle}
    \node[] (\nodenum) at (\xpos,\ypos) {\name};
    
	\renewcommand{\xpos}{\xcircle}
	\renewcommand{\ypos}{.6*\yrel}
	\draw [decorate,decoration={brace,amplitude=5pt},xshift=\xrel,yshift=-\rowspace]	 (\xpos,\ypos)--(\xpos+1.75*\xcircle,\ypos)  node [black,midway,xshift=9pt] {};
    %
	\renewcommand{\xpos}{4*\xcircle}
	\renewcommand{\ypos}{.6*\yrel}
	\draw [decorate,decoration={brace,amplitude=5pt},xshift=\xrel,yshift=-\rowspace]	 (\xpos,\ypos)--(\xpos+\xcircle,\ypos)  node [black,midway,xshift=9pt] {};
    \draw [line width=\linethick,color=dullred]  
    (r1c0)--++(-\xcircle,-\ycircle)
    (r4c0)--(r1c1)
    (r2c1)--(r1c2)
    (r2c2)--++(.5*\xcircle,.5*\ycircle)
    (r1c3)--++(-.5*\xcircle,-.5*\ycircle)
    (r2c3)--++(.5*\xcircle,.5*\ycircle)
    (r1c4)--++(-.5*\xcircle,-.5*\ycircle)
    (r2c4)--++(.5*\xcircle,.5*\ycircle);
    
    \draw [line width=\linethick,color=dullblue]  
    (r1c0)--(r1s1);
    
     \draw [line width=\linethick,color=dullblue]  
    (r4c0)--(S1)
    (r2c1)--(S2)
    (r2c2)--(S3)
    (r2c3)--(O)--(S4)
    (r2c4)--(S5);
    
    \draw [line width=\linethick, color=black]  
    (r2c0)--(G)--(r3c0);
    
    \draw [line width=\linethick, dashed, color=black]  
    (r1c0)--(r2c0)
    (r3c0)--(r4c0)
    (r1c1)--(r2c1)
    (r1c2)--(r2c2)
    (r1c3)--(r2c3)
    (r1c4)--(r2c4);
\end{tikzpicture}
.
\end{equation}
This evaluates to
\begin{equation}
	\mathrm{Var}[\partial_k C]	=\frac{C_4\eta^{\Delta}}{q^2}\Bigg[\epsilon(O)\left(-\frac{1}{d}+D\xi \Gamma_{n-\Delta-1}+D^2\eta^{n-\Delta-1}\right)+\frac{1}{d}\ptr{d}{O}^2(D^2-1)\eta^{n-\Delta-1}\Bigg].
\end{equation}
In the large $n$ limit,
\begin{equation}
\begin{split}
	\mathrm{Var}[\partial_k C]
	=&\epsilon(O)\frac{C_4\eta^{\Delta}}{q^2}\left(-\frac{1}{d}+\frac{D\xi}{1-\eta} \right).
\end{split}
\end{equation}

\subsection{On-site case}
We now compute the variance in the on-site case, where the derivative acts on the same site as $O$. When only $U_-$ forms a 2-design, the variance is

\begin{equation}
\begin{tikzpicture}	
	
     \renewcommand{\nodenum}{u-edge}
	\renewcommand{\xpos}{-.5*\heightsingle}
    \renewcommand{\ypos}{0}
    \coordinate (\nodenum) at (\xpos,\ypos) {};
    
     \renewcommand{\nodenum}{u+edge}
	\renewcommand{\xpos}{.5*\heightsingle}
    \renewcommand{\ypos}{-2*\rowspace}
    \coordinate (\nodenum) at (\xpos,\ypos) {};
    
   
   \renewcommand{\nodenum}{r1s1}
    \renewcommand{\name}{}
	\renewcommand{\xpos}{0}
    \renewcommand{\ypos}{\ycircle+\yleg}
    \renewcommand{\height}{\heightstate}
    \renewcommand{\width}{\widthstate}
    \node[rectangle, fill=lavender, rounded corners=.2em, minimum width=\width, minimum height=\height, draw] (\nodenum) at (\xpos,\ypos) {\name};
   
   \renewcommand{\nodenum}{Var}
    \renewcommand{\name}{$\mathrm{Var}[\partial_k C]=$}
	\renewcommand{\xpos}{-1.5*\xrel}
    \renewcommand{\ypos}{-\rowspace}
    \node[] (\nodenum) at (\xpos,\ypos) {\name};
    
   \renewcommand{\nodenum}{S1}
    \renewcommand{\name}{$S$}
	\renewcommand{\xpos}{0}
    \renewcommand{\ypos}{-3*\rowspace-\yleg}
    \node[] (\nodenum) at (\xpos,\ypos) {\name};
    
   \renewcommand{\nodenum}{S2}
    \renewcommand{\name}{$S$}
	\renewcommand{\xpos}{\xcircle}
    \renewcommand{\ypos}{-\rowspace-\yleg}
    \node[] (\nodenum) at (\xpos,\ypos) {\name};
        
   \renewcommand{\nodenum}{S3}
    \renewcommand{\name}{$S$}
	\renewcommand{\xpos}{2*\xcircle}
    \renewcommand{\ypos}{-\rowspace-\yleg}
    \node[] (\nodenum) at (\xpos,\ypos) {\name};
            
   \renewcommand{\nodenum}{S4}
    \renewcommand{\name}{$S$}
	\renewcommand{\xpos}{3.5*\xcircle}
    \renewcommand{\ypos}{-\rowspace-\yleg}
    \renewcommand{\height}{\heightstate}
    \renewcommand{\width}{\widthstate}
    \node[] (\nodenum) at (\xpos,\ypos) {\name};
    
    \renewcommand{\nodenum}{r1c0}
    \renewcommand{\name}{}
	\renewcommand{\xpos}{0}
    \renewcommand{\ypos}{\ycircle}
    \node[circle, scale=\circlescale, fill=evergreen, draw] (\nodenum) at (\xpos,\ypos) {\name};
        
    \renewcommand{\nodenum}{r2c0}
    \renewcommand{\name}{}
	\renewcommand{\xpos}{0}
    \renewcommand{\ypos}{0}
    \node[circle, scale=\circlescale, fill=evergreen, draw] (\nodenum) at (\xpos,\ypos) {\name};
    
    \renewcommand{\nodenum}{G}
    \renewcommand{\name}{$G$}
	\renewcommand{\xpos}{0}
    \renewcommand{\ypos}{-\rowspace}
    \renewcommand{\height}{\heightsingle}
    \renewcommand{\width}{\widthsingle}
    \node[rectangle, fill=lavender, rounded corners, minimum width=\width, minimum height=\height, draw] (\nodenum) at (\xpos,\ypos) {\name};
    
    \renewcommand{\nodenum}{u+}
    \renewcommand{\name}{$U_{+}$}
	\renewcommand{\xpos}{0}
    \renewcommand{\ypos}{-2*\rowspace}
    \renewcommand{\height}{\heightsingle}
    \renewcommand{\width}{\widthsingle}
    \node[rectangle, fill=evergreen, rounded corners, minimum width=\width, minimum height=\height, draw] (\nodenum) at (\xpos,\ypos) {\name};
    	    
    	\renewcommand{\nodenum}{O}
    \renewcommand{\name}{$O$}
	\renewcommand{\xpos}{0}
    \renewcommand{\ypos}{-3*\rowspace}
    \renewcommand{\height}{\heightsingle}
    \renewcommand{\width}{\widthsingle}
    \node[rectangle, fill=lavender, rounded corners, minimum width=\width, minimum height=\height, draw] (\nodenum) at (\xpos,\ypos) {\name};
    
    \renewcommand{\nodenum}{r1c1}
    \renewcommand{\name}{}
	\renewcommand{\xpos}{\xcircle}
    \renewcommand{\ypos}{0}
    \node[circle, scale=\circlescale, fill=evergreen, draw] (\nodenum) at (\xpos,\ypos) {\name};
    
    \renewcommand{\nodenum}{r2c1}
    \renewcommand{\name}{}
	\renewcommand{\xpos}{\xcircle}
    \renewcommand{\ypos}{-\ycircle}
    \node[circle, scale=\circlescale, fill=evergreen, draw] (\nodenum) at (\xpos,\ypos) {\name};
        
    \renewcommand{\nodenum}{r1c2}
    \renewcommand{\name}{}
	\renewcommand{\xpos}{2*\xcircle}
    \renewcommand{\ypos}{0}
    \node[circle, scale=\circlescale, fill=evergreen, draw] (\nodenum) at (\xpos,\ypos) {\name};
    
    \renewcommand{\nodenum}{r2c2}
    \renewcommand{\name}{}
	\renewcommand{\xpos}{2*\xcircle}
    \renewcommand{\ypos}{-\ycircle}
    \node[circle, scale=\circlescale, fill=evergreen, draw] (\nodenum) at (\xpos,\ypos) {\name};
        
    \renewcommand{\nodenum}{r1c3}
    \renewcommand{\name}{}
	\renewcommand{\xpos}{3.5*\xcircle}
    \renewcommand{\ypos}{0}
    \node[circle, scale=\circlescale, fill=evergreen, draw] (\nodenum) at (\xpos,\ypos) {\name};
    
    \renewcommand{\nodenum}{r2c3}
    \renewcommand{\name}{}
	\renewcommand{\xpos}{3.5*\xcircle}
    \renewcommand{\ypos}{-\ycircle}
    \node[circle, scale=\circlescale, fill=evergreen, draw] (\nodenum) at (\xpos,\ypos) {\name};

   \renewcommand{\nodenum}{Ellipses}
    \renewcommand{\name}{$\bm{\cdots}$}
	\renewcommand{\xpos}{2.8*\xcircle}
    \renewcommand{\ypos}{-1.5*\yrel}
    \node[] (\nodenum) at (\xpos,\ypos) {\name};

    \draw [line width=\linethick,color=dullred]  
    (r1c0)--++(-\xcircle,-\ycircle)
    (u+edge)--(r1c1)
    (r2c1)--(r1c2)
    (r2c2)--++(.5*\xcircle,.5*\ycircle)
    (r1c3)--++(-.5*\xcircle,-.5*\ycircle)
    (r2c3)--++(.5*\xcircle,.5*\ycircle);
    
    \draw [line width=\linethick,color=dullblue]  
    (r1c0)--(r1s1);
    
     \draw [line width=\linethick,color=dullblue]  
    (u+)--(O)--(S1)
    (r2c1)--(S2)
    (r2c2)--(S3)
    (r2c3)--(S4);
    
    \draw [line width=\linethick, color=black]  
    (r2c0)--(G)--(u+);
    
    \draw [line width=\linethick, dashed, color=black]  
    (r1c0)--(r2c0)
    (r1c1)--(r2c1)
    (r1c2)--(r2c2)
    (r1c3)--(r2c3);
\end{tikzpicture}.
\end{equation}
This evaluates to
\begin{equation}
	\mathrm{Var}[\partial_k C]=\frac{1}{q}\left(C_5\left(-\frac{1}{Dd}+\xi\Gamma_{n-1}\right)+C_6\eta^{n-1}\right).
\end{equation}
The variance can be upper bounded by:
\begin{equation}
\begin{split}
\mathrm{Var}[\partial_k C]
	&\leq \epsilon(O)\frac{4 \norm{G}_{\infty}^2}{q}\left(1+\frac{Dd\xi}{1-\eta}\right).
\end{split}
\end{equation}

In the case where only $U_+$ forms a 2-design, the variance is
\begin{equation}
\begin{tikzpicture}	
	
     \renewcommand{\nodenum}{u-edge}
	\renewcommand{\xpos}{-.5*\heightsingle}
    \renewcommand{\ypos}{0}
    \coordinate (\nodenum) at (\xpos,\ypos) {};
    
     \renewcommand{\nodenum}{u+edge}
	\renewcommand{\xpos}{.5*\heightsingle}
    \renewcommand{\ypos}{-2*\rowspace}
    \coordinate (\nodenum) at (\xpos,\ypos) {};
    
   
   \renewcommand{\nodenum}{r1s1}
    \renewcommand{\name}{}
	\renewcommand{\xpos}{0}
    \renewcommand{\ypos}{\yleg}
    \renewcommand{\height}{\heightstate}
    \renewcommand{\width}{\widthstate}
    \node[rectangle, fill=lavender, rounded corners=.2em, minimum width=\width, minimum height=\height, draw] (\nodenum) at (\xpos,\ypos) {\name};
   
   \renewcommand{\nodenum}{Var}
    \renewcommand{\name}{$\mathrm{Var}[\partial_k C]=$}
	\renewcommand{\xpos}{-1.5*\xrel}
    \renewcommand{\ypos}{-\rowspace}
    \node[] (\nodenum) at (\xpos,\ypos) {\name};
    
   \renewcommand{\nodenum}{S1}
    \renewcommand{\name}{$S$}
	\renewcommand{\xpos}{0}
    \renewcommand{\ypos}{-4*\rowspace-\yleg}
    \node[] (\nodenum) at (\xpos,\ypos) {\name};
    
   \renewcommand{\nodenum}{S2}
    \renewcommand{\name}{$S$}
	\renewcommand{\xpos}{\xcircle}
    \renewcommand{\ypos}{-\rowspace-\yleg}
    \node[] (\nodenum) at (\xpos,\ypos) {\name};
        
   \renewcommand{\nodenum}{S3}
    \renewcommand{\name}{$S$}
	\renewcommand{\xpos}{2*\xcircle}
    \renewcommand{\ypos}{-\rowspace-\yleg}
    \node[] (\nodenum) at (\xpos,\ypos) {\name};
            
   \renewcommand{\nodenum}{S4}
    \renewcommand{\name}{$S$}
	\renewcommand{\xpos}{3.5*\xcircle}
    \renewcommand{\ypos}{-\rowspace-\yleg}
    \renewcommand{\height}{\heightstate}
    \renewcommand{\width}{\widthstate}
    \node[] (\nodenum) at (\xpos,\ypos) {\name};
    
    \renewcommand{\nodenum}{u-}
    \renewcommand{\name}{$U_{-}$}
	\renewcommand{\xpos}{0}
    \renewcommand{\ypos}{0}
    \renewcommand{\height}{\heightsingle}
    \renewcommand{\width}{\widthsingle}
    \node[rectangle, fill=evergreen, rounded corners, minimum width=\width, minimum height=\height, draw] (\nodenum) at (\xpos,\ypos) {\name};
    
    \renewcommand{\nodenum}{G}
    \renewcommand{\name}{$G$}
	\renewcommand{\xpos}{0}
    \renewcommand{\ypos}{-\rowspace}
    \renewcommand{\height}{\heightsingle}
    \renewcommand{\width}{\widthsingle}
    \node[rectangle, fill=lavender, rounded corners, minimum width=\width, minimum height=\height, draw] (\nodenum) at (\xpos,\ypos) {\name};
   
    \renewcommand{\nodenum}{r1c0}
    \renewcommand{\name}{}
	\renewcommand{\xpos}{0}
    \renewcommand{\ypos}{-2*\rowspace}
    \node[circle, scale=\circlescale, fill=evergreen, draw] (\nodenum) at (\xpos,\ypos) {\name};
    
    \renewcommand{\nodenum}{r2c0}
    \renewcommand{\name}{}
	\renewcommand{\xpos}{0}
    \renewcommand{\ypos}{-2*\rowspace-\ycircle}
    \node[circle, scale=\circlescale, fill=evergreen, draw] (\nodenum) at (\xpos,\ypos) {\name};
    	    
    	\renewcommand{\nodenum}{O}
    \renewcommand{\name}{$O$}
	\renewcommand{\xpos}{0}
    \renewcommand{\ypos}{-4*\rowspace}
    \renewcommand{\height}{\heightsingle}
    \renewcommand{\width}{\widthsingle}
    \node[rectangle, fill=lavender, rounded corners, minimum width=\width, minimum height=\height, draw] (\nodenum) at (\xpos,\ypos) {\name};
    
    \renewcommand{\nodenum}{r1c1}
    \renewcommand{\name}{}
	\renewcommand{\xpos}{\xcircle}
    \renewcommand{\ypos}{0}
    \node[circle, scale=\circlescale, fill=evergreen, draw] (\nodenum) at (\xpos,\ypos) {\name};
    
    \renewcommand{\nodenum}{r2c1}
    \renewcommand{\name}{}
	\renewcommand{\xpos}{\xcircle}
    \renewcommand{\ypos}{-\ycircle}
    \node[circle, scale=\circlescale, fill=evergreen, draw] (\nodenum) at (\xpos,\ypos) {\name};
        
    \renewcommand{\nodenum}{r1c2}
    \renewcommand{\name}{}
	\renewcommand{\xpos}{2*\xcircle}
    \renewcommand{\ypos}{0}
    \node[circle, scale=\circlescale, fill=evergreen, draw] (\nodenum) at (\xpos,\ypos) {\name};
    
    \renewcommand{\nodenum}{r2c2}
    \renewcommand{\name}{}
	\renewcommand{\xpos}{2*\xcircle}
    \renewcommand{\ypos}{-\ycircle}
    \node[circle, scale=\circlescale, fill=evergreen, draw] (\nodenum) at (\xpos,\ypos) {\name};
        
    \renewcommand{\nodenum}{r1c3}
    \renewcommand{\name}{}
	\renewcommand{\xpos}{3.5*\xcircle}
    \renewcommand{\ypos}{0}
    \node[circle, scale=\circlescale, fill=evergreen, draw] (\nodenum) at (\xpos,\ypos) {\name};
    
    \renewcommand{\nodenum}{r2c3}
    \renewcommand{\name}{}
	\renewcommand{\xpos}{3.5*\xcircle}
    \renewcommand{\ypos}{-\ycircle}
    \node[circle, scale=\circlescale, fill=evergreen, draw] (\nodenum) at (\xpos,\ypos) {\name};

   \renewcommand{\nodenum}{Ellipses}
    \renewcommand{\name}{$\bm{\cdots}$}
	\renewcommand{\xpos}{2.8*\xcircle}
    \renewcommand{\ypos}{-1.5*\yrel}
    \node[] (\nodenum) at (\xpos,\ypos) {\name};

    \draw [line width=\linethick,color=dullred]  
    (u-edge)--++(-.5*\xcircle,-.5*\ycircle)
    (r2c0)--(r1c1)
    (r2c1)--(r1c2)
    (r2c2)--++(.5*\xcircle,.5*\ycircle)
    (r1c3)--++(-.5*\xcircle,-.5*\ycircle)
    (r2c3)--++(.5*\xcircle,.5*\ycircle);
    
    \draw [line width=\linethick,color=dullblue]  
    (u-)--(r1s1);
    
     \draw [line width=\linethick,color=dullblue]  
    (r2c0)--(O)--(S1)
    (r2c1)--(S2)
    (r2c2)--(S3)
    (r2c3)--(S4);
    
    \draw [line width=\linethick, color=black]  
    (u-)--(G)--(r1c0);
    
    \draw [line width=\linethick, dashed, color=black]  
    (r1c0)--(r2c0)
    (r1c1)--(r2c1)
    (r1c2)--(r2c2)
    (r1c3)--(r2c3);
\end{tikzpicture}.
\end{equation}
This evaluates to
\begin{equation}
\begin{split}
	\mathrm{Var}[\partial_k C]
	&=\frac{1}{q^2}\Bigg[\epsilon(O)\Bigg(Dd^2C_2-C_3+\left(-\frac{C_2d}{D}+C_3d\right)\left(\xi\Gamma_{n-2}D+\eta^{n-2}D^2\right)\Bigg)\\
	&\hspace{12mm}+\ptr{d}{O}^2(D^2-1)\eta^{n-2}\left(-\frac{C_2}{D}+C_3\right)\Bigg].
\end{split}
\end{equation}
In the large $n$ limit, this simplifies to
\begin{equation}
\begin{split}
	\mathrm{Var}[\partial_k C]
	=&\frac{\epsilon(O)}{q}\Bigg(Dd^2C_2-C_3+\left(-\frac{C_2}{D}+C_3\right)\frac{\xi Dd}{1-\eta}\Bigg).
\end{split}
\end{equation}

In the on-site case where both $U_-$ and $U_+$ form 2-designs, the variance can be written as:

\begin{equation}
\begin{tikzpicture}	
	
     \renewcommand{\nodenum}{u-edge}
	\renewcommand{\xpos}{-.5*\heightsingle}
    \renewcommand{\ypos}{0}
    \coordinate (\nodenum) at (\xpos,\ypos) {};
    
     \renewcommand{\nodenum}{u+edge}
	\renewcommand{\xpos}{.5*\heightsingle}
    \renewcommand{\ypos}{-2*\rowspace}
    \coordinate (\nodenum) at (\xpos,\ypos) {};
    
   
   \renewcommand{\nodenum}{r1s1}
    \renewcommand{\name}{}
	\renewcommand{\xpos}{0}
    \renewcommand{\ypos}{\ycircle+\yleg}
    \renewcommand{\height}{\heightstate}
    \renewcommand{\width}{\widthstate}
    \node[rectangle, fill=lavender, rounded corners=.2em, minimum width=\width, minimum height=\height, draw] (\nodenum) at (\xpos,\ypos) {\name};
   
   \renewcommand{\nodenum}{Var}
    \renewcommand{\name}{$\mathrm{Var}[\partial_k C]=$}
	\renewcommand{\xpos}{-1.5*\xrel}
    \renewcommand{\ypos}{-\rowspace}
    \node[] (\nodenum) at (\xpos,\ypos) {\name};
    
   \renewcommand{\nodenum}{S1}
    \renewcommand{\name}{$S$}
	\renewcommand{\xpos}{0}
    \renewcommand{\ypos}{-4*\rowspace-\yleg}
    \node[] (\nodenum) at (\xpos,\ypos) {\name};
    
   \renewcommand{\nodenum}{S2}
    \renewcommand{\name}{$S$}
	\renewcommand{\xpos}{\xcircle}
    \renewcommand{\ypos}{-\rowspace-\yleg}
    \node[] (\nodenum) at (\xpos,\ypos) {\name};
        
   \renewcommand{\nodenum}{S3}
    \renewcommand{\name}{$S$}
	\renewcommand{\xpos}{2*\xcircle}
    \renewcommand{\ypos}{-\rowspace-\yleg}
    \node[] (\nodenum) at (\xpos,\ypos) {\name};
            
   \renewcommand{\nodenum}{S4}
    \renewcommand{\name}{$S$}
	\renewcommand{\xpos}{3.5*\xcircle}
    \renewcommand{\ypos}{-\rowspace-\yleg}
    \renewcommand{\height}{\heightstate}
    \renewcommand{\width}{\widthstate}
    \node[] (\nodenum) at (\xpos,\ypos) {\name};

    \renewcommand{\nodenum}{r1c0}
    \renewcommand{\name}{}
	\renewcommand{\xpos}{0}
    \renewcommand{\ypos}{\ycircle}
    \node[circle, scale=\circlescale, fill=evergreen, draw] (\nodenum) at (\xpos,\ypos) {\name};
    
    \renewcommand{\nodenum}{r2c0}
    \renewcommand{\name}{}
	\renewcommand{\xpos}{0}
    \renewcommand{\ypos}{0}
    \node[circle, scale=\circlescale, fill=evergreen, draw] (\nodenum) at (\xpos,\ypos) {\name};

    \renewcommand{\nodenum}{G}
    \renewcommand{\name}{$G$}
	\renewcommand{\xpos}{0}
    \renewcommand{\ypos}{-\rowspace}
    \renewcommand{\height}{\heightsingle}
    \renewcommand{\width}{\widthsingle}
    \node[rectangle, fill=lavender, rounded corners, minimum width=\width, minimum height=\height, draw] (\nodenum) at (\xpos,\ypos) {\name};

    \renewcommand{\nodenum}{r3c0}
    \renewcommand{\name}{}
	\renewcommand{\xpos}{0}
    \renewcommand{\ypos}{-2*\rowspace}
    \node[circle, scale=\circlescale, fill=evergreen, draw] (\nodenum) at (\xpos,\ypos) {\name};
    
    \renewcommand{\nodenum}{r4c0}
    \renewcommand{\name}{}
	\renewcommand{\xpos}{0}
    \renewcommand{\ypos}{-2*\rowspace-\ycircle}
    \node[circle, scale=\circlescale, fill=evergreen, draw] (\nodenum) at (\xpos,\ypos) {\name};
    	    
    	\renewcommand{\nodenum}{O}
    \renewcommand{\name}{$O$}
	\renewcommand{\xpos}{0}
    \renewcommand{\ypos}{-4*\rowspace}
    \renewcommand{\height}{\heightsingle}
    \renewcommand{\width}{\widthsingle}
    \node[rectangle, fill=lavender, rounded corners, minimum width=\width, minimum height=\height, draw] (\nodenum) at (\xpos,\ypos) {\name};
    
    \renewcommand{\nodenum}{r1c1}
    \renewcommand{\name}{}
	\renewcommand{\xpos}{\xcircle}
    \renewcommand{\ypos}{0}
    \node[circle, scale=\circlescale, fill=evergreen, draw] (\nodenum) at (\xpos,\ypos) {\name};
    
    \renewcommand{\nodenum}{r2c1}
    \renewcommand{\name}{}
	\renewcommand{\xpos}{\xcircle}
    \renewcommand{\ypos}{-\ycircle}
    \node[circle, scale=\circlescale, fill=evergreen, draw] (\nodenum) at (\xpos,\ypos) {\name};
        
    \renewcommand{\nodenum}{r1c2}
    \renewcommand{\name}{}
	\renewcommand{\xpos}{2*\xcircle}
    \renewcommand{\ypos}{0}
    \node[circle, scale=\circlescale, fill=evergreen, draw] (\nodenum) at (\xpos,\ypos) {\name};
    
    \renewcommand{\nodenum}{r2c2}
    \renewcommand{\name}{}
	\renewcommand{\xpos}{2*\xcircle}
    \renewcommand{\ypos}{-\ycircle}
    \node[circle, scale=\circlescale, fill=evergreen, draw] (\nodenum) at (\xpos,\ypos) {\name};
        
    \renewcommand{\nodenum}{r1c3}
    \renewcommand{\name}{}
	\renewcommand{\xpos}{3.5*\xcircle}
    \renewcommand{\ypos}{0}
    \node[circle, scale=\circlescale, fill=evergreen, draw] (\nodenum) at (\xpos,\ypos) {\name};
    
    \renewcommand{\nodenum}{r2c3}
    \renewcommand{\name}{}
	\renewcommand{\xpos}{3.5*\xcircle}
    \renewcommand{\ypos}{-\ycircle}
    \node[circle, scale=\circlescale, fill=evergreen, draw] (\nodenum) at (\xpos,\ypos) {\name};

   \renewcommand{\nodenum}{Ellipses}
    \renewcommand{\name}{$\bm{\cdots}$}
	\renewcommand{\xpos}{2.8*\xcircle}
    \renewcommand{\ypos}{-1.5*\yrel}
    \node[] (\nodenum) at (\xpos,\ypos) {\name};

    \draw [line width=\linethick,color=dullred]  
    (r1c0)--++(-\xcircle,-\ycircle)
    (r4c0)--(r1c1)
    (r2c1)--(r1c2)
    (r2c2)--++(.5*\xcircle,.5*\ycircle)
    (r1c3)--++(-.5*\xcircle,-.5*\ycircle)
    (r2c3)--++(.5*\xcircle,.5*\ycircle);
    
    \draw [line width=\linethick,color=dullblue]  
    (r1c0)--(r1s1);
    
     \draw [line width=\linethick,color=dullblue]  
    (r4c0)--(O)--(S1)
    (r2c1)--(S2)
    (r2c2)--(S3)
    (r2c3)--(S4);
    
    \draw [line width=\linethick, color=black]  
    (r2c0)--(G)--(r3c0);
    
    \draw [line width=\linethick, dashed, color=black]  
    (r1c0)--(r2c0)
    (r3c0)--(r4c0)
    (r1c1)--(r2c1)
    (r1c2)--(r2c2)
    (r1c3)--(r2c3);
\end{tikzpicture}.
\end{equation}
This evaluates to
\begin{equation}\label{Eq:VarNumerics}
	\mathrm{Var}[\partial_k C]=\frac{C_4}{q^2}\Bigg[\epsilon(O)\left(-\frac{1}{d}+D\xi\Gamma_{n-1}+D^2\eta^{n-1}\right)+\ptr{d}{O}^2\frac{D^2-1}{d}\eta^{n-1}\Bigg].
\end{equation}
In the large $n$-limit,
\begin{equation}
	\mathrm{Var}[\partial_k C]=\epsilon(O)\frac{C_4}{q^2}\left(-\frac{1}{d}+\frac{D\xi}{1-\eta}\right).
\end{equation}

In all six cases, in the large $n$ limit, we have an upper bound on the variance of the following form:
\begin{equation}
	\mathrm{Var}[\partial_k C]\leq \epsilon(O)\mathcal{O}\left(\frac{P(D,d)}{Q(D,d)}\right),
\end{equation}
where $P(D,d)$ and $Q(D,d)$ are polynomial functions of $D$ and $d$ which are independent of $n$.

%% file: ProofTheorem2.tex
\section{Proof of Theorem~\ref{Thm:XEB}}
\label{Appendix:Theorem2}
We first compute $\epsilon(O_\chi)$, noting that $\tr{O_\chi}=0$:
\begin{equation}
\begin{split}
	\epsilon(O_\chi)
	&=\norm{O_\chi-\tr{O_\chi}\frac{I}{2}}_{\mathrm{HS}}^2\\
	&=\norm{O_\chi}_{\mathrm{HS}}^2\\
	&=\tr{O_\chi^2}\\
	&=\tr{\left(\sum_x \left(2p(V,x)-1\right)\ket{x}\bra{x}\right)^2}\\
	&=\sum_{x,x'} (2p(V,x)-1)(2p(V,x')-1)\tr{\ket{x}\langle x\ket{x'}\bra{x'}}\\
	&=\sum_{x,x'} (2p(V,x)-1)(2p(V,x')-1)\delta_{x,x'}\\
	&=\sum_{x} (2p(V,x)-1)^2\\
	&=\sum_{x} (4p(V,x)^2-4p(V,x)+1)\\
	&=4\sum_{x} p(V,x)^2-4+2\\
	&=4\sum_{x} p(V,x)^2-2.
\end{split}
\end{equation}
We compute its average:
\begin{equation}
\begin{split}
	\int_{\mathrm{Haar}} dV\epsilon(O_\chi)
	&=\int_{\mathrm{Haar}} dV\left(4\sum_{x} p(V,x)^2-2\right)\\
	&=4\sum_{x} \int_{\mathrm{Haar}} dV \tr{V \ket{\psi_0}\bra{\psi_0}V^\dagger (\ket{x}\bra{x}\otimes I^{\otimes n-1})}^2-2\\
	&=4\sum_{x} \int_{\mathrm{Haar}} dV \tr{V^{\otimes 2} \ket{\psi_0}\bra{\psi_0}^{\otimes 2}V^{\dagger\otimes 2} (\ket{x}\bra{x}\otimes I^{\otimes n-1})^{\otimes 2}}-2\\
	&=4\sum_{x} \frac{1}{2^{n}(2^{n}+1)}\tr{ (\mathbb{I}+\mathbb{S})(\ket{x}\bra{x}\otimes I^{\otimes n-1})^{\otimes 2}}-2\\
	&=4\sum_{x} \frac{1}{2^{n}(2^{n}+1)}\left(\tr{ \ket{x}\bra{x}\otimes I^{\otimes n-1}}^2+\tr{ \ket{x}\bra{x}\otimes I^{\otimes n-1}}\right)-2\\
	&=4\sum_{x} \frac{1}{2^{n}(2^{n}+1)}\left(\left(2^{n-1}\right)^2+2^{n-1}\right)-2\\
	&=8\left(\frac{2^{n-1}\left(2^{n-1}+1\right)}{2^n(2^{n}+1)}\right)-2\\
	&=2\left(\frac{2^{n}\left(2^{n}+2\right)}{2^n(2^{n}+1)}-1\right)\\
	&=2\left(\frac{\left(2^{n}+2\right)}{2^{n}+1}-\frac{2^n+1}{2^n+1}\right)\\
	&=\frac{2}{2^{n}+1}.
\end{split}
\end{equation}
From the above calculations, one can verify that ${\int_{\mathrm{Haar}}dV \norm{O_{\chi}}_{\infty}^2\leq \int_{\mathrm{Haar}}dV \norm{O_{\chi}}_{\mathrm{HS}}^2=\frac{2}{2^n+1}}$. Also, ${\int_{\mathrm{Haar}}dV  \tr{O_{\chi}}^2=0}$. Therefore,  $\int_{\mathrm{Haar}}dV \norm{O_{\chi}}_{\infty}^2$ and $\int_{\mathrm{Haar}}dV  \tr{O_{\chi}}^2$ do not grow exponentially in $n$. Hence, $\mathrm{Var}_{\bm{\theta},V}[\partial_k \chi]$ vanishes at least exponentially in $n$ by Lemma~\ref{Lemma:Scrambling}, inducing a barren plateau.

%% file: ProofObservation1.tex
\section{Condition for Observation~\ref{Observation}}
\label{Appendix:Observation}
We show that ${\int_{\mathrm{Haar}}dV\tr{O_E}^2}$ and $\int_{\mathrm{Haar}}dV\norm{O_E}_{\infty}^2$ grow slower than exponentially in $n$ if $\int_{\mathrm{Haar}}dV\tr{O_E^2}$ grows slower than exponentially in $n$.
This implies that $O_E$ meets the conditions of Lemma~\ref{Lemma:Scrambling}. First, we can show that ${\norm{O_E}_{\infty}^2\leq \norm{O_E}_{\mathrm{HS}}^2=\tr{O_E^2}}$. Therefore, $\int_{\mathrm{Haar}}dV\norm{O_E}_{\infty}^2\leq \int_{\mathrm{Haar}}dV\tr{O_E^2}$. 

Defining $\langle \cdot \rangle_{\mathrm{max}} =\tr{\frac{I}{2}\cdot}$, we derive the following inequality:
\begin{equation}
\begin{split}
\tr{O_E}^2 
	&=\tr{2\frac{I}{2}O_E}^2 \\
	&=4\langle O_E\rangle_{\mathrm{max}}^2 \\
	&\leq 4\langle O_E^2\rangle_{\mathrm{max}}\\
	&= 2\tr{O_E^2}.
\end{split}
\end{equation}
Therefore, $\int_{\mathrm{Haar}}dV \tr{O_E}^2 \leq 2\int_{\mathrm{Haar}}dV \tr{O_E^2} $.

\begin{figure}[t!]
    \centering
   \includegraphics[scale=.75]{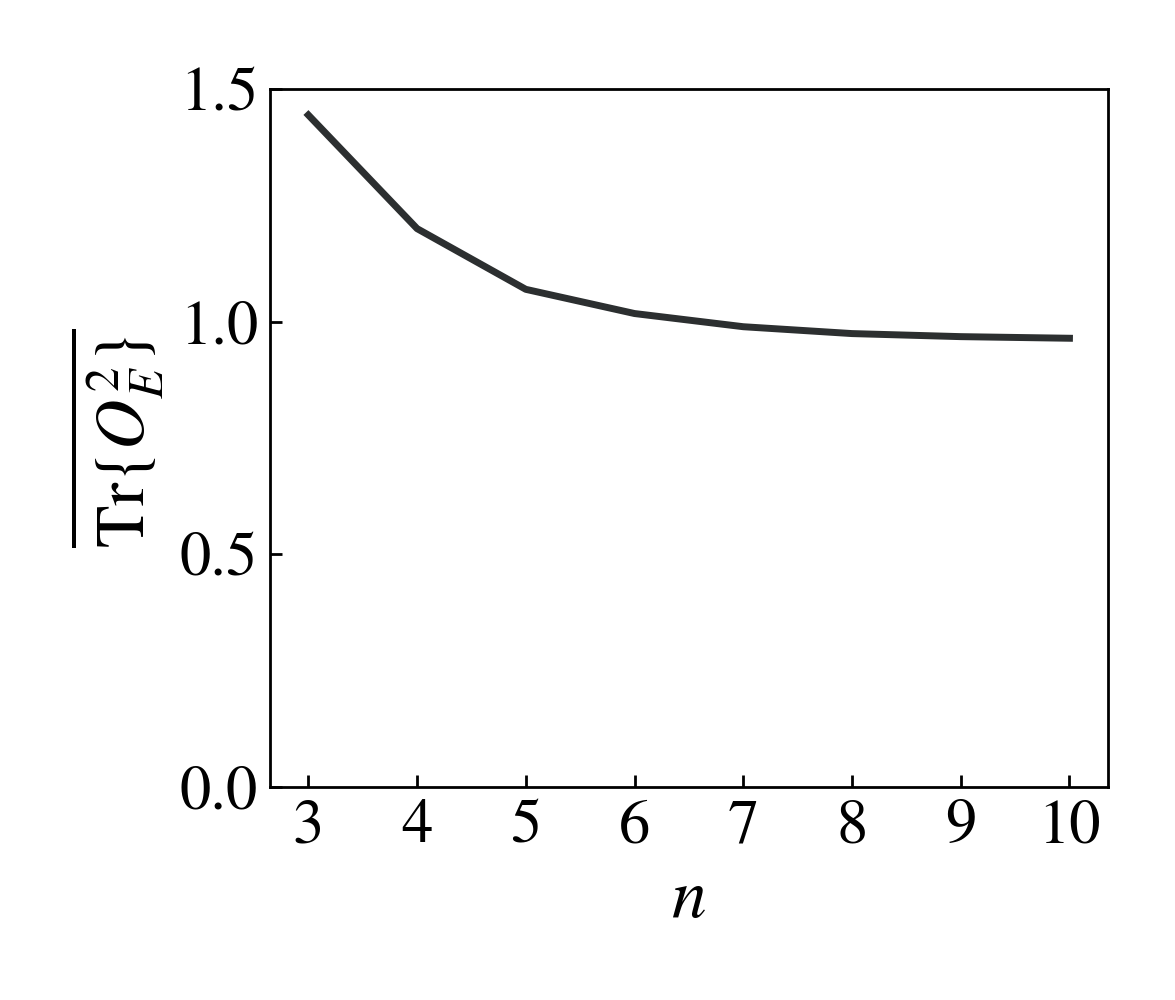}
    \caption{Plot of $\overline{\tr{O_E^2}}=\int_{\mathrm{Haar}}dV \tr{O_E^2}$ against the qubit number, $n$. The quantity $\overline{\tr{O_E^2}}$ is empirically computed by averaging $\tr{O_E^2}$ over 2000 unitaries randomly sampled from the Haar measure on the unitary group. The plot indicates that $\overline{\tr{O_E^2}}$ does not increase exponentially in $n$.}
    \label{Fig:Condition}
\end{figure}

We now give an explicit expression for $\tr{O_E^2}$, which is useful for numerical computations:
\begin{equation}
\begin{split}
\tr{O_E^2}
	&=\tr{\left(-\sum_{x}\mathrm{ln}[p(V,x)]\ket{x}\bra{x}\right)^2}\\
	&=\tr{\sum_{x,x'}\mathrm{ln}[p(V,x)]\mathrm{ln}[p(V,x')]\ket{x}\bra{x}x'\rangle \bra{x'}}\\
	&=\sum_{x,x'}\mathrm{ln}[p(V,x)]\mathrm{ln}[p(V,x')]\delta_{x,x'}\\
	&=\sum_{x}\mathrm{ln}[p(V,x)]^2.
\end{split}
\end{equation}
Figure~\ref{Fig:Condition} shows that $\overline{\tr{O_E^2}}=\int_{\mathrm{Haar}}dV \tr{O_E^2}$ decreases with $n$. Hence,  ${\int_{\mathrm{Haar}}dV\tr{O_E}^2}$ and $\int_{\mathrm{Haar}}dV\norm{O_E}_{\infty}^2$ do not grow exponentially in $n$. Lemma~\ref{Lemma:Scrambling} can therefore be applied by using $E$ as the cost function.

%% file: ProofProposition1.tex
\section{Proof of Proposition~\ref{Prop:Circuit}}
\label{Appendix:Prop}
Define the cost function
\begin{equation}
	C_c=\tr{U\ket{\psi_0}\bra{\psi_0}U^\dagger O_A},
\end{equation}
where $O_A=O\otimes I^{A'}$ is a local operator which acts $O$ on the system $A$ and the identity on $A'$, the complement system. $U$ is a parameterized unitary and $\ket{\psi_0}$ is an $n$-qubit initial state. We assume that $U$ has the general form $U=\prod_{i=1}^{L}U_{S_i}$ where $U_{S_i}=(e^{-i \theta_i^{\mathrm{polyn}(n)} V_i^{\mathrm{poly}(n)}}\cdots e^{-i \theta_i^1 V_i^1})\otimes I^{S'_i}$ acts non-trivially on system $S_i$ and acts the identity on its complement $S_i'$. We assume that each $\theta_i$ is random such that $U_{S_i}$ forms a 2-design on system $S_i$. This model is general, as many parameterized circuits have this structure. Assume that the derivative is taken with respect to a parameter $\theta_k$. We can write $U=U_-U_+$, where $U_+$  is the collection of unitaries to the right of and including $e^{-i\theta_k V_k}$, the unitary on which the derivative acts. Unitary $U_-$ is the collection of unitaries to the left of $e^{-i\theta_k V_k}$. Assume that the derivative is not taken on the layer $U_{S_O}$. The derivative of $C_c$ with respect to the parameter $\theta_k$ is
\begin{equation}\label{Eq:PartialCc}
\begin{split}
	\partial_k C_c
	&=\tr{\partial_k U\ket{\psi_0}\bra{\psi_0}U^\dagger O_A+U\ket{\psi_0}\bra{\psi_0}\partial_kU^\dagger O_A}\\
	&=\mathrm{Tr}\Big\{(U_-(-iV_k\otimes I^{S^{'}_k})U_+)\ket{\psi_0}\bra{\psi_0}U_+^\dagger U_-^\dagger O_A\\ &\hspace{10mm}+U_-U_+\ket{\psi_0}\bra{\psi_0}(U_+^\dagger (iV_k\otimes I^{S_k^{'}}) U_- ^\dagger )O_A\Big\}\\
	&=\tr{U_-\left[U_+\ket{\psi_0}\bra{\psi_0}U_+^\dagger,(iV_k\otimes I^{S^{'}_k})\right]\ U_-^\dagger O_A}\\
	&=\tr{U_-^{'}\left[U_+\ket{\psi_0}\bra{\psi_0}U_+^\dagger,(iV_k\otimes I^{S^{'}_k})\right]\ U_-^{'\dagger} U_{S_O}^\dagger O_AU_{S_O}}\\
	&=\tr{QU_{S_O}^\dagger O_AU_{S_O}}.
\end{split}
\end{equation}
In the above, we define $Q=U_-^{'}\left[U_+\ket{\psi_0}\bra{\psi_0}U_+^\dagger,(iV_k\otimes I^{S^{'}_k})\right]\ U_-^{'\dagger}$.  We also make the assumption that there exists a unitary $U_{S_{O}}\in \{U_{S_i}\}_i$ such that $A\subseteq S_{O}$ and $U_-=U_{S_O}U_{-}^{'}$, where $U_-^{'}$ contains the remaining unitaries in $U_-$. In other words, $U_{S_O}$ acts on $O$ entirely. This assumption is compatible with, for example, the alternating layer ansatz and the QCNN. 

We now compute the average over all unitaries:
\begin{equation}
\begin{split}
	\langle \partial_k C_c\rangle_{U}
	&=\int dU_+ dU_-^{'}  dU_{S_O} \tr{QU_{S_O}^\dagger O_AU_{S_O}}\\
	&=\int dU_+ dU_-^{'} \frac{1}{d_{S_O}} \tr{Q}\ptr{S_O}{O\otimes I^{S_{O}\backslash A}}\\
	&=0.
\end{split}
\end{equation}
In the above, we use that $U_{S_O}$ forms a 2-design and that the trace of a commutator is zero.

We now compute the variance:
\begin{equation}\label{Eq:VarSimple}
\begin{split}
\mathrm{Var}_{U}[\partial_k C_c]
	&=\langle (\partial_k C_c)^2\rangle_{U}\\
	&=\int dU_+ dU_-^{'}  dU_{S_O} \tr{QU_{S_O}^\dagger O_AU_{S_O}}^2\\
	&=\int dU_+ dU_-^{'}  dU_{S_O} \tr{Q^{\otimes 2}U_{S_O}^{\dagger\otimes  2} O_A^{\otimes 2}U_{S_O}^{\otimes 2}}\\
	&=\frac{1}{d_{S_O}^2-1}\int dU_+ dU_-^{'}   \Big[\tr{Q^{\otimes 2}(\mathbb{I}_{S_O'} \otimes \mathbb{I}_{S_O})} \ptr{S_O}{\mathbb{I}_{S_O}(O\otimes I^{S_O\backslash A})^{\otimes 2}}\\
&\hspace{40mm} 	+\tr{Q^{\otimes 2}(\mathbb{I}_{S_O'} \otimes \mathbb{S}_{S_O}) }\ptr{S_O}{\mathbb{S}_{S_O}(O\otimes I^{S_O\backslash A})^{\otimes 2}}\\
&\hspace{40mm} -\frac{1}{d_{S_O}}\tr{Q^{\otimes 2}(\mathbb{I}_{S_O'} \otimes \mathbb{I}_{S_O} )}\ptr{S_O}{\mathbb{S}_{S_O}(O\otimes I^{S_O\backslash A})^{\otimes 2}}\\
&\hspace{40mm} -\frac{1}{d_{S_O}} \tr{Q^{\otimes 2}(\mathbb{I}_{S_O'} \otimes \mathbb{S}_{S_O}) }\ptr{S_O}{\mathbb{I}_{S_O}(O\otimes I^{S_O\backslash A})^{\otimes 2}}\Big]\\
	&=\frac{1}{d_{S_O}^2-1}\int dU_+ dU_-^{'}   
	\Big[\tr{Q}^2\ptr{S_O}{(O\otimes I^{S_O\backslash A})}^2\\
&\hspace{40mm} + \ptr{S_O}{\ptr{S_O'}{Q}^2}\ptr{S_O}{(O\otimes I^{S_O\backslash A})^2}\\
&\hspace{40mm} -\frac{1}{d_{S_O}}\tr{Q}^2\ptr{S_O}{(O\otimes I^{S_O\backslash A})^2}\\
&\hspace{40mm} -\frac{1}{d_{S_O}} \ptr{S_O}{\ptr{S_O'}{Q}^2}\ptr{S_O}{(O\otimes I^{S_O\backslash A})}^2
	\Big]\\
&=\frac{1}{d_{S_O}^2-1}\int dU_+ dU_-^{'}   
\ptr{S_O}{\ptr{S_O'}{Q}^2}
\Big[\ptr{S_O}{(O\otimes I^{S_O\backslash A})^2}\\
 &\hspace{73mm}-\frac{1}{d_{S_O}} \ptr{S_O}{(O\otimes I^{S_O\backslash A})}^2\Big]\\
	&=\frac{1}{d_{S_O}^2-1}\int dU_+ dU_-^{'}   
	\ptr{S_O}{\ptr{S_O'}{Q}^2}
	\Bigg[\ptr{S_O}{O^2\otimes I^{S_O\backslash A}}\\
 &\hspace{73mm}-\frac{1}{d_{S_O}} \left(\ptr{A}{O}\frac{d_{S_O}}{d_A}\right)^2\Bigg]\\
 	&=\frac{1}{d_{S_O}^2-1}\int dU_+ dU_-^{'}   
	\ptr{S_O}{\ptr{S_O'}{Q}^2}
	\left[\ptr{A}{O^2}\left(\frac{d_{S_O}}{d_A}\right)
 -\frac{d_{S_{O}}}{d_A^2}\ptr{A}{O}^2\right]\\
 	&=\frac{d_{S_O}}{d_A(d_{S_O}^2-1)}\int dU_+ dU_-^{'}   
	\ptr{S_O}{\ptr{S_O'}{Q}^2}
	\left[\ptr{A}{O^2}
 -\frac{1}{d_A}\ptr{A}{O}^2\right]\\
 	&=\frac{\epsilon(O)d_{S_O}}{d_A(d_{S_O}^2-1)}
	\left\langle \ptr{S_O}{\ptr{S_O'}{Q}^2}\right \rangle_{U_+,U_-'}\\
 	&=\epsilon(O)F_1.
\end{split}
\end{equation}
Line four follows from the Weingarten calculus. We define 
\begin{equation}
	\epsilon(O)=\norm{O-\frac{I^A}{d_A}\tr{O}}_{\mathrm{HS}}^2=\ptr{A}{O^2}
 -\frac{1}{d_A}\ptr{A}{O}^2
\end{equation}
and
\begin{equation}
	F_1=\frac{d_{S_O}}{d_A(d_{S_O}^2-1)}
	\left\langle \ptr{S_O}{\ptr{S_O'}{Q}^2}\right \rangle_{U_+,U_-'}.
\end{equation}
We adopt the notation: $\langle \cdot \rangle_{U_+,U_-'}=\int dU_+ dU_-'(\cdot)$. In an abuse of notation, we let $\mathrm{Tr}_{S_O}\{\cdot\}$ denote the trace over system $S_O$ \textit{and} its copy. We also let $\mathbb{I}_{S_O}$ and $\mathbb{S}_{S_O}$ denote the identity and swap permutations between $S_O$ and its copy. We let $\mathbb{I}_{S_O'}$ denote the identity permutation on the complement of $S_O$ and its copy.

Now take the case where the derivative is taken with respect to a parameter $\theta_k$ within layer $U_{S_O}$. The unitary $U$ can be written as $U=U_{S_O-}U_{S_O+}U_+$, where $U_{S_O+}U_+$ is the collection of unitaries to the right of and including $e^{-i\theta_k V_k}$. Unitary $U_-$ is the collection of unitaries to the left of $e^{-i\theta_k V_k}$. More explicitly, $U_{S_O}=U_{S_O-}U_{S_O+}$. The partial derivative of $C_c$ is:

\begin{equation}
\begin{split}
	\partial_k C_c
	&=\tr{\partial_k U\ket{\psi_0}\bra{\psi_0}U^\dagger O_A+U\ket{\psi_0}\bra{\psi_0}\partial_kU^\dagger O_A}\\
	&=\mathrm{Tr}\{(U_{S_O-}(-iV_k\otimes I^{S^{'}_k})U_{S_O+}U_+)\ket{\psi_0}\bra{\psi_0}U_+^\dagger U_{S_O+}^\dagger U_{S_O-}^\dagger O_A\\
	&\hspace{5mm}+U_{S_O-}U_{S_O+}U_{+}\ket{\psi_0}\bra{\psi_0}(U_{+}^\dagger U_{S_O+}^\dagger  (iV_k\otimes I^{S_k^{'}}) U_{S_O-}^\dagger )O_A\}\\
	&=\tr{U_{S_O-}\left[U_{S_O+}U_+\ket{\psi_0}\bra{\psi_0}U_+^\dagger U_{S_O+}^\dagger ,(iV_k\otimes I^{S^{'}_k})\right]\ U_{S_O-}^\dagger O_A}\\
	&=\tr{\left[U_{S_O+}U_+\ket{\psi_0}\bra{\psi_0}U_+^\dagger U_{S_O+}^\dagger ,(iV_k\otimes I^{S^{'}_k})\right]\ U_{S_O-}^\dagger O_AU_{S_O-}}\\
	&=\tr{Q_2\ U_{S_O-}^\dagger O_AU_{S_O-}}.
\end{split}
\end{equation}
We define $Q_2=\left[U_{S_O+}U_+\ket{\psi_0}\bra{\psi_0}U_+^\dagger U_{S_O+}^\dagger ,(iV_k\otimes I^{S^{'}_k})\right]$. This has the same form as Eq.~\eqref{Eq:PartialCc}. Take the case where $U_{S_O-}$ forms a 2-design. Then
\begin{equation}
\begin{split}
\langle \partial_k C_c \rangle_U
	&=\int dU_+ dU_{S_O+}dU_{S_O-}\tr{Q_2\ U_{S_O-}^\dagger O_AU_{S_O-}}\\
	&=\int dU_+ dU_{S_O+} \frac{1}{d_{S_O}} \tr{Q_2}\ptr{S_O}{O\otimes I^{S_{O}\backslash A}}\\
	&=0.
\end{split}
\end{equation}
We use that $Q_2$ is traceless. Similar to Eq.~\eqref{Eq:VarSimple}, the variance is
\begin{equation}
	\begin{split}
		\mathrm{Var}_U[\partial_k C_c]=\epsilon(O)F_2,
	\end{split}
\end{equation}
where
\begin{equation}
	F_2=\frac{d_{S_O}}{d_A(d_{S_O}^2-1)}
	\left\langle \ptr{S_O}{\ptr{S_O'}{Q_2}^2}\right \rangle_{U_{S_O+},U_+}.
\end{equation}

Now take the case where $U_{S_O+}$ forms a 2-design, but $U_{S_O-}$ does not. First rewrite $\partial_k C_c$:

\begin{equation}
\begin{split}
	\partial_k C_c
	&=\tr{\partial_k U\ket{\psi_0}\bra{\psi_0}U^\dagger O_A+U\ket{\psi_0}\bra{\psi_0}\partial_kU^\dagger O_A}\\
	&=\mathrm{Tr}\{(U_{S_O-}(-iV_k\otimes I^{S^{'}_k})U_{S_O+}U_+)\ket{\psi_0}\bra{\psi_0}U_+^\dagger U_{S_O+}^\dagger U_{S_O-}^\dagger O_A\\
	&\hspace{5mm}+U_{S_O-}U_{S_O+}U_{+}\ket{\psi_0}\bra{\psi_0}(U_{+}^\dagger U_{S_O+}^\dagger  (iV_k\otimes I^{S_k^{'}}) U_{S_O-}^\dagger )O_A\}\\
	&=\mathrm{Tr}\{U_+\ket{\psi_0}\bra{\psi_0}U_+^\dagger U_{S_O+}^\dagger U_{S_O-}^\dagger O_AU_{S_O-}(-iV_k\otimes I^{S^{'}_k})U_{S_O+}\\
	&\hspace{5mm}+U_{+}\ket{\psi_0}\bra{\psi_0}U_{+}^\dagger U_{S_O+}^\dagger  (iV_k\otimes I^{S_k^{'}}) U_{S_O-}^\dagger O_AU_{S_O-}U_{S_O+}\}\\
	&=\tr{U_+\ket{\psi_0}\bra{\psi_0}U_+^\dagger U_{S_O+}^\dagger [(iV_k\otimes I^{S^{'}_k}),U_{S_O-}^\dagger O_AU_{S_O-}]U_{S_O+}}\\
	&=\tr{\tilde{\rho} U_{S_O+}^\dagger B U_{S_O+}},
\end{split}
\end{equation}
where we define $\tilde{\rho}=U_+\ket{\psi_0}\bra{\psi_0}U_+^\dagger$ and $B=[(iV_k\otimes I^{S^{'}_k}),U_{S_O-}^\dagger O_AU_{S_O-}]$. The average of $\partial_k C_c$ is
\begin{equation}
\begin{split}
\langle \partial_k C_c \rangle_{U}
	&=\int dU_{+}dU_{S_O+}dU_{S_O-}\tr{\tilde{\rho} U_{S_O+}^\dagger B U_{S_O+}}\\
	&=\int dU_{+}dU_{S_O-}\frac{1}{d_{S_O}}\tr{\tilde{\rho}}\ptr{S_O}{[iV_k,U_{S_O-}^\dagger O_AU_{S_O-}]}\\
	&=0.
\end{split}
\end{equation}
The variance of $\partial_k C_c$ is
\begin{equation}
\begin{split}
\mathrm{Var}_{U}&[\partial_k C_c]\\
&=\langle (\partial_k C_c)^2\rangle_{U}\\
&=\int dU_+ dU_{S_O+}  dU_{S_O-} \tr{\tilde{\rho}U_{S_O+}^\dagger BU_{S_O+}}^2\\
&=\int dU_+ dU_{S_O+}  dU_{S_O-}  \tr{{\tilde{\rho}}^{\otimes 2}U_{S_O+}^{\dagger\otimes  2} B^{\otimes 2}U_{S_O+}^{\otimes 2}}\\
&=\int \frac{dU_+ dU_{S_O-}}{d_{S_O}^2-1} \Big[\tr{{\tilde{\rho}}^{\otimes 2}(\mathbb{I}_{S_O'} \otimes \mathbb{I}_{S_O})} \ptr{S_O}{\mathbb{I}_{S_O}([iV_k,U_{S_O-}^\dagger O_AU_{S_O-}])^{\otimes 2}}\\
    &\hspace{30mm} 	+\tr{\tilde{\rho}^{\otimes 2}(\mathbb{I}_{S_O'} \otimes \mathbb{S}_{S_O}) }\ptr{S_O}{\mathbb{S}_{S_O}([iV_k,U_{S_O-}^\dagger O_AU_{S_O-}])^{\otimes 2}}\\
    &\hspace{30mm} -\frac{1}{d_{S_O}}\tr{\tilde{\rho}^{\otimes 2}(\mathbb{I}_{S_O'} \otimes \mathbb{I}_{S_O} )}\ptr{S_O}{\mathbb{S}_{S_O}([iV_k,U_{S_O-}^\dagger O_AU_{S_O-}])^{\otimes 2}}\\
    &\hspace{30mm} -\frac{1}{d_{S_O}} \tr{\tilde{\rho}^{\otimes 2}(\mathbb{I}_{S_O'} \otimes \mathbb{S}_{S_O}) }\ptr{S_O}{\mathbb{I}_{S_O}([iV_k,U_{S_O-}^\dagger O_AU_{S_O-}])^{\otimes 2}}\Big]\\
&=\int \frac{dU_+ dU_{S_O-}}{d_{S_O}^2-1}  
	\Big[\tr{\tilde{\rho}}^2\ptr{S_O}{[iV_k,U_{S_O-}^\dagger O_AU_{S_O-}]}^2\\
    &\hspace{30mm} +\ptr{S_O}{\ptr{S_O'}{\tilde{\rho}}^2}\ptr{S_O}{([iV_k,U_{S_O-}^\dagger O_AU_{S_O-}])^2}\\
    &\hspace{30mm} -\frac{1}{d_{S_O}}\tr{\tilde{\rho}}^2\ptr{S_O}{([iV_k,U_{S_O-}^\dagger O_AU_{S_O-}])^2}\\
    &\hspace{30mm} -\frac{1}{d_{S_O}} \ptr{S_O}{\ptr{S_O'}{\tilde{\rho}}^2}\ptr{S_O}{[iV_k,U_{S_O-}^\dagger O_AU_{S_O-}]}^2\Big]\\
&=\int \frac{dU_+ dU_{S_O-}}{d_{S_O}^2-1}  
\left(\ptr{S_O}{\ptr{S_O'}{\tilde{\rho}}^2}-\frac{1}{d_{S_O}}\tr{\tilde{\rho}}^2\right)\ptr{S_O}{([iV_k,U_{S_O-}^\dagger O_AU_{S_O-}])^2}\\
    &\leq \int \frac{dU_+ dU_{S_O-}}{d_{S_O}^2-1} \abs{\ptr{S_O}{\ptr{S_O'}{\tilde{\rho}}^2}-\frac{1}{d_{S_O}}}\abs{\ptr{S_O}{([iV_k,U_{S_O-}^\dagger O_AU_{S_O-}])^2}}\\
    &\leq \epsilon(O)\frac{4d_{S_O}\norm{V_k}_{\infty}^2}{d_{S_O}^2-1}\int dU_+ dU_{S_O-}  
\abs{\ptr{S_O}{\ptr{S_O'}{\tilde{\rho}}^2}-\frac{1}{d_{S_O}}}\\
	&=\epsilon(O)F_3,
\end{split}
\end{equation}
where we define
\begin{equation}
	F_3=\frac{4d_{S_O}\norm{V_k}_{\infty}^2}{d_{S_O}^2-1} \int dU_+ dU_{S_O-}  
\abs{\ptr{S_O}{\ptr{S_O'}{\tilde{\rho}}^2}-\frac{1}{d_{S_O}}}.
\end{equation}
In line four, we let $U_{S_O-}$ denote the operator defined on only the $S_O$ system. In the last inequality, we use:
\begin{equation}
\begin{split}
\Big|\mathrm{Tr}_{S_O}\Big\{&([iV_k,U_{S_O-}^\dagger O_AU_{S_O-}])^2\Big\}\Big|\\
&=\abs{\ptr{S_O}{\left(\left[iV_k,U_{S_O-}^\dagger O_AU_{S_O-}-\frac{I^{S_O}}{d_{A}}\tr{O}\right]\right)^2}}\\
&=\abs{\ptr{S_O}{\left(\left[iV_k,\tilde{O}_A\right]\right)^2}}\\
&=\abs{\ptr{S_O}{\left(iV_k\tilde{O}_A-\tilde{O}_AiV_k\right)^2}}\\
&=2\abs{\ptr{S_O}{iV_k\tilde{O}_AiV_k\tilde{O}_A-\tilde{O}_A(iV_k)^2\tilde{O}_A}}\\
&\leq 2\left[\abs{\ptr{S_O}{iV_k\tilde{O}_AiV_k\tilde{O}_A}}+\abs{\ptr{S_O}{\tilde{O}_A(iV_k)^2\tilde{O}_A}}\right]\\
&=2d_{S_O}\left[\abs{\ptr{S_O}{\frac{I^{S_O}}{d_{S_O}}V_k\tilde{O}_AV_k\tilde{O}_A}}+\abs{\ptr{S_O}{\frac{I^{S_O}}{d_{S_O}}\tilde{O}_A(V_k)^2\tilde{O}_A}}\right]\\
&\leq 2d_{S_O}\left[\norm{V_k\tilde{O}_AV_k\tilde{O}_A}_{\infty}+\norm{\tilde{O}_A(V_k)^2\tilde{O}_A}_{\infty}\right]\\
&\leq 4d_{S_O}\norm{V_k}_{\infty}^2\norm{\tilde{O}_A}_{\infty}^2\\
&=4d_{S_O}\norm{V_k}_{\infty}^2\norm{U^\dagger_{S_O-}(O\otimes I^{S_O\backslash A})U_{S_O-}-\frac{I^{S_O}}{d_A}\tr{O}}_{\infty}^2\\
&=4d_{S_O}\norm{V_k}_{\infty}^2\norm{U^\dagger_{S_O-}\left(\left(O-\frac{I^{A}}{d_A}\tr{O}\right)\otimes  I^{S_O\backslash A}\right)U_{S_O-}}_{\infty}^2\\
&\leq 4d_{S_O}\norm{V_k}_{\infty}^2\norm{O-\frac{I^{A}}{d_A}\tr{O}}_{\mathrm{HS}}^2\\
&=4d_{S_O}\norm{V_k}_{\infty}^2\epsilon(O).
\end{split}
\end{equation}
In line two, we define $\tilde{O}_A=U_{S_O-}^\dagger O_AU_{S_O-}-\frac{I^{S_O}}{d_{A}}\tr{O}$.

In all three cases, $\langle \partial_k C_c\rangle_{U}=0$ and $\mathrm{Var}_U[\partial_k C_c]=\epsilon(O)F$, where $F\in\{F_1,F_2,F_3\}$.

%% file: Identities.tex
\section{Identities}\label{Sec:Identities}
We present identities which are useful in proving Theorem~\ref{Theorem}. These identities are derived simply by evaluating the diagrams. The first set of identities is:
\begin{align*}

\end{align}
We also have
\begin{align*}
	%
\end{align}
From Eq.~\eqref{Eq:Tree}, we can derive
\begin{align*}
%
\end{align}
This, combined with Eq.~\eqref{Eq:OTree}, produces the identities:

\begin{align}\label{Eq:ManyOTrees1}
%
\end{align}
We now provide identities which will help us evaluate the variance diagrams when $U_-$, $U_+$, or both form 2-designs. In the off-site case, when only $U_-$ forms a 2-design, the following identities follow:
\begin{align}\label{Eq:OffU-}
%
\end{align}
where
\begin{equation}
	C_1=2\int dU_+ \Big[-\ptr{d}{\ptr{D}{U_+^\dagger GU_+}^2}+D\tr{G^2}\Big].
\end{equation}
When only $U_+$ forms a 2-design in the off-site case, we obtain the useful identities: 

\begin{align*}
%
\end{align}
where
\begin{equation}
\begin{split}
	C_2&=2\int dU_-  \left[-\tr{\rho G\rho G}+\tr{ G^2 \rho^2}\right],\\
	C_3&=2\int dU_-  \left[-\tr{\rho G}^2+D\tr{\rho G^2 }\right].\\
\end{split}
\end{equation}
We define $\rho=U_-^\dagger (I_D\otimes \ket{0}\bra{0})U_-$ as an unnormalized state with a trace of $\tr{\rho}=D$. In the off-site case, when both $U_-$ and $U_+$ form 2-designs, we obtain the following useful identities,
\begin{align}\label{Eq:OffBoth}
%
\end{align}
where
\begin{equation}
	C_4=2\left[-\tr{G}^2+Dd\tr{G^2}\right].
\end{equation}
Now consider  the on-site case, where $O$ acts on the derivative site. Then when only $U_-$ forms a 2-design, we have the following identities:
\begin{align}\label{Eq:OnU-}
%
\end{align}
where 
\begin{equation}
\begin{split}
	C_5&=2\int dU_+  \tr{\sigma G[G,\sigma]},\\
	C_6&=2\int dU_+ \Big[-\ptr{d}{\left(\ptr{D}{U_+^\dagger GU_+}O\right)^2}+D\ptr{d}{\ptr{D}{U_+^\dagger G^2U_+}O^2}\Big].
\end{split}
\end{equation}
In the above, we define $\sigma=U_+ (I_D\otimes O)U_+^\dagger$. 

In the on-site case, when only $U_+$ forms a 2-design, we have the following identities: 

\begin{align}\label{Eq:OnU+1}
%
\end{align}

In the on-site case, when both $U_-$ and $U_+$ form 2-designs, we have the following identities:
\begin{align*}
%
\end{align}

%% file: ExtendedProofTheorem1.tex
\section{Extended proof of Theorem~\ref{Theorem}}\label{Appendix:Extended}

\subsection{Off-site case}
Take the off-site case where only $U_-$ forms a 2-design. We use periodic boundary conditions for the open red legs. Using the identities from Eq.~\eqref{Eq:OffU-} and Eqs.~\eqref{Eq:ManyOTrees1} to \eqref{Eq:ManyOTrees4}, we compute the variance: 

\begin{equation}
%
.
\end{equation}

Upon evaluating these diagrams, we get the following expression for the variance
\begin{equation}
\begin{split}
	\mathrm{Var}[\partial_k C]
	=&\left(-\frac{C_1}{Ddq}\right)\left(\frac{D\eta^{\Delta-1}}{q}\left[-\frac{1}{d}\ptr{d}{O}^2+\ptr{d}{O^2}\right]\right)\\
	&+\left(\frac{C_1}{q}\right)\Bigg(\frac{1}{q}\Bigg(-\ptr{d}{O}^2\left[\frac{D}{d}\eta^{\Delta-1}\xi\Gamma_{n-\Delta-1}+\frac{1}{d}\eta^{\Delta-1}\eta^{n-\Delta-1}\right]\\
		&\hspace{26mm}+\ptr{d}{O^2}\left[D\eta^{\Delta-1}\xi\Gamma_{n-\Delta-1}+D^2\eta^{\Delta-1}\eta^{n-\Delta-1}\right]\Bigg)\Bigg)\\
	=&\frac{C_1\eta^{\Delta-1}}{q^2}\Bigg(\left[\frac{1}{d^2}\ptr{d}{O}^2-\frac{1}{d}\ptr{d}{O^2}\right]-\ptr{d}{O}^2\left[\frac{D}{d}\xi\Gamma_{n-\Delta-1}+\frac{1}{d}\eta^{n-\Delta-1}\right]\\
		&\hspace{17mm}+\ptr{d}{O^2}\left[D\xi\Gamma_{n-\Delta-1}+D^2\eta^{n-\Delta-1}\right]\Bigg)\\
	=&\frac{C_1\eta^{\Delta-1}}{q^2}\Bigg(\ptr{d}{O}^2\left(\frac{1}{d^2}-\frac{D}{d}\xi\Gamma_{n-\Delta-1}-\frac{1}{d}\eta^{n-\Delta-1}\right)\\
	&\hspace{16mm}+\ptr{d}{O^2}\left(-\frac{1}{d}+D\xi\Gamma_{n-\Delta-1}+D^2\eta^{n-\Delta-1}\right)\Bigg)\\
	=&\frac{C_1\eta^{\Delta-1}}{q^2}\Bigg(-\frac{1}{d}\ptr{d}{O}^2\left(-\frac{1}{d}+D\xi\Gamma_{n-\Delta-1}+\eta^{n-\Delta-1}\right)\\
	&\hspace{16mm}+\ptr{d}{O^2}\left(-\frac{1}{d}+D\xi\Gamma_{n-\Delta-1}+D^2\eta^{n-\Delta-1}\right)\Bigg)\\
	=&\frac{C_1\eta^{\Delta-1}}{q^2}\Bigg(-\frac{1}{d}\ptr{d}{O}^2\left(-\frac{1}{d}+D\xi\Gamma_{n-\Delta-1}+D^2\eta^{n-\Delta-1}+(1-D^2)\eta^{n-\Delta-1}\right)\\
	&\hspace{16mm}+\ptr{d}{O^2}\left(-\frac{1}{d}+D\xi\Gamma_{n-\Delta-1}+D^2\eta^{n-\Delta-1}\right)\Bigg)\\
	=&\frac{C_1\eta^{\Delta-1}}{q^2}\Bigg(\left(\ptr{d}{O^2}-\frac{1}{d}\ptr{d}{O}^2\right)\left(-\frac{1}{d}+D\xi\Gamma_{n-\Delta-1}+D^2\eta^{n-\Delta-1}\right)\\
	&\hspace{16mm}+\ptr{d}{O}^2\frac{(D^2-1)\eta^{n-\Delta-1}}{d}\Bigg)\\
	=&\frac{C_1\eta^{\Delta-1}}{q^2}\Bigg(\epsilon(O)\left(-\frac{1}{d}+D\xi\Gamma_{n-\Delta-1}+D^2\eta^{n-\Delta-1}\right)+\ptr{d}{O}^2\frac{(D^2-1)\eta^{n-\Delta-1}}{d}\Bigg).
\end{split}
\end{equation}
In the last line, we use
\begin{equation}
\begin{split}
	\epsilon(O)
	&=\norm{O-\ptr{d}{O}\frac{I_d}{d}}_\mathrm{HS}^2\\
	&=\ptr{d}{\left(O-\ptr{d}{O}\frac{I_d}{d}\right)^2}\\
	&=\ptr{d}{O^2}-\frac{2}{d}\ptr{d}{O}^2+\ptr{d}{O}^2\frac{\ptr{d}{I_d}}{d^2}\\
	&=\ptr{d}{O^2}-\frac{1}{d}\ptr{d}{O}^2.
\end{split}
\end{equation}

Taking the large $n$ limit and fixing $m$ (thereby fixing $\Delta$),
\begin{equation}
\begin{split}
	\lim_{n\rightarrow \infty}\Gamma_{n-\Delta-1}
	&=\lim_{n\rightarrow \infty}\frac{1-\eta^{n-\Delta-1}}{1-\eta}=\frac{1}{1-\eta},\\
	\lim_{n\rightarrow \infty}\ptr{d}{O}^2\eta^{n-\Delta-1}&=0,
\end{split}
\end{equation}
where we used $\eta<1$ and the assumption that $\ptr{d}{O}^2$ grows slower than exponentially in $n$. Therefore, in the large $n$ limit, the variance is
\begin{equation}
	\mathrm{Var}[\partial_k C]=\epsilon(O)\frac{C_1\eta^{\Delta-1}}{q^2}\left(-\frac{1}{d}+\frac{D\xi}{1-\eta}\right).
\end{equation}
Using the identities from Eq.~\eqref{Eq:OffU+}, we compute the variance in the off-site case where only $U_+$ forms a 2-design: 

\begin{equation}
.
\end{equation}

For simplicity, we define $L_1=\Delta-1$ and $L_2=n-\Delta-2$. The variance evaluates to
\begin{equation}
\begin{split}
	\mathrm{Var}[\partial_k C]
	=&\left(\frac{D\eta^{L_1}}{q}\left[-\frac{1}{d}\ptr{d}{O}^2+\ptr{d}{O^2}\right]\right)\left(\frac{\eta}{q}\left(d^2C_2-\frac{C_3}{D}\right)\right)\\
		&+\Big(\frac{1}{q}\Big(-\ptr{d}{O}^2\Big(\frac{D}{d}\eta^{L_1}\xi \Gamma_{L_2}+\frac{1}{d}\eta^{L_1}\eta^{L_2}\Big)\\
	&\hspace{12mm}+\ptr{d}{O^2}\Big(D\eta^{L_1}\xi \Gamma_{L_2}+D^2\eta^{L_1}\eta^{L_2}\Big)\Big)\Big)\left(\frac{\eta}{q}\left(-\frac{C_2 d}{D}+C_3d\right)\right)\\
	=&\frac{\eta^{\Delta}}{q^2}\left(C_2 Dd^2-C_3\right)\left[-\frac{1}{d}\ptr{d}{O}^2+\ptr{d}{O^2}\right]\\
&+\frac{\eta^{\Delta}}{q^2}\left(-\frac{C_2d}{D}+C_3d\right)\Big(-\ptr{d}{O}^2\Big(\frac{D}{d}\xi \Gamma_{L_2}+\frac{1}{d}\eta^{L_2}\Big)\\
&\hspace{40mm}+\ptr{d}{O^2}\Big(D\xi \Gamma_{L_2}+D^2\eta^{L_2}\Big)\Big)\\
	=&\frac{\eta^{\Delta}}{q^2}\Bigg[-\ptr{d}{O}^2\left(\frac{C_2Dd^2-C_3}{d}+\left(-\frac{C_2d}{D}+C_3d\right)\Big(\frac{D}{d}\xi \Gamma_{L_2}+\frac{1}{d}\eta^{L_2}\Big)\right)\\
&\hspace{8mm}+\ptr{d}{O^2}\left((C_2Dd^2-C_3)+\left(-\frac{C_2d}{D}+C_3d\right)\Big(D\xi \Gamma_{L_2}+D^2\eta^{L_2}\Big)\right)\Bigg]\\
	=&\frac{\eta^{\Delta}}{q^2}\Bigg[-\frac{1}{d}\ptr{d}{O}^2\Big(C_2Dd^2-C_3\\
&\hspace{34mm}+\left(-\frac{C_2d}{D}+C_3d\right)\Big(D\xi \Gamma_{L_2}+(D^2-D^2+1)\eta^{L_2}\Big)\Big)\\
&\hspace{8mm}+\ptr{d}{O^2}\left(C_2Dd^2-C_3+\left(-\frac{C_2d}{D}+C_3d\right)\Big(D\xi \Gamma_{L_2}+D^2\eta^{L_2}\Big)\right)\Bigg]\\
	=&\frac{\eta^{\Delta}}{q^2}\Bigg[\left(\ptr{d}{O^2}-\frac{1}{d}\ptr{d}{O}^2\right)\\
&\hspace{10mm}\cdot\left(C_2Dd^2-C_3+\left(-\frac{C_2d}{D}+C_3d\right)\Big(D\xi \Gamma_{L_2}+D^2\eta^{L_2}\Big)\right)\\
&\hspace{10mm}+\ptr{d}{O}^2\left(-\frac{C_2}{D}+C_3\right)(D^2-1)\eta^{L_2}\Bigg]\\
	=&\frac{\eta^{\Delta}}{q^2}\Bigg[\epsilon(O)
\left(C_2Dd^2-C_3+\left(-\frac{C_2d}{D}+C_3d\right)\Big(D\xi \Gamma_{n-\Delta-2}+D^2\eta^{n-\Delta-2}\Big)\right)\\
&\hspace{10mm}+\ptr{d}{O}^2\left(-\frac{C_2}{D}+C_3\right)(D^2-1)\eta^{n-\Delta-2}\Bigg].
\end{split}
\end{equation}
In the large $n$ limit, this becomes

\begin{equation}
\mathrm{Var}[\partial_k C]
	=\epsilon(O)
\frac{\eta^{\Delta}}{q^2}\left(C_2Dd^2-C_3+\left(-\frac{C_2d}{D}+C_3d\right)\frac{D\xi}{1-\eta}\right).
\end{equation}

Using the identities from Eq.~\eqref{Eq:OffBoth}, we compute the variance in the off-site case where both $U_-$ and $U_+$ form a 2-design: 

\begin{equation}
.
\end{equation}

For simplicity, we define $L_1=\Delta-1$ and $L_2=n-\Delta-1$. The variance is

\begin{equation}
\begin{split}
	\mathrm{Var}[\partial_k C]
	=&\left(-\frac{C_4}{Ddq}\eta\right)\left(\frac{D\eta^{L_1}}{q}\left[-\frac{1}{d}\ptr{d}{O}^2+\ptr{d}{O^2}\right]\right)\\
&+\left(\frac{C_4}{q}\eta\right)\Big(\frac{1}{q}\Big(-\ptr{d}{O}^2\Big(\frac{D}{d}\eta^{L_1}\xi \Gamma_{L_2}+\frac{1}{d}\eta^{L_1}\eta^{L_2}\Big)\\
&\hspace{12mm}+\ptr{d}{O^2}\Big(D\eta^{L_1}\xi \Gamma_{L_2}+D^2\eta^{L_1}\eta^{L_2}\Big)\Big)\Big)\\
	=&\frac{C_4\eta^{\Delta}}{q^2}\Bigg[-\frac{1}{d}\left[-\frac{1}{d}\ptr{d}{O}^2+\ptr{d}{O^2}\right]\\
&\hspace{12mm}-\ptr{d}{O}^2\Big(\frac{D}{d}\xi \Gamma_{L_2}+\frac{1}{d}\eta^{L_2}\Big)+\ptr{d}{O^2}\Big(D\xi \Gamma_{L_2}+D^2\eta^{L_2}\Big)\Bigg]\\
	=&\frac{C_4\eta^{\Delta}}{q^2}\Bigg[\ptr{d}{O}^2\left(\frac{1}{d^2}-\Big(\frac{D}{d}\xi \Gamma_{L_2}+\frac{1}{d}\eta^{L_2}\Big)\right)\Bigg]\\
&\hspace{12mm}+\ptr{d}{O^2}\left(-\frac{1}{d}+\Big(D\xi \Gamma_{L_2}+D^2\eta^{L_2}\Big)\right)\\
	=&\frac{C_4\eta^{\Delta}}{q^2}\Bigg[-\frac{1}{d}\ptr{d}{O}^2\left(-\frac{1}{d}+D\xi \Gamma_{L_2}+(D^2-D^2+1)\eta^{L_2}\right)\\
&\hspace{12mm}+\ptr{d}{O^2}\left(-\frac{1}{d}+D\xi \Gamma_{L_2}+D^2\eta^{L_2}\right)\Bigg]\\
	=&\frac{C_4\eta^{\Delta}}{q^2}\Bigg[\left(\ptr{d}{O^2}-\frac{1}{d}\ptr{d}{O}^2\right)\left(-\frac{1}{d}+D\xi \Gamma_{L_2}+D^2\eta^{L_2}\right)\\
&\hspace{12mm}+\frac{1}{d}\ptr{d}{O}^2(D^2-1)\eta^{L_2}\Bigg]\\
	=&\frac{C_4\eta^{\Delta}}{q^2}\Bigg[\epsilon(O)\left(-\frac{1}{d}+D\xi \Gamma_{n-\Delta-1}+D^2\eta^{n-\Delta-1}\right)+\frac{1}{d}\ptr{d}{O}^2(D^2-1)\eta^{n-\Delta-1}\Bigg].
\end{split}
\end{equation}

In the large $n$ limit,
\begin{equation}
\begin{split}
	\mathrm{Var}[\partial_k C]
	=&\epsilon(O)\frac{C_4\eta^{\Delta}}{q^2}\left(-\frac{1}{d}+\frac{D\xi}{1-\eta} \right).
\end{split}
\end{equation}

\subsection{On-site case}
We compute the variance in the on-site case when only $U_-$ forms a 2-design:

\begin{equation}
.
\end{equation}

This evaluates to 
\begin{equation}
\begin{split}
	\mathrm{Var}[\partial_k C]
	&=\left(-\frac{C_5}{Ddq}\right)\left(1\right)+\left(-\frac{C_6}{Ddq}\right)(0)+\left(\frac{C_5}{q}\right)(\xi\Gamma_{n-1})+\left(\frac{C_6}{q}\right)(\eta^{n-1})\\
	&=\frac{1}{q}\left(C_5\left(-\frac{1}{Dd}+\xi\Gamma_{n-1}\right)+C_6\eta^{n-1}\right).
\end{split}
\end{equation}

We construct an upper bound on the variance using $\epsilon(O)$. Using the triangle inequality, we get the following upper bound:
\begin{equation}
\begin{split}\label{Eq:VarC5C6}
\mathrm{Var}[\partial_k C]
	&\leq \frac{1}{q}\left(\abs{C_5}\abs{-\frac{1}{Dd}+\xi\Gamma_{n-1}}+\abs{C_6}\eta^{n-1}\right)\\
	&\leq \frac{1}{q}\left(\abs{C_5}\left(\frac{1}{Dd}+\xi\Gamma_{n-1}\right)+\abs{C_6}\eta^{n-1}\right).
\end{split}
\end{equation}
We now bound $\abs{C_5}$ and $\abs{C_6}$. Before proceeding, it will help to define the following operators: $O_d=I_D\otimes O$, $\tilde{O}_d=O_d-\ptr{d}{O}\frac{I_{Dd}}{d}$, and $G_+=U_+^\dagger GU_+$. In order to bound $\abs{C_5}$, we first rewrite the following:
\begin{equation}
\begin{split}
\tr{\sigma G[G,\sigma]}
	&=\tr{\sigma GG\sigma}-\tr{\sigma G\sigma G}\\
	&=\tr{U_+ O_d U_+^\dagger  GU_+U_+^\dagger  GU_+ O_d U_+^\dagger  }-\tr{U_+ O_d U_+^\dagger  GU_+ O_d U_+^\dagger  G}\\
	&= \tr{O_dG_+G_+O_d}-\tr{ O_dG_+O_dG_+}\\
	&=\tr{O_dG_+[G_+,O_d]}\\
	&=  \tr{O_dG_+\left[G_+,O_d-\ptr{d}{O}\frac{I_{Dd}}{d}\right]}\\
	&= \tr{O_dG_+\left[G_+,\tilde{O}_d\right]}\\
	&=\tr{O_dG_+G_+\tilde{O}_d}-\tr{O_dG_+\tilde{O}_dG_+}\\
	&=\tr{O_dG_+G_+\tilde{O}_d}-\tr{G_+O_dG_+\tilde{O}_d}\\
	&= \tr{[O_d,G_+]G_+\tilde{O}_d}\\
	&= \tr{\left[O_d-\ptr{d}{O}\frac{I_{Dd}}{d},G_+\right]G_+\tilde{O}_d}\\
	&=\tr{[\tilde{O}_d,G_+]G_+\tilde{O}_d}\\
	&= \tr{\tilde{O}_dG_+G_+\tilde{O}_d}-\tr{G_+\tilde{O}_dG_+\tilde{O}_d}.
\end{split}
\end{equation}

Defining $\left< \cdot\right>_{\mathrm{max}}=\tr{(\cdot) \frac{I_{Dd}}{Dd}}$ as the expectation value over the maximally mixed state, we can bound $\abs{C_5}$:

\begin{equation}
\begin{split}
	\abs{C_5}
	&=\abs{2\int dU_- \tr{\sigma G[G,\sigma]}}\\
	&=\abs{2\int dU_-\left[ \tr{\tilde{O}_dG_+G_+\tilde{O}_d}-\tr{G_+\tilde{O}_dG_+\tilde{O}_d}\right]}\\
	&=\abs{2Dd\int dU_- \left[\left<\tilde{O}_dG_+G_+\tilde{O}_d\right>_{\mathrm{max}}-\left<G_+\tilde{O}_dG_+\tilde{O}_d\right>_{\mathrm{max}}\right]}\\
	&\leq 2Dd\int dU_- \left[\abs{ \left<\tilde{O}_dG_+G_+\tilde{O}_d\right>_{\mathrm{max}}}+\abs{\left<G_+\tilde{O}_dG_+\tilde{O}_d\right>_{\mathrm{max}}}\right]\\
	&\leq 2Dd\int dU_-\left[\norm{ \tilde{O}_dG_+G_+\tilde{O}_d}_{\infty}+\norm{G_+\tilde{O}_dG_+\tilde{O}_d}_{\infty}\right]\\
	&\leq 4Dd\int dU_-\norm{ \tilde{O}_d}_{\infty}^{2}\norm{G_+}_{\infty}^2\\
	&= 4Dd\int dU_-\norm{I_D\otimes \left(O-\frac{\ptr{d}{O}I_d}{d}\right)}_{\infty}^{2}\norm{G}_{\infty}^2\\
	&= 4Dd\norm{O-\frac{\ptr{d}{O}I_d}{d}}_{\infty}^{2}\norm{G}_{\infty}^2\\
	&\leq 4Dd\norm{O-\frac{\ptr{d}{O}I_d}{d}}_{\mathrm{HS}}^{2}\norm{G}_{\infty}^2\\
	&= \epsilon(O)4Dd\norm{G}_{\infty}^2.
\end{split}
\end{equation}

We now bound $\abs{C_6}$:
\begin{equation}
\begin{split}
\abs{C_6}
&=\abs{2\int dU_+ \Big[-\ptr{d}{\left(\ptr{D}{U_+^\dagger  GU_+}O\right)^2}+D\ptr{d}{\ptr{D}{U_+^\dagger  G^2U_+}O^2}\Big]}\\
&=\abs{2\int dU_+ \Big[-\ptr{d}{\left(\ptr{D}{G_+}O\right)^2}+D\tr{G_+^2O_d^2}\Big]}\\
&=\Big|2\int dU_+ \Big[-\av{P_D}D^2d\tr{(I_D\otimes O_dG_+)(\overline{P}_D\otimes P_D \otimes I_d)(I_D\otimes O_dG_+)\rho_{\mathrm{BD}}}\\
	&\hspace{23mm}+D^2d\left<G_+^2O_d^2\right>_{\mathrm{max}}\Big]\Big|\\
&\leq2\int dU_+ \Big[\av{P_D}D^2d\abs{\tr{(I_D\otimes O_dG_+)(\overline{P}_D\otimes P_D \otimes I_d)(I_D\otimes O_dG_+)\rho_{\mathrm{BD}}}}\\
	&\hspace{22mm}+D^2d\abs{\left<G_+^2O_d^2\right>_{\mathrm{max}}}\Big]\\
&\leq2\int dU_+ \Big[\av{P_D}D^2d\norm{(I_D\otimes O_dG_+)(\overline{P}_D\otimes P_D \otimes I_d)(I_D\otimes O_dG_+)}_{\infty}+D^2d\norm{G_+^2O_d^2}_{\infty}\Big]\\
&\leq2\int dU_+ \Big[\av{P_D}D^2d\norm{I_D\otimes O_dG_+}_{\infty}\norm{\overline{P}_D\otimes P_D \otimes I_d}_{\infty}\norm{I_D\otimes O_dG_+}_{\infty}\\
&\hspace{22mm}+D^2d\norm{G_+}_{\infty}^2\norm{O_d}^2_{\infty}\Big]\\
&=2\int dU_+ \Big[\av{P_D}D^2d\norm{ O_d}_\infty^2\norm{G_+}_{\infty}^2\norm{\overline{P}_D}_\infty\norm{P_D }_{\infty}+D^2d\norm{G_+}_{\infty}^2\norm{O_d}^2_{\infty}\Big]\\
&=2\int dU_+ \Big[\av{P_D}D^2d\norm{ O_d}_\infty^2\norm{G_+}_{\infty}^2+D^2d\norm{G_+}_{\infty}^2\norm{O_d}^2_{\infty}\Big]\\
&=2\int dU_+ 2D^2d\norm{ O_d}_\infty^2\norm{G}_{\infty}^2\\
&=4D^2d\norm{ O_d}_\infty^2\norm{G}_{\infty}^2\\
&=4D^2d\norm{ I_D\otimes O}_\infty^2\norm{G}_{\infty}^2\\
&=4D^2d\norm{O}_\infty^2\norm{G}_{\infty}^2.
\end{split}
\end{equation}

In the above, we used:
\begin{equation}
\begin{tikzpicture}	
    \renewcommand{\xglobalshift}{0}
    \renewcommand{\yglobalshift}{0}
    %
%
	\renewcommand{\nodenum}{O}
	\renewcommand{\xpos}{\xglobalshift+.28*\xrel}
    \renewcommand{\ypos}{\yglobalshift-\rowspace}
    \coordinate (\nodenum) at (\xpos,\ypos) {};
	\renewcommand{\nodenum}{G+1Red}
	\renewcommand{\xpos}{\xglobalshift-.28*\xrel}
    \renewcommand{\ypos}{\yglobalshift-2*\rowspace}
    \coordinate (\nodenum) at (\xpos,\ypos) {}; 
	\renewcommand{\nodenum}{G+2Red}
	\renewcommand{\xpos}{\xglobalshift-.28*\xrel}
    \renewcommand{\ypos}{\yglobalshift-4*\rowspace}
    \coordinate (\nodenum) at (\xpos,\ypos) {}; 
	%
	%
	\draw [line width=\linethick,color=dullred]  
	(G+1Red)--++(0,.75*\yleg)--++(-.75*\xleg,0)--++(0,-1.5*\yleg)--++(.75*\xleg,0)--(G+1Red)
	(G+2Red)--++(0,.75*\yleg)--++(-.75*\xleg,0)--++(0,-1.5*\yleg)--++(.75*\xleg,0)--(G+2Red);
	\draw [line width=\linethick,color=dullblue]  
	(O)--++(0,\yleg)--++(.75*\xleg,0)--++(0,-5*\rowspace)--++(-.75*\xleg,0)--(O);
	%
	%
     \renewcommand{\nodenum}{Equation}
    \renewcommand{\name}{$\ptr{d}{(\ptr{D}{G_+}O)^2}=$}
	\renewcommand{\xpos}{\xglobalshift-2.5*\xrel}
    \renewcommand{\ypos}{\yglobalshift-2*\rowspace}
    \renewcommand{\height}{\heightsingle}
    \renewcommand{\width}{\widthsingle}
    \node[] (\nodenum) at (\xpos,\ypos) {\name};
     \renewcommand{\nodenum}{G+1}
    \renewcommand{\name}{$G_+$}
	\renewcommand{\xpos}{\xglobalshift}
    \renewcommand{\ypos}{\yglobalshift-2*\rowspace}
    \renewcommand{\height}{\heightsingle}
    \renewcommand{\width}{\widthdouble}
    \node[rectangle, fill=egg, rounded corners, minimum width=\width, minimum height=\height, draw] (\nodenum) at (\xpos,\ypos) {\name};
     \renewcommand{\nodenum}{G+2}
    \renewcommand{\name}{$G_+$}
	\renewcommand{\xpos}{\xglobalshift}
    \renewcommand{\ypos}{\yglobalshift-4*\rowspace}
    \renewcommand{\height}{\heightsingle}
    \renewcommand{\width}{\widthdouble}
    \node[rectangle, fill=egg, rounded corners, minimum width=\width, minimum height=\height, draw] (\nodenum) at (\xpos,\ypos) {\name};
	\renewcommand{\nodenum}{O1}
    \renewcommand{\name}{$O$}
	\renewcommand{\xpos}{\xglobalshift+.28*\xrel}
    \renewcommand{\ypos}{\yglobalshift-\rowspace}
    \renewcommand{\height}{\heightsingle}
    \renewcommand{\width}{\widthsingle}
    \node[rectangle, fill=egg, rounded corners, minimum width=\width, minimum height=\height, draw] (\nodenum) at (\xpos,\ypos) {\name};
	\renewcommand{\nodenum}{O1}
    \renewcommand{\name}{$O$}
	\renewcommand{\xpos}{\xglobalshift+.28*\xrel}
    \renewcommand{\ypos}{\yglobalshift-3*\rowspace}
    \renewcommand{\height}{\heightsingle}
    \renewcommand{\width}{\widthsingle}
    \node[rectangle, fill=egg, rounded corners, minimum width=\width, minimum height=\height, draw] (\nodenum) at (\xpos,\ypos) {\name};
%
    \renewcommand{\xglobalshift}{1*\xrel}
    \renewcommand{\yglobalshift}{-5.5*\rowspace}
    %
%
	\renewcommand{\nodenum}{O}
	\renewcommand{\xpos}{\xglobalshift+.28*\xrel}
    \renewcommand{\ypos}{\yglobalshift-\rowspace}
    \coordinate (\nodenum) at (\xpos,\ypos) {};
	\renewcommand{\nodenum}{G+1RedRight}
	\renewcommand{\xpos}{\xglobalshift-.28*\xrel}
    \renewcommand{\ypos}{\yglobalshift-2*\rowspace}
    \coordinate (\nodenum) at (\xpos,\ypos) {}; 
	\renewcommand{\nodenum}{G+1RedLeft}
	\renewcommand{\xpos}{\xglobalshift-.28*\xrel-.55*\xrel}
    \renewcommand{\ypos}{\yglobalshift-2*\rowspace}
    \coordinate (\nodenum) at (\xpos,\ypos) {}; 
	\renewcommand{\nodenum}{G+2Red}
	\renewcommand{\xpos}{\xglobalshift-.28*\xrel}
    \renewcommand{\ypos}{\yglobalshift-4*\rowspace}
    \coordinate (\nodenum) at (\xpos,\ypos) {}; 
	%
	%
	\draw [line width=\linethick,color=dullred]  
	(G+1RedRight)--++(0,2*\yleg)--++(-1.35*\xleg,0)--++(0,-6*\rowspace)--++(1.35*\xleg,0)--(G+1RedRight)
	(G+1RedLeft)--++(0,1.75*\yleg)--++(-.45*\xleg,0)--++(0,-5.5*\rowspace)--++(.45*\xleg,0)--(G+1RedLeft);
	\draw [line width=\linethick,color=dullblue]  
	(O)--++(0,\yleg)--++(.75*\xleg,0)--++(0,-6*\rowspace)--++(-.75*\xleg,0)--(O);
	%
	%
     \renewcommand{\nodenum}{Equation}
    \renewcommand{\name}{$=\av{P_D}D^2d$}
	\renewcommand{\xpos}{\xglobalshift-1.95*\xrel}
    \renewcommand{\ypos}{\yglobalshift-2*\rowspace}
    \renewcommand{\height}{\heightsingle}
    \renewcommand{\width}{\widthsingle}
    \node[] (\nodenum) at (\xpos,\ypos) {\name};
     \renewcommand{\nodenum}{G+1}
    \renewcommand{\name}{$G_+$}
	\renewcommand{\xpos}{\xglobalshift}
    \renewcommand{\ypos}{\yglobalshift-2*\rowspace}
    \renewcommand{\height}{\heightsingle}
    \renewcommand{\width}{\widthdouble}
    \node[rectangle, fill=egg, rounded corners, minimum width=\width, minimum height=\height, draw] (\nodenum) at (\xpos,\ypos) {\name};
     \renewcommand{\nodenum}{G+2}
    \renewcommand{\name}{$G_+$}
	\renewcommand{\xpos}{\xglobalshift}
    \renewcommand{\ypos}{\yglobalshift-5*\rowspace}
    \renewcommand{\height}{\heightsingle}
    \renewcommand{\width}{\widthdouble}
    \node[rectangle, fill=egg, rounded corners, minimum width=\width, minimum height=\height, draw] (\nodenum) at (\xpos,\ypos) {\name};
     \renewcommand{\nodenum}{rho}
    \renewcommand{\name}{$\rho_{\mathrm{BD}}$}
	\renewcommand{\xpos}{\xglobalshift-.28*\xrel}
    \renewcommand{\ypos}{\yglobalshift-3*\rowspace}
    \renewcommand{\height}{\heightsingle}
    \renewcommand{\width}{1.5*\widthdouble}
    \node[rectangle, fill=egg, rounded corners, minimum width=\width, minimum height=\height, draw] (\nodenum) at (\xpos,\ypos) {\name};
	\renewcommand{\nodenum}{P_D*}
    \renewcommand{\name}{$\overline{P}_D$}
	\renewcommand{\xpos}{\xglobalshift+.28*\xrel-.55*\xrel-.55*\xrel}
    \renewcommand{\ypos}{\yglobalshift-\rowspace}
    \renewcommand{\height}{\heightsingle}
    \renewcommand{\width}{\widthsingle}
    \node[rectangle, fill=egg, rounded corners, minimum width=\width, minimum height=\height, draw] (\nodenum) at (\xpos,\ypos) {\name};
	\renewcommand{\nodenum}{P_D}
    \renewcommand{\name}{$P_D$}
	\renewcommand{\xpos}{\xglobalshift+.28*\xrel-.55*\xrel}
    \renewcommand{\ypos}{\yglobalshift-\rowspace}
    \renewcommand{\height}{\heightsingle}
    \renewcommand{\width}{\widthsingle}
    \node[rectangle, fill=egg, rounded corners, minimum width=\width, minimum height=\height, draw] (\nodenum) at (\xpos,\ypos) {\name};
	\renewcommand{\nodenum}{O1}
    \renewcommand{\name}{$O$}
	\renewcommand{\xpos}{\xglobalshift+.28*\xrel}
    \renewcommand{\ypos}{\yglobalshift-\rowspace}
    \renewcommand{\height}{\heightsingle}
    \renewcommand{\width}{\widthsingle}
    \node[rectangle, fill=egg, rounded corners, minimum width=\width, minimum height=\height, draw] (\nodenum) at (\xpos,\ypos) {\name};
	\renewcommand{\nodenum}{O1}
    \renewcommand{\name}{$O$}
	\renewcommand{\xpos}{\xglobalshift+.28*\xrel}
    \renewcommand{\ypos}{\yglobalshift-4*\rowspace}
    \renewcommand{\height}{\heightsingle}
    \renewcommand{\width}{\widthsingle}
    \node[rectangle, fill=egg, rounded corners, minimum width=\width, minimum height=\height, draw] (\nodenum) at (\xpos,\ypos) {\name};
%
    \renewcommand{\xglobalshift}{-2.1*\xrel}
    \renewcommand{\yglobalshift}{-11*\rowspace}
     \renewcommand{\nodenum}{Equation}
    \renewcommand{\name}{$=\av{P_D}D^2d\tr{(I_D\otimes O_dG_+)(\overline{P}_D\otimes P_D \otimes I_d)(I_D\otimes O_dG_+)\rho_{\mathrm{BD}}},$}
	\renewcommand{\xpos}{\xglobalshift+4.25*\xrel}
    \renewcommand{\ypos}{\yglobalshift-2*\rowspace}
    \renewcommand{\height}{\heightsingle}
    \renewcommand{\width}{\widthsingle}
    \node[] (\nodenum) at (\xpos,\ypos) {\name};
\end{tikzpicture}
\end{equation}
where we define the state $\rho_{\mathrm{BD}}=\frac{I_d}{d}\otimes \rho_{\mathrm{Bell,D}}$, and where $\rho_{\mathrm{Bell,D}}$ is the Bell state between the $D$-dimensional subsystem and a reference system. 

Using these bounds on $\abs{C_5}$ and $\abs{C_6}$, Ineq.~\eqref{Eq:VarC5C6} becomes
\begin{equation}
\begin{split}
\mathrm{Var}[\partial_k C]
	&\leq \frac{1}{q}\left(\abs{C_5}\left(\frac{1}{Dd}+\xi\Gamma_{n-1}\right)+\abs{C_6}\eta^{n-1}\right)\\
	&\leq \frac{1}{q}\left(\left(\epsilon(O)4Dd\norm{G}_{\infty}^2\right)\left(\frac{1}{Dd}+\xi\Gamma_{n-1}\right)+\left(4D^2d\norm{G}_{\infty}^2\norm{O}_{\infty}^2\right)\eta^{n-1}\right)\\
	&=\frac{4Dd \norm{G}_{\infty}^2}{q}\left(\epsilon(O)\left(\frac{1}{Dd}+\xi\Gamma_{n-1}\right)+D\norm{O}_{\infty}^2\eta^{n-1}\right).
\end{split}
\end{equation}
In the large $n$ limit, assuming $\norm{O}_{\infty}^2$ grows slower than exponentially in $n$, we have the following inequality

\begin{equation}
\begin{split}
\mathrm{Var}[\partial_k C]
	&\leq \epsilon(O)\frac{4 \norm{G}_{\infty}^2}{q}\left(1+\frac{Dd\xi}{1-\eta}\right).
\end{split}
\end{equation}

We now compute the variance for the on-site case where only $U_+$ forms a 2-design:

\begin{equation}
.
\end{equation}
This can be written as
\begin{equation}
%
.
\end{equation}

Using the identities from Eqs.~\eqref{Eq:OnU+1} to \eqref{Eq:OnU+4}, this evaluates to
\begin{equation}
\begin{split}
	\mathrm{Var}[\partial_k C]
	=&(1)\left(\frac{1}{q^2}\left(Dd^2C_2-C_3\right)\left(-\frac{1}{d}\ptr{d}{O}^2+\ptr{d}{O^2}\right)\right)\\
	&+(\xi\Gamma_{n-2})\left(\frac{1}{q^2}\left(-dC_2+DdC_3\right)\left(-\frac{1}{d}\ptr{d}{O}^2+\ptr{d}{O^2}\right)\right)\\
	&+\left(0\right)\left(\frac{1}{q^2}\left(d^2C_2-\frac{C_3}{D}\right)\left(-\frac{1}{d}\ptr{d}{O}^2+D^2\ptr{d}{O^2}\right)\right)\\
	&+\left(\eta^{n-2}\right)\left(\frac{1}{q^2}\left(-\frac{dC_2}{D}+dC_3\right)\left(-\frac{1}{d}\ptr{d}{O}^2+D^2\ptr{d}{O^2}\right)\right)\\
	=&\frac{1}{q^2}\Big[-\frac{1}{d}\ptr{d}{O}^2\left(Dd^2C_2-C_3+\xi\Gamma_{n-2}(-dC_2+DdC_3)+\eta^{n-2}\left(-\frac{dC_2}{D}+dC_3\right)\right)\\
	&+\ptr{d}{O^2}\left(Dd^2C_2-C_3+\xi\Gamma_{n-2}(-dC_2+DdC_3)+\eta^{n-2}D^2\left(-\frac{dC_2}{D}+dC_3\right)\right)\Big]\\
	=&\frac{1}{q^2}\Big[-\frac{1}{d}\ptr{d}{O}^2\Big(Dd^2C_2-C_3+\xi\Gamma_{n-2}(-dC_2+DdC_3)\\
&\hspace{32mm}+(D^2-D^2+1)\eta^{n-2}\Big(-\frac{dC_2}{D}+dC_3\Big)\Big)\\
&+\ptr{d}{O^2}\left(Dd^2C_2-C_3+\xi\Gamma_{n-2}(-dC_2+DdC_3)+\eta^{n-2}D^2\left(-\frac{dC_2}{D}+dC_3\right)\right)\Big]\\
	=&\frac{1}{q^2}\Big[\left(\ptr{d}{O^2}-\frac{1}{d}\ptr{d}{O}^2\right)\Big(Dd^2C_2-C_3+\xi\Gamma_{n-2}(-dC_2+DdC_3)\\
	&\hspace{52mm}+\eta^{n-2}D^2\left(-\frac{dC_2}{D}+dC_3\right)\Big)\\
	&\hspace{5mm}+\frac{1}{d}\ptr{d}{O}^2(D^2-1)\eta^{n-2}\left(-\frac{dC_2}{D}+dC_3\right)\Big]\\
	=&\frac{1}{q^2}\Bigg[\epsilon(O)\Bigg(Dd^2C_2-C_3+\xi\Gamma_{n-2}Dd\left(-\frac{C_2}{D}+C_3\right)+\eta^{n-2}D^2d\left(-\frac{C_2}{D}+C_3\right)\Bigg)\\
	&\hspace{5mm}+\ptr{d}{O}^2(D^2-1)\eta^{n-2}\left(-\frac{C_2}{D}+C_3\right)\Bigg]\\
	=&\frac{1}{q^2}\Bigg[\epsilon(O)\Bigg(Dd^2C_2-C_3+\left(-\frac{C_2}{D}+C_3\right)\left(\xi\Gamma_{n-2}Dd+\eta^{n-2}D^2d\right)\Bigg)\\
	&\hspace{5mm}+\ptr{d}{O}^2(D^2-1)\eta^{n-2}\left(-\frac{C_2}{D}+C_3\right)\Bigg].
\end{split}
\end{equation}
In the large $n$ limit, assuming $\ptr{d}{O}^2$ grows slower than exponentially in $n$, the variance is 

\begin{equation}
\begin{split}
	\mathrm{Var}[\partial_k C]
	=&\frac{\epsilon(O)}{q^2}\Bigg(Dd^2C_2-C_3+\left(-\frac{C_2}{D}+C_3\right)\frac{\xi Dd}{1-\eta}\Bigg).
\end{split}
\end{equation}

We now compute the variance for the on-site case where both $U_-$ and $U_+$ form 2-designs:
\begin{equation}
.
\end{equation}

Using the identities from Eq.~\eqref{Eq:OnBoth}, the variance evaluates to
\begin{equation}
\begin{split}
\mathrm{Var}[\partial_k C]
	=&\left(\frac{C_4}{q^2}\left(\frac{1}{d^2}\ptr{d}{O}^2-\frac{1}{d}\ptr{d}{O^2}\right)\right)(1)\\
&+\left(\frac{C_4}{q^2}\left(\frac{1}{Dd^2}\ptr{d}{O}^2-\frac{D}{d}\ptr{d}{O^2}\right)\right)(0)\\
&+\left(\frac{C_4}{q^2}\left(-\frac{D}{d}\ptr{d}{O}^2+D\ptr{d}{O^2}\right)\right)(\xi\Gamma_{n-1})\\
&+\left(\frac{C_4}{q^2}\left(-\frac{1}{d}\ptr{d}{O}^2+D^2\ptr{d}{O^2}\right)\right)(\eta^{n-1})\\
	=&\frac{C_4}{q^2}\Bigg[\ptr{d}{O}^2\left(\frac{1}{d^2}-\frac{D\xi\Gamma_{n-1}}{d}-\frac{\eta^{n-1}}{d}\right)\\
&\hspace{8mm}+\ptr{d}{O^2}\left(-\frac{1}{d}+D\xi\Gamma_{n-1}+D^2\eta^{n-1}\right)\Bigg]\\
	=&\frac{C_4}{q^2}\Bigg[-\frac{1}{d}\ptr{d}{O}^2\left(-\frac{1}{d}+D\xi\Gamma_{n-1}+(D^2-D^2+1)\eta^{n-1}\right)\\
&\hspace{8mm}+\ptr{d}{O^2}\left(-\frac{1}{d}+D\xi\Gamma_{n-1}+D^2\eta^{n-1}\right)\Bigg]\\
	=&\frac{C_4}{q^2}\Bigg[\left(\ptr{d}{O^2}-\frac{1}{d}\ptr{d}{O}^2\right)\left(-\frac{1}{d}+D\xi\Gamma_{n-1}+D^2\eta^{n-1}\right)\\
&\hspace{8mm}+\ptr{d}{O}^2\frac{D^2-1}{d}\eta^{n-1}\Bigg]\\
	=&\frac{C_4}{q^2}\Bigg[\epsilon(O)\left(-\frac{1}{d}+D\xi\Gamma_{n-1}+D^2\eta^{n-1}\right)+\ptr{d}{O}^2\frac{D^2-1}{d}\eta^{n-1}\Bigg].
\end{split}
\end{equation}

In the large $n$ limit, this becomes
\begin{equation}
\begin{split}
\mathrm{Var}[\partial_k C]
	=&\epsilon(O)\frac{C_4}{q^2}\left(-\frac{1}{d}+\frac{D\xi}{1-\eta}\right).
\end{split}
\end{equation}

%% file: ProofIdentities.tex
\section{Proof of identities}
\label{Appendix:ProofIdentities}
We prove the identities in appendix~\ref{Sec:Identities}. We first prove the identities in Eq.~\eqref{Eq:Tree}:
\begin{align}\label{Eq:Tree1}

\end{align}
We now prove the identities from Eq.~\eqref{Eq:OTree}:
\begin{align}
	%
\end{align}
In the above, we use the following relations:
\begin{align}
%
\end{align}

We derive the identities in Eq.~\eqref{Eq:ManyTrees}:
\begin{align}
%
\end{align}
In the first line, the second term vanishes by Eq.~\eqref{Eq:Tree3}. The fourth line follows from Eq.~\eqref{Eq:Tree1}. We prove the third identity in Eq.~\eqref{Eq:ManyTrees}:

\begin{align}
%
\end{align}
We prove the fourth identity in Eq.~\eqref{Eq:ManyTrees}:

\begin{align}
%
\end{align}
The first term in the first line vanishes by Eq.~\eqref{Eq:Tree3}.
We prove the second identity in Eq.~\eqref{Eq:ManyTrees}:
\begin{align}
%
\end{align}
The second line follows from Eqs.~\eqref{Eq:Tree2} and \eqref{Eq:Tree4}. The second line is a recursion relation, which produces the sum in the third line. The fourth line follows from an identity for a  geometric series, and the fifth line follows from the definition of $\Gamma_{L}$.

We now prove the identities in Eqs.~\eqref{Eq:ManyOTrees1} to \eqref{Eq:ManyOTrees4}:

\begin{align}
%
\end{align}

Before proving the identities in Eq.~\eqref{Eq:OffU-}, we first prove the following:
\begin{align}
%
\end{align}
Using the above results, we prove the identities in Eq.~\eqref{Eq:OffU-}:
\begin{align}
%
\end{align}

Before proving the identities in Eq.~\eqref{Eq:OffU+}, we prove the following relations:
\begin{align}\label{Eq:U+short1}
%
\end{align}
In the above, we define the unnormalized state $\rho=U_-^\dagger (I_D\otimes \ket{0}\bra{0})U_-$ with a trace of $\tr{\rho}=D$. Using the above, along with the identities in Eq.~\eqref{Eq:Tree}, we prove the identities in Eq.~\eqref{Eq:OffU+}: 

\begin{align}
%
\end{align}

Before proving the identities in Eq.~\eqref{Eq:OffBoth}, we first prove the following:
\begin{align}\label{Eq:G1}
%
\end{align}

Using the above, along with the identities in Eq.~\eqref{Eq:Tree}, we prove the relations in Eq.~\eqref{Eq:OffBoth}:
\begin{align}
%
\end{align}

Before proving the identities in Eq.~\eqref{Eq:OnU-}, we prove the following:

\begin{align}
%
\end{align}
In the above, we define $\sigma=U_+(I_D\otimes O)U_+^\dagger$. Using the above, we prove the identities in Eq.~\eqref{Eq:OnU-}:

\begin{align}
%
\end{align}

Using the relations from Eqs.~\eqref{Eq:U+short1} to \eqref{Eq:U+short4} and the identities from Eq.~\eqref{Eq:OTree}, we prove the identities from Eqs.~\eqref{Eq:OnU+1} to \eqref{Eq:OnU+4}: 

\begin{align}
%
\end{align}

Using the relations from Eqs.~\eqref{Eq:G1} to \eqref{Eq:G4} and the  identities from Eq.~\eqref{Eq:OTree}, we prove the identities from Eq.~\eqref{Eq:OnBoth}:

\begin{align}
%
\end{align}